\documentclass[]{JHEP3}
\usepackage{graphicx, epsfig}

\newcommand{\be}{\begin{equation}}
\newcommand{\ee}{\end{equation}}
\newcommand{\bea}{\begin{eqnarray}}
\newcommand{\eea}{\end{eqnarray}}

\def\cw{c_{\rm w}}

\def\gsim{\gtrsim}

\newcommand{\ptmiss}{p_T\!\!\!\!\!\! /\,\,}

\title{Smoking-gun signatures of little Higgs models}

\author{Tao Han$^{1,2}$, Heather E. Logan$^{1}$ and Lian-Tao Wang$^{1,3}$ \\
$^1$ {\it Department of Physics, University of Wisconsin, Madison, 
Wisconsin 53706 USA}\\
$^2$ {\it Institute of Theoretical Physics, Academia Sinica,
Beijing 100080, China}\\
$^3$ {\it Jefferson Laboratory of Physics, Harvard University,
Cambridge, Massachusetts 02138 USA}\\
E-mail: \email{than@physics.wisc.edu}, \email{logan@physics.wisc.edu},
\email{liantaow@schwinger.harvard.edu}
}

\preprint{
 MADPH--04--1409\\
 hep-ph/0506313}

\abstract{
Little Higgs models predict new gauge bosons, fermions and scalars
at the TeV scale that stabilize the Higgs mass against quadratically 
divergent one-loop radiative corrections.
We categorize the many little Higgs models into two classes based on
the structure of the extended electroweak gauge group and examine
the experimental signatures that identify the little Higgs
\emph{mechanism} in addition to those that identify the particular
little Higgs \emph{model}.
We find that by 
examining the properties of the new heavy fermion(s) at the LHC, one can
distinguish the structure of the top quark mass generation mechanism
and test the little Higgs mechanism in the top sector.
Similarly, by studying the couplings of the new gauge bosons to
the light Higgs boson and to the Standard Model fermions, 
one can confirm the little Higgs mechanism and determine the
structure of the extended electroweak gauge group. 
}

\keywords{Beyond Standard Model, Higgs Physics}

\begin{document}

\section{Introduction}

Elucidating the mechanism of electroweak symmetry breaking (EWSB) is the
central goal of particle physics today.  A full understanding of EWSB
will include a solution to the hierarchy or naturalness problem --
that is, why the weak scale is so much lower than the Planck scale.
Whatever is responsible for EWSB and its hierarchy, it must manifest
experimentally at or below the TeV energy scale.

A wide variety of models have been introduced over the past three
decades to address EWSB and the hierarchy problem: supersymmetry, extra
dimensions, strong dynamics leading to a composite Higgs boson, and
the recent ``little Higgs'' models 
\cite{LHidea,MinMoose,Littlest,SU6Sp6,KS,ChangWacker,
SkibaTerning,ChangSU2,Schmaltznote}
in which the Higgs is a pseudo-Goldstone
boson.  In this paper we consider this last possibility.

In the little Higgs models, the Standard Model (SM) Higgs doublet appears as a
pseudo-Goldstone boson of an approximate global symmetry that is
spontaneously broken at the TeV scale.
The low energy degrees of freedom are described by 
nonlinear sigma models, with a cutoff  at an energy scale one loop factor
above the spontaneous symmetry breaking scale.
Thus the little Higgs models require an ultraviolet (UV) completion 
\cite{HongJian,UVcompletions} at roughly the 10~TeV scale.  

The explicit breaking of the global symmetry, by gauge,
Yukawa and scalar interactions, gives the Higgs a mass and non-derivative
interactions, as required of the SM Higgs doublet.  The little Higgs models are
constructed in such a way that no \emph{single} interaction breaks
\emph{all} of the symmetry forbidding a mass term for the SM Higgs doublet.
This collective symmetry breaking 
guarantees the cancellation of the one-loop quadratically divergent
radiative corrections to the Higgs boson mass.  
Quadratic sensitivity of the Higgs mass to the cutoff scale then 
arises only at the
\emph{two}-loop level, so that a Higgs mass at the 100~GeV scale, two
loop factors below the 10~TeV cutoff, is natural.  Little Higgs models
can thus stabilize the ``little hierarchy'' between the electroweak scale
and the 10 TeV scale at which strongly-coupled new physics is allowed
by electroweak precision constraints.

Little    
Higgs models contain new gauge bosons, a heavy top-like quark, and new
scalars, which cancel the quadratically divergent one-loop 
contributions to the Higgs boson mass from the SM gauge bosons, top quark,
and Higgs self-interaction, respectively.  Thus the ``smoking gun'' feature
of the little Higgs mechanism is the existence of these new gauge
bosons, heavy top-like quark, and new scalars, with the appropriate 
couplings to the Higgs boson to cancel the one-loop quadratic divergence.

Since the little Higgs idea was introduced \cite{LHidea}, many explicit models
\cite{MinMoose,Littlest,SU6Sp6,KS,ChangWacker,
SkibaTerning,ChangSU2,Schmaltznote} have been constructed.  
Since the little Higgs idea could be implemented in a number of ways, 
it is crucial to pick
out the experimental signatures that identify the little Higgs \emph{mechanism}
in addition to those that identify the particular little Higgs \emph{model}.
Detailed phenomenological \cite{Burdman,LHPheno,Peskin}
and experimental \cite{ATLAS,Garcia-talk}
studies of little Higgs physics at the CERN Large Hadron Collider (LHC)
have so far been carried out only within the ``Littlest Higgs'' 
model \cite{Littlest}.\footnote{The LHC phenomenology of the Littlest
Higgs model with $T$-parity \cite{Tparity1,Tparity12,Tparity2} 
was studied in Ref.~\cite{Tparity-pheno};
models with $T$-parity will be briefly discussed in Sec.~\ref{sec:classes}.}
Fortunately, this effort need not be repeated for each of the many little
Higgs models, because the models can be grouped into two classes
that share many phenomenological features, including the crucial ``smoking
gun'' signatures that identify the little Higgs mechanism.

In this paper we categorize the little Higgs models into two classes
based on the structure of the extended electroweak gauge group:
models in which the SM SU(2)$_L$ gauge group arises from the diagonal
breaking of two or more gauge groups, called ``product group'' models
\cite{LHidea,MinMoose,Littlest,SU6Sp6,ChangWacker,ChangSU2},
and models in which the SM SU(2)$_L$ gauge group arises from the
breaking of a single larger gauge group down to an SU(2) subgroup, called
``simple group'' models \cite{KS,SkibaTerning,Schmaltznote}.  
(This categorization and nomenclature was introduced in Ref.~\cite{KS}.)
These two classes of models also exhibit an important difference 
in the implementation of the little Higgs mechanism in the fermion
sector.
As representatives of the two classes, we study the Littlest Higgs model
\cite{Littlest} and the SU(3) simple group model \cite{KS,Schmaltznote},
respectively.
We find that by examining the properties of the new heavy fermion(s),
one can distinguish the structure of the top quark mass generation
mechanism and test the little Higgs mechanism in the top sector.
Furthermore, by measuring the couplings of the new TeV-scale gauge bosons to
the Higgs, SM gauge bosons, and fermions, one can determine the gauge 
structure of the extended theory and test the little Higgs 
mechanism in the gauge sector.  To emphasize the ``smoking gun'' nature
of the signals, we also compare our results with other models that give
rise to similar signatures.  For the heavy top partner, we compare the
little Higgs signatures with the signatures of a fourth generation top-prime
and of the top quark see-saw model.  For the TeV-scale gauge bosons, we 
compare with the $Z^{\prime}$ signatures in $E_6$, left-right symmetric,
and sequential $Z^{\prime}$ models.  In each case, we point out the features
of the little Higgs model that distinguish it from competing interpretations.

The rest of this paper is organized as follows. In the next section we 
describe the basic features of the two representative models.
Specific  little Higgs models that fall into each of the two classes are
surveyed in Appendix~\ref{survey}.
In Sec.~\ref{sec:top}, we discuss the top quark mass generation 
and the quadratic divergence cancellation mechanism 
in the two classes of models, describe the resulting differences in
phenomenology, and show how to test the little Higgs mechanism in the
top sector.  We also comment on the phenomenological differences between
little Higgs models and other models with extended top sectors.
In Sec.~\ref{sec:gauge}, we discuss the gauge sectors in the two classes
of models and identify features common to the models in each class.
We discuss techniques for determining the structure of the extended
gauge sector and for testing the little Higgs mechanism in the 
gauge sector.
In Sec.~\ref{sec:other} we collect some additional features of the 
phenomenology of the SU(3) simple group model.
We conclude in Sec.~\ref{sec:conclusions}.
Technical details of the SU(3) simple group model are given in 
Appendix~\ref{appendixB}.

\section{\label{sec:classes}Two classes of little Higgs models}

If the little Higgs mechanism is realized in nature, 
it will be of ultimate importance to verify it at the LHC, by discovering 
the predicted new particles and determining their specific couplings to
the SM fields that guarantee the cancellation of the Higgs mass quadratic 
divergence.
The most important characteristics of implementations of the little
Higgs idea are ($i$) the structure of the extended gauge symmetry and 
its breaking pattern, and ($ii$) the
treatment of the new heavy fermion sector necessary to cancel the Higgs
mass quadratic divergence coming from the top quark. 
As we will see, the distinctive features of both the gauge and top sectors
of little Higgs models
separate naturally into the product group and simple group classes.

The majority of little Higgs models are product group models.
In addition to the Littlest Higgs, these include
the theory space models (the Big Moose \cite{LHidea} and the Minimal
Moose \cite{MinMoose}), the SU(6)/Sp(6) model of Ref.~\cite{SU6Sp6},
and two extensions of the Littlest Higgs with built-in custodial
SU(2) symmetry \cite{ChangWacker,ChangSU2}.
The product group models have the following generic features.  
First, the models
all contain a set of SU(2) gauge bosons at the TeV scale, obtained from
the diagonal breaking of two or more gauge groups down to SU(2)$_L$, and thus
contain free parameters in the gauge sector from the independent gauge
couplings. Second, since the
collective symmetry breaking in the gauge sector is achieved by
multiple gauged subgroups of the global symmetry, models can be built in which
the SM Higgs doublet is embedded within a single non-linear sigma model
field; many product group models make this simple choice.
Third, the fermion sector of this class of models can usually be chosen to 
be very simple, involving only a single new vector-like quark.

The simplest incarnation of the product group 
class is the so-called Littlest Higgs model
\cite{Littlest}, which we briefly review here. It features a 
[SU(2)$\times$U(1)]$^2$ gauge 
symmetry\footnote{Strictly speaking, it is not necessary to gauge
  two factors of U(1) in order to stablize the little hierarchy, because
  the hypercharge gauge coupling is rather small and does not contribute 
  significantly to the Higgs mass quadratic divergence below a scale of 
  several TeV.  Thus, there is an alternate
  version of the Littlest Higgs model \cite{GrahamEW2}
  in which only SU(2)$^2 \times$U(1)$_Y$ is gauged.}
embedded in an SU(5) global symmetry. The gauge symmetry
is broken by a single vacuum condensate $f \sim$ TeV down to the
SM SU(2)$_L \times$U(1)$_Y$ gauge symmetry. 
The SM Higgs doublet is contained in the resulting
Goldstone bosons, whose interactions are parameterized by a nonlinear
sigma model.  The gauge and Yukawa couplings radiatively 
generate a Higgs potential and trigger EWSB.

The new heavy quark sector in the Littlest Higgs model consists of a 
pair of vectorlike
SU(2)-singlet quarks that couple to the top sector.  The Lagrangian is 
\begin{equation}
	\mathcal{L}_Y = \frac{i}{2} \lambda_1 f \epsilon_{ijk} \epsilon_{xy} 
	\chi_i \Sigma_{jx} \Sigma_{ky} u^{\prime c}_3
	+ \lambda_2 f \tilde{t} \tilde{t}^{\prime c}
	+ {\rm h.c.},
	\label{eq:LYLH}
\end{equation}
where $\chi_i = (b_3, t_3, i \tilde t)$ and the factors of $i$ in 
Eq.~(\ref{eq:LYLH}) and $\chi_i$ are inserted to make the masses
and mixing angles real.  
The summation indices
are $i,j,k=1,2,3$ and $x,y = 4,5$, and $\epsilon_{ijk}$, $\epsilon_{xy}$
are antisymmetric tensors.  The vacuum expectation value (vev) 
$\langle \Sigma \rangle \equiv \Sigma_0$ marries $\tilde t$ to 
a linear combination of $u^{\prime c}_3$ and $\tilde t^{\prime c}$,
giving it a mass of order $f \sim$ TeV.  The resulting new charge 2/3 quark
$T$ is an isospin singlet up to its small mixing with the SM top quark
(generated after EWSB).
The orthogonal linear combination of $u^{\prime c}_3$ and $\tilde t^{\prime c}$
becomes the right-handed top quark and marries $t_3$.  The scalar
interactions of the up-type quarks of the first two generations can be
chosen to take the
same form as Eq.~(\ref{eq:LYLH}), except that there is no need for an 
extra $\tilde t$, $\tilde t^{\prime c}$ since the contribution to the 
Higgs mass quadratic divergence from quarks other than top is numerically 
insignificant below the nonlinear sigma model cutoff 
$\Lambda \sim 4 \pi f \sim 10$ TeV. 

In contrast, the simple group models share two features that distinguish 
them from the product group models.  First, the simple group 
models all contain an
SU($N$)$\times$U(1) gauge symmetry that is broken down to 
SU(2)$_L \times$U(1)$_Y$, yielding a set of TeV-scale gauge bosons.
The two gauge couplings of the SU($N$)$\times$U(1) are 
fixed in terms of the two SM SU(2)$_L \times$U(1)$_Y$ gauge couplings, 
leaving no free parameters in the gauge sector once the symmetry-breaking
scale is fixed.
This gauge structure also forbids mixing between the SM $W^{\pm}$ bosons 
and the TeV-scale gauge bosons, again in contrast to the product group models.
Second, in order to implement the collective symmetry breaking, simple-group
models require at least two sigma-model multiplets.  The SM Higgs doublet is
embedded as a linear combination of the Goldstone bosons from these
multiplets.  This introduces at least one additional model parameter, which 
can be chosen as the ratio of the vevs of the sigma-model multiplets.
Moreover, due to the enlarged SU($N$) gauge symmetry, all
SM fermion representations have to be extended to transform as 
fundamental (or antifundamental) representations of SU($N$), giving rise
to additional heavy fermions in all three generations.
The existence of multiple sigma-model multiplets generically results in a 
more complicated structure for the fermion couplings to scalars.  On the 
other hand, the existence of heavy fermion states in all three generations as
required by the enlarged gauge symmetry provides extra
experimental observables that in principle allow one to disentangle this
more complicated structure.

The simplest incarnation of the simple group class is the SU(3) simple
group model \cite{KS,Schmaltznote}.  We briefly review its
construction here; additional details are presented in
Appendix~\ref{appendixB}.  The electroweak gauge structure is
SU(3)$\times$U(1)$_X$.  There are two sigma-model fields, $\Phi_1$ and
$\Phi_2$, transforming as $\mathbf{3}$s under SU(3).  Vacuum
condensates $\langle \Phi_{1,2} \rangle=(0,0,f_{1,2})^T$ 
break SU(3)$\times$U(1)$_X$
down to the SM SU(2)$_L \times$U(1)$_Y$.  The TeV-scale gauge sector consists
of an SU(2)$_L$ doublet $(Y^0, X^-)$ of gauge bosons corresponding to
the broken off-diagonal generators of SU(3), and a $Z^{\prime}$ gauge boson
corresponding to the broken linear combination of the $T^8$ generator 
of SU(3) and the U(1)$_X$.  The model also contains a singlet 
pseudoscalar $\eta$.

The top quark mass is generated by the Lagrangian
\begin{equation}
	\mathcal{L}_Y = i \lambda_1^t u_1^c \Phi_1^{\dagger} Q_3
	+ i \lambda_2^t u_2^c \Phi_2^{\dagger} Q_3,
	\label{eq:LYSU3}
\end{equation}
where $Q_3^T = (t, b, i T)$ and the factors of $i$ in Eq.~(\ref{eq:LYSU3})
and $Q_3$ are again inserted to make the masses and mixing angles real.  The 
$\Phi$ vevs marry $T$ to a linear combination of $u_1^c$ and $u_2^c$,
giving it a mass of order $f \sim$ TeV.  The new charge 2/3 quark $T$ is
a singlet under SU(2)$_L$ up to its small mixing with the SM top quark
(generated after EWSB).  The orthogonal linear
combination of $u_1^c$ and $u_2^c$ becomes the right-handed top
quark. For the rest of the quarks, the scalar interactions depend on the choice
of their embedding into SU(3).  The most straightforward choice is to
embed all three generations in a universal way, $Q_m^T = (u, d, iU)_m$,
so that each quark generation contains a new heavy charge 2/3 quark.
This embedding leaves the SU(3) and U(1)$_X$ gauge groups anomalous;
the anomalies can be canceled by adding new spectator fermions at the
cutoff scale $\Lambda \sim 4 \pi f$.  An alternate, anomaly-free embedding
\cite{Kong} puts the quarks of the first two generations into 
antifundamentals of SU(3), $Q_m^T = (d, -u, iD)_m$, with $m=1,2$,
so that the first two quark generations each contain a new heavy charge
$-1/3$ quark. Interestingly, an anomaly-free embedding of the SM fermions into 
SU(3)$_c \times$SU(3)$\times$U(1)$_X$ is only possible if the number
of generations is a multiple of three \cite{Kong,331}.\footnote{This 
rule can be violated in models containing fermion generations with 
non-SM quantum numbers, e.g., mirror families \cite{Rodolfo}.}

Electroweak precision observables provide strong constraints on any
extensions of the SM.  The constraints on the little Higgs 
models have been studied extensively 
\cite{GrahamEW2,GrahamEW1,JoAnneEW,MuChun,Barbieri:2004qk,Marandella:2005wd,SkibaEW}.
Of course, any phenomenological study of a particular model must take 
these constraints into account.  However, in this paper we 
study the generic phenomenology of classes of little 
Higgs models, using specific models only as prototypes.
We focus on features of the phenomenology that are expected to persist 
in all models within a given class, in spite of variations in the model that 
can give rise to very different constraints from electroweak precision 
observables.
For exmaple, variations of the model that improve the electroweak fit 
will not in general change the generic features of the new heavy top-partner 
phenomenology.  Thus, in order to maintain applicability to a wide range 
of models in each class, we will not limit our presentation of results 
to the parameter space allowed by electroweak precision fits in the 
specific models under consideration.

For completeness, we now briefly summarize the results of electroweak
precision fits in the models under consideration. 
The most up-to-date
studies are Refs.~\cite{Barbieri:2004qk,Marandella:2005wd,SkibaEW}, 
which include LEP-2 data above the $Z$ pole.
In most little Higgs models, particularly the product group models,
the electroweak data mostly set lower bounds on
the masses of the heavy vector bosons due to their contributions to four-Fermi
operators and their mixing with the $W$ and $Z$ bosons.
On the other hand, the most important contributions to the Higgs mass
quadratic divergence cancellation come from the top quark partner $T$,
which should be as light as possible to minimize the fine-tuning.
These competing desires dictate the favored parameter regions of the
little Higgs models.

\begin{itemize}

\item {\it Littlest Higgs model:}
The Littlest Higgs model with [SU(2)$\times$U(1)]$^2$ gauged
contains a new U(1) boson, $A_H$, which is relatively light and
tends to give rise to large corrections to electroweak precision 
observables.  Assigning the fermions to transform under SU(2)$_1$
and U(1)$_1$ only, Ref.~\cite{Marandella:2005wd} finds a stringent constraint
$f \geq 5$ TeV.  However, allowing the fermions to transform under 
both U(1) groups (as required in order to write down gauge invariant Yukawa
couplings in a straightforward way) tends to reduce this constraint;
Refs.~\cite{GrahamEW2,GrahamEW1}, which do not include LEP-2 data in
their fit, found the constraint on $f$ reduced from 4 TeV to about 1 TeV;
similarly, Ref.~\cite{Marandella:2005wd} found the constraint reduced from
5 TeV to about 2--3 TeV. 
Gauging only SU(2)$^2 \times$U(1)$_Y$, Ref.~\cite{Barbieri:2004qk}
found that $f > {\rm max}(6.5 c^2,3.7 c) $ TeV [$c$ is defined below
Eq.~(\ref{eq:gaugemasses})].
Thus, for example, $f > 1$ TeV for $c \sim 1/3$; this yields a 
lower bound on the heavy gauge boson mass of $M_{W_H} = M_{Z_H} \geq 2$ TeV.
The mass of the $T$ quark is constrained to be $M_T \geq \sqrt{2} f$, 
or in this most favorable case $M_T \geq 1.4$ TeV.

\item {\it SU(3) simple group model:}
Reference~\cite{SkibaEW} expands on the analysis of 
Ref.~\cite{Marandella:2005wd} for this model by including the effect
of the TeV-scale fermions in the universal fermion embedding.  For our
choice of parameterization, the constraint on 
$f \equiv \sqrt{f_1^2 + f_2^2}$ is relaxed by going to 
$t_{\beta} \equiv f_2/f_1 > 1$ \cite{Skibaemail}.  For $t_{\beta} = 3$, 
$f \geq 3.9$ TeV \cite{Skibaemail}, corresponding to 
$M_{Z^{\prime}} \geq 2.2$ TeV.  The mass of the $T$ quark in this model
is bounded by $M_T \geq f \sin 2\beta$; this constraint then
translates into $M_T \geq 2.3$ TeV.
Reference~\cite{Schmaltznote} found that the anomaly-free fermion embedding 
is somewhat favored over the universal embedding by electroweak precision 
constraints.

\end{itemize}

Finally, we mention briefly a different approach to alleviating the
electroweak precision constraints on little Higgs models.  Because the
little Higgs mechanism for canceling the quadratically divergent 
radiative corrections to the Higgs mass operates at one-loop, 
it is possible to impose
an additional symmetry, dubbed $T$-parity \cite{Tparity1,Tparity12,Tparity2},
under which the new gauge bosons and scalars are odd.  This eliminates
tree-level contributions of the new particles to electroweak precision 
observables, thereby essentially eliminating the electroweak precision 
constraints\footnote{Although $T$-parity suppresses the contributions of
  heavy gauge bosons and heavy top partners to electroweak oblique
  parameters, there
 is a contribution to four fermion operators through a box diagram
 involving mirror fermions and
 Goldstone bosons that is not suppressed by the same mechanism and 
 does not decouple as the mirror fermions become heavy.  The mirror
 fermions must be kept light (i.e., be introduced into the low energy 
 spectrum) in order to
 suppress the relevant couplings \cite{Tparity12,Tparity-pheno}.}.
It also changes the collider phenomenology drastically, 
by eliminating signals from single production of the new particles that 
are odd under $T$-parity: in particular, the heavy gauge bosons can only be 
produced in pairs, eliminating the distinctive Drell-Yan signal.
The heavy top-partners remain even under $T$-parity, however, so that
their signals are robust.
It was shown in Ref.~\cite{Tparity2} how to add 
$T$-parity to any product group little Higgs model.  Ref.~\cite{Tparity2}
also concluded that in simple group models, one cannot find a consistent
definition of $T$-parity under which all heavy gauge bosons are odd.


\section{The heavy quark sector}
\label{sec:top}

The SM top quark gives rise to the largest quadratically divergent 
correction to the Higgs mass. A characteristic
feature of all little Higgs models is the existence of new TeV-scale
quark state(s) with specific couplings to the Higgs so that the loops
involving the TeV-scale quark(s) cancel the quadratic divergence from 
the SM top quark loop. Therefore, we begin with a study of the extended
top sector of little Higgs models. 

\subsection{Top sector masses and parameters}
\label{sec:topdiv}

The masses of the top quark $t$ and its heavy partner $T$ are given
in terms of the model parameters by
\begin{eqnarray}
\nonumber
&& m_t  =  \lambda_t v  =  \left\{  
\begin{array}{ll} \displaystyle
	\frac{\lambda_1 \lambda_2 }{\sqrt{\lambda_1^2 + \lambda_2^2}} v
	& \qquad  {\rm  in \ the \ Littlest\ Higgs\ model}, \\  [.45cm]  
\displaystyle
\frac{\lambda_1 \lambda_2 }
	{\sqrt{2}\sqrt{\lambda_1^2 c_{\beta}^2 + \lambda_2^2 s_{\beta}^2}} v
	& \qquad   {\rm  in\  the\  SU(3) \ simple\  group\ model};
\end{array}
\right.  \\ [.35cm]  
\nonumber
&& M_T  =  \left\{  
\begin{array}{ll} \displaystyle
  \sqrt{\lambda_1^2 + \lambda_2^2}\ f 
  = (x_\lambda + x_\lambda^{-1}){m_t\over v}f
	& \qquad   {\rm  in\  the\  Littlest\ Higgs\ model}, \\ [.35cm]  
\displaystyle
 \sqrt{\lambda_1^2 c_{\beta}^2 + \lambda_2^2 s_{\beta}^2} f
 = \sqrt{2}\  \frac{t_{\beta}^2 + x_{\lambda}^2}{(1 + t_{\beta}^2) x_{\lambda}}
  \frac{m_t}{v} \ f 
	& \qquad   {\rm  in\  the\  SU(3) \ simple\  group\ model}.
\end{array}
\right.
\end{eqnarray}
Fixing the top quark mass $m_t$ leaves two free parameters in the Littlest 
Higgs model, 
which can be chosen to be $f$ and $x_{\lambda} \equiv \lambda_1/\lambda_2$.
We see that the SU(3) simple group model contains one additional 
parameter, $t_\beta \equiv \tan \beta =f_2/f_1$.  In the SU(3) 
simple group model, we define $f \equiv \sqrt{f_1^2 + f_2^2}$.

To reduce fine-tuning in the Higgs mass, the top-partner $T$ should be
as light as possible.  The lower bound on $M_T$ is obtained for certain 
parameter choices:
\begin{eqnarray}
	\nonumber
	M_T  \geq   \left\{  
\begin{array}{ll} \displaystyle
 2 {m_t\over v}f \approx \sqrt 2 f
& \qquad   {\rm for}\ x_{\lambda} = 1\ 
	{\rm  in\  the\  Littlest\ Higgs\ model}, \\ [.35cm]  
\displaystyle
 2 \sqrt{2}  s_{\beta} c_{\beta} \frac{m_t}{v} f \approx f \sin 2 \beta
& \qquad   {\rm for}\ x_{\lambda} = t_{\beta}\   
	{\rm  in\  the\  SU(3)\ simple \ group \ model},
	\end{array}
	\right.
\end{eqnarray}
where in the last step we used $m_t/v \approx 1/\sqrt{2}$.  The $T$ mass
can be lowered in the SU(3) model for fixed $f$ by choosing 
$t_{\beta} \neq 1$, 
thereby introducing a mild hierarchy between $f_1$ and $f_2$.  With our
parameter definitions, the choice $t_{\beta} > 1$ reduces the mixing 
between the light SM fermions and their TeV-scale partners, thereby 
reducing constraints from $W$ coupling universality.

\subsection{Heavy $T$ couplings to Higgs and gauge bosons}

The couplings of the Higgs doublet to the $t$ and $T$ mass eigenstates
can be written in terms of an effective Lagrangian,
\begin{equation}
	\mathcal{L}_Y \supset \lambda_t H t^c t + 
	 \lambda_T H T^c t +
	  \frac{\lambda_T^{\prime}}{2 M_T} 
	H H T^c T + {\rm h.c.},
	\label{eq:lambdaTLH}
\end{equation}
where the four-point coupling arises from the expansion of the nonlinear
sigma model field.
This effective Lagrangian leads to three diagrams contributing to the
Higgs mass corrections at one-loop level, shown in Fig.~\ref{fig:toploops}: 
(a) the SM top quark diagram, which depends on the well-known SM top Yukawa 
coupling $\lambda_t$; 
(b) the diagram involving a top quark and a top-partner $T$, which
depends on the $HTt$ coupling $\lambda_T$; and (c) the diagram involving 
a $T$ loop coupled to the Higgs doublet via the dimension-five $HHTT$ 
coupling.
The couplings in the three diagrams of Fig.~\ref{fig:toploops} must 
satisfy the following relation \cite{Peskin} in order for the quadratic 
divergences to cancel:
\begin{equation}
	\lambda_T^{\prime} = \lambda_t^2 + \lambda_T^2.
	\label{eq:topdivrel}
\end{equation}
This equation embodies the cancellation of the Higgs mass quadratic divergence
in any little Higgs theory.  It is of course satisfied by the couplings 
in both the Littlest Higgs and the SU(3) simple group models, as can be seen 
by plugging in the explicit couplings given in
Table \ref{table-T}.  Note that in the SU(3) simple group model,
$\lambda_T$ vanishes when $x_\lambda=1$.
If the little Higgs mechanism is realized in nature,
it will be of fundamental importance to establish the relation in 
Eq.~(\ref{eq:topdivrel}) experimentally.

\FIGURE{
\includegraphics{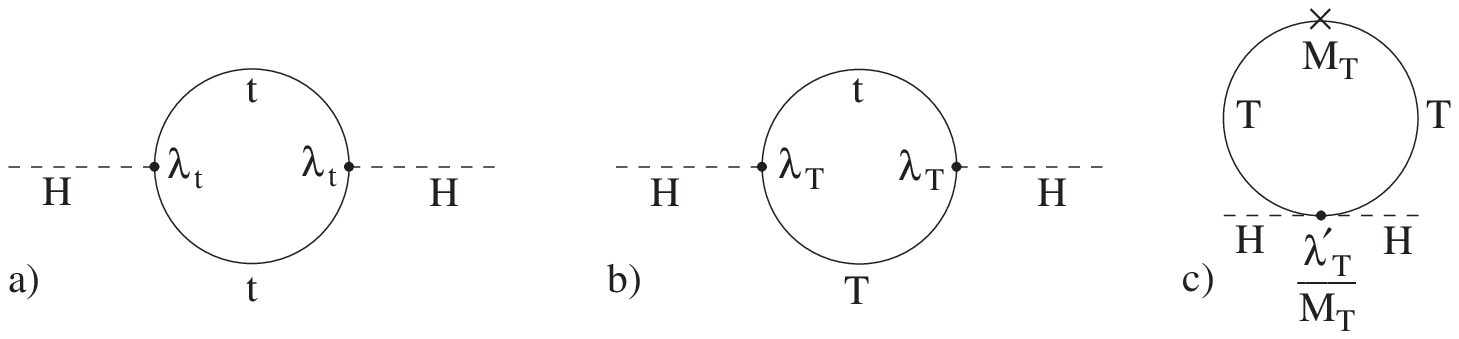}
\caption{\label{fig:toploops} 
Quadratically divergent one-loop contributions to the Higgs
boson mass-squared from the top sector in little Higgs models.}
}

\TABLE{
\begin{tabular}{|c|c|c|}
\hline \hline 
          & Littlest Higgs     & SU(3) simple group  \\
\hline
$\lambda_t = $  & $  m_t/v =
\frac{\lambda_1 \lambda_2 }{\sqrt{\lambda_1^2 + \lambda_2^2}}$ &  
$ m_t/v =
\frac{\lambda_1 \lambda_2 }
	{\sqrt{2}\sqrt{\lambda_1^2 c_{\beta}^2 + \lambda_2^2 s_{\beta}^2}} $ \\
$\lambda_T =$ & $x_{\lambda} {m_t}/{v} $   &   
 $s_{\beta}c_{\beta}(x_{\lambda} - x_{\lambda}^{-1})  {m_t}/{v}$ \\
$\lambda_T^{\prime}= $ & $ ( x_{\lambda}^2 + 1 ) {m_t^2}/{v^2}$   &  
$ \left[ s_{\beta}^2 c_{\beta}^2 
(x_{\lambda} - x_{\lambda}^{-1})^2 + 1 \right] {m_t^2}/{v^2} $ \\
\hline
 $H t_R \bar t_L$: &  $i \lambda_t $  &  $i \lambda_t $  \\
 $ H T_R \bar t_L $: & $ i \lambda_T$  & $ i \lambda_T$   \\
 $H H T_R \bar T_L$: & $i \lambda_T^{\prime}/M_T$ 
 & $i \lambda_T^{\prime}/M_T$  \\
\hline
$W^+_{\mu} \overline{T} b: \  i\delta_T \frac{ g }{\sqrt{2} } 
\gamma_{\mu} P_L$; \  $\delta_T =$ & $\lambda_T{v / M_T}= x_\lambda m_t/M_T$   
&  $ \lambda_T{v / M_T}=s_{\beta}c_{\beta}(x_{\lambda} - x_{\lambda}^{-1}) 
	m_t/M_T $   \\
$Z_{\mu} \overline{T} t: \  i\delta_T \frac{ g }{ 2\cw } 
\gamma_{\mu} P_L$; \  $\delta_T =$ & same as above & 
 same as above \\
\hline \hline
\end{tabular}
\caption{\label{table-T} Heavy $T$ couplings and Feynman rules in the 
Littlest Higgs and SU(3) simple group models.}
}

After EWSB, the coupling $\lambda_T$
induces a small mixing of electroweak doublet into $T$,
\begin{equation}
	T = T_0 - \delta_T  t_0,\quad \delta_T = \lambda_T {v \over M_T},
	\label{eq:Tmixing}
\end{equation}
where $T_0,t_0$ stand for the electroweak eigenstates before the
mass diagonalization at the order of $v/f$.
This mixing gives rise to the couplings of $T$ to the SM states
$bW$ and $tZ$ with the same form as the corresponding SM couplings of the
top quark except suppressed by the mixing factor $\delta_T$. The Feynman 
rules are given in Table~\ref{table-T}.

\subsection{Additional heavy quark couplings in the SU(3) simple group model}
\label{sec:SU3fermions}

Expanding the SU(2)$_L$ gauge symmetry to SU(3) forces the introduction 
of a heavy partner associated with each SU(2)$_L$ fermion doublet of the SM.
The first two generations of quarks are therefore enlarged to contain two
new TeV-scale quarks $Q_{1,2}$.  
We consider both the universal and the anomaly-free
fermion embeddings, as discussed in more detail in 
Sec.~\ref{sec:fermionappendix}.  The universal
embedding gives rise to two charge 2/3 quarks, $U$ and $C$, while the
anomaly-free embedding gives rise to two charge $-1/3$ quarks, $D$ and $S$.  

The masses of the two heavy quarks $Q_{1,2}$ are given, for either fermion 
embedding, by 
\begin{equation}
	M_{Q_m} = s_{\beta} \lambda_{Q_m} f  \qquad \qquad (m=1,2),
	\label{eq:MQ}
\end{equation}
where we have neglected the masses of the quarks of the 
first two generations and chosen $\lambda_{Q_m}$ to be the Yukawa 
coupling involving $\Phi_2$ (see Sec.~\ref{AFmq} and \ref{UNmq} for further
details). 
The heavy quark couplings to the Higgs boson are proportional
to the Yukawa couplings $\lambda_{Q_m}$ as expected, and can be rewritten 
in terms of the heavy quark mass $M_Q$ (see Table \ref{table-Q}).

\TABLE{
\begin{tabular}{|c|c|}
\hline \hline
          &  SU(3) simple group  \\
\hline
 $H \overline{U_R} u_L$: &  $ -i c_\beta \lambda_U  /  \sqrt{2} = 
  - i{M_{U}}/{\sqrt{2} f t_{\beta}} $ \\
$W^+_{\mu} \overline{U} d:$ & $ i\delta_\nu \frac{ g }{\sqrt{2} } 
\gamma_{\mu} P_L$  \\
$Z_{\mu} \overline{U} u: $  &  $i\delta_\nu \frac{ g }{ 2\cw } 
\gamma_{\mu} P_L$ \\
\hline
 $H \overline{D_R} d_L$: & $ i c_\beta \lambda_D  /  \sqrt{2} = 
   i{M_{D}}/{\sqrt{2} f t_{\beta}} $ \\
$W^-_{\mu} \overline{D} u:$  &  $-i\delta_\nu \frac{ g }{\sqrt{2} } 
\gamma_{\mu} P_L$ \\
$Z_{\mu} \overline{D} d:$ &  $-i\delta_\nu \frac{ g }{ 2\cw } 
\gamma_{\mu} P_L$ \\
\hline
$X^-_{\mu} \overline{b} T:$   & $ \frac{g}{\sqrt{2}} \gamma_{\mu} P_L$ \\
$Y^0_{\mu} \overline{t} T:$   & $ \frac{g}{\sqrt{2}} \gamma_{\mu} P_L$ \\
\hline
$\eta  \overline{t} T:$   & $ -m_t/v  P_L$ \\
\hline
$X^-_{\mu} \overline{d} U:$  & $ \frac{g}{\sqrt{2}} \gamma_{\mu} P_L $ \\
$Y^0_{\mu} \overline{u} U:$  & $ \frac{g}{\sqrt{2}} \gamma_{\mu} P_L $ \\
$X^-_{\mu} \overline{D} u:$  & $-\frac{g}{\sqrt{2}} \gamma_{\mu} P_L $ \\
$Y^0_{\mu} \overline{D} d:$  & $-\frac{g}{\sqrt{2}} \gamma_{\mu} P_L $ \\
$X^-_{\mu} \overline{e} N:$  & $ \frac{g}{\sqrt{2}} \gamma_{\mu} P_L $ \\
$Y^0_{\mu} \overline{\nu} N:$ & $ \frac{g}{\sqrt{2}} \gamma_{\mu} P_L $ \\
\hline
$Z^{\prime} \overline{T} T:$ & $-\frac{ig}{c_W \sqrt{3-4s_W^2}}
		[(-1 + \frac{5}{3}s_W^2)P_L + \frac{2}{3}s_W^2P_R]$ \\
$Z^{\prime} \overline{U} U:$ & $-\frac{ig}{c_W \sqrt{3-4s_W^2}}
		[(-1 + \frac{5}{3}s_W^2)P_L + \frac{2}{3}s_W^2P_R]$ \\
$Z^{\prime} \overline{D} D:$ & $-\frac{ig}{c_W \sqrt{3-4s_W^2}}
		[(-1 + \frac{5}{3}s_W^2)P_L - \frac{1}{3}s_W^2P_R]$ \\
$Z^{\prime} \overline{N} N:$ & $-\frac{ig}{c_W \sqrt{3-4s_W^2}}
		(-1 + s_W^2)P_L$ \\
\hline \hline
\end{tabular}
\caption{\label{table-Q}
Feynman rules for $T$ and $Q$ in the SU(3) simple group model.
Note that $U=U,C$ in the universal embedding and $D=D,S$ in the
anomaly-free embedding.  $\delta_{\nu}$ is defined in Eq.~(\ref{eq:deltaq}).
The extra $i$s in the couplings of $X,Y$ are due to our phase choice.}
}

After EWSB, the Yukawa couplings $\lambda_{Q_m}$ lead to mixing 
between the heavy quarks $Q$ and the corresponding SM quarks of like 
charge given by 
$Q = Q_0 - \delta_q q_0$, where as usual $Q_0,q_0$ denote the electroweak
eigenstates of each generation.  The mixing angle $\delta_q$ is given
to order $v/f$ by
\begin{equation}
	\delta_q = \pm \frac{v}{\sqrt{2} f t_{\beta}}
	\equiv \mp \delta_{\nu},
	\label{eq:deltaq}
\end{equation}
where the upper sign is for the anomaly-free embedding ($Q = D,S$)
and the lower sign is for the universal embedding ($Q = U,C$).

The mixing between SM quarks and their heavy counterparts causes isospin 
violation at order $\delta_{\nu}^2$ in processes involving only SM 
fermions.  This isospin violation can be suppressed 
by choosing $t_{\beta} \gsim 1$.
As in the top sector, the mixing due to $\delta_q$ gives rise to 
the couplings of $Q$ to $q^{\prime}W$ and $q Z$; the Feynman rules
are given in Table~\ref{table-Q}.

Although the new heavy quarks $Q_{1,2}$ of the first two generations
do not play a significant role in the cancellation of the Higgs mass 
quadratic divergence (they take part in the cancellation of the numerically
insignificant Higgs mass quadratic divergence from their SM partners in the
first two generations), they share the common 
parameters $f$ and $t_\beta$ with the top sector, providing additional
experimental observables that can be used to test the little Higgs
structure of the couplings.  The new heavy quarks of the first two
generations introduce two further parameters, which can be chosen as
their masses $M_{Q_m}$ or equivalently their Yukawa couplings
$\lambda_{Q_m}$, as related by Eq.~(\ref{eq:MQ}).
The couplings between the new heavy quarks and the TeV-scale gauge bosons are
fixed by the gauge symmetry; they are summarized in Table
\ref{table-Q}. We will not comment on them further here since they
will not play a significant role in our phenomenological analysis.

\subsection{Heavy quark production and decay at the LHC}

\subsubsection{$T$ production and decay}
\label{sec:Tprod}

The top-partner $T$ can be pair-produced via QCD interactions
at the LHC; however,
because the final state contains two heavy particles, the pair-production 
cross section falls quickly with increasing $M_T$.
Instead, single $T$ production via $Wb$ fusion yields a larger cross 
section in both the Littlest Higgs model and the SU(3) simple group model,
as shown in Figs.~\ref{fig:TLH} and \ref{fig:TSU3}, respectively.

\FIGURE{
\resizebox{0.7\textwidth}{!}{\includegraphics{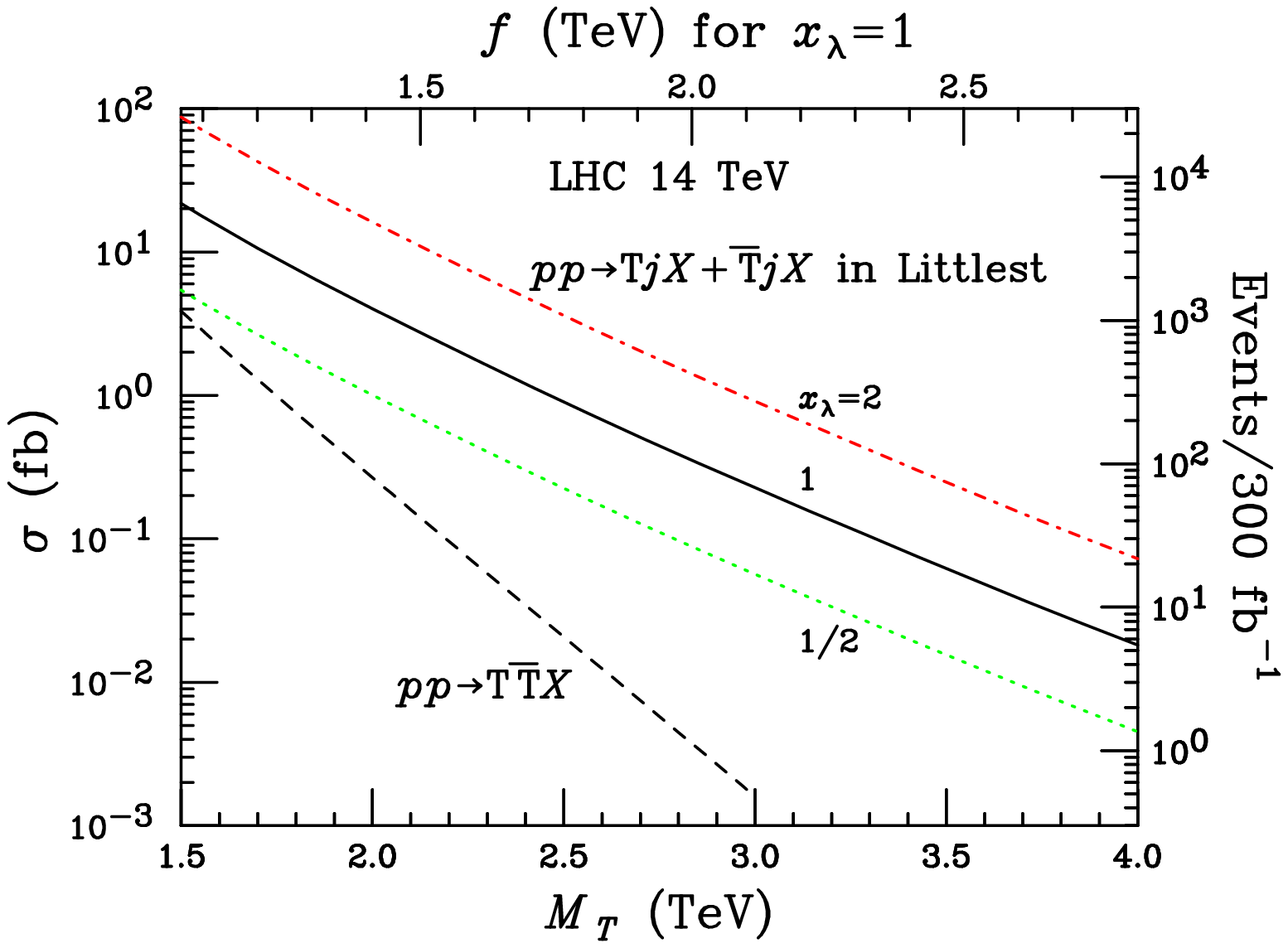}}
\caption{\label{fig:TLH}
Production cross sections for $T$ in the Littlest Higgs model.
The top axis shows the value of $f$ corresponding to $M_T$ for 
$x_{\lambda} = 1$.}
}

\FIGURE{
\resizebox{0.7\textwidth}{!}{\includegraphics{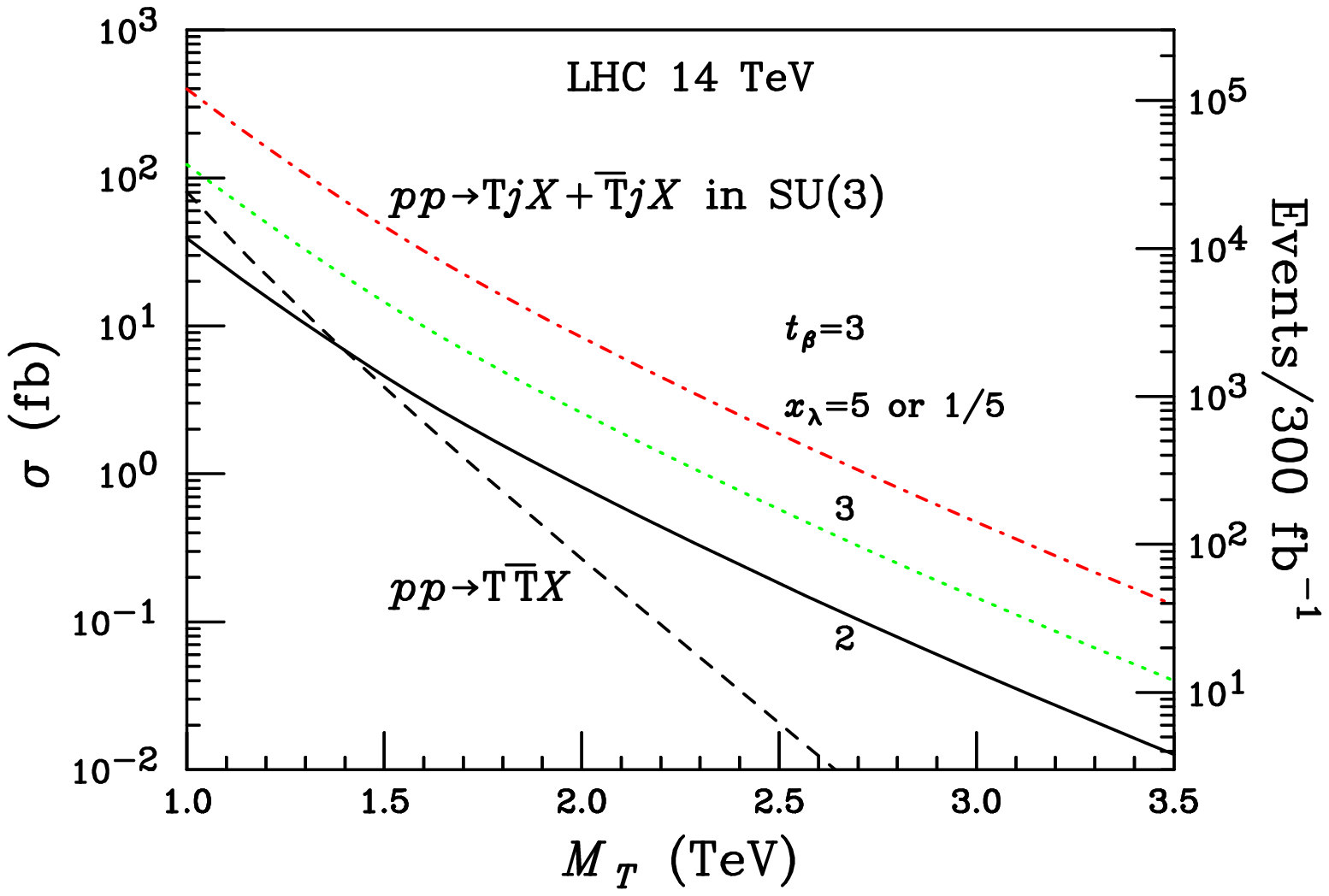}}
\caption{\label{fig:TSU3}
Production cross sections for $T$ in the SU(3) simple group model.
Single $T$ production is plotted for $t_{\beta} = 3$ and various values
of $x_{\lambda}$.  The single $T$ production cross section is invariant under 
$x_{\lambda} \to 1/x_{\lambda}$ and vanishes at $x_{\lambda} = 1$.}
}

In the Littlest Higgs model, the single $T$ production cross section 
at fixed $M_T$ depends on only one model parameter, $x_{\lambda}$,
as shown in Fig.~\ref{fig:TLH}.
In particular, the cross section is proportional to $x_{\lambda}^2$,
as can be seen by examining the $W^+ \overline{T} b$ coupling in 
Table~\ref{table-T} while holding $M_T$ fixed.  
We see that the cross section is typically in the range 0.01--100 fb 
for $M_T = 1.5$--3.5 TeV.

In the SU(3) simple group model, the single $T$ production cross section
at fixed $M_T$ depends on two model parameters, $x_{\lambda}$ and $t_{\beta}$.
From the $W^+ \overline{T} b$ coupling in Table~\ref{table-T} one can see 
that at fixed $M_T$, the cross section scales with $\lambda_T^2$:
\begin{equation}
	\sigma \propto \lambda_T^2 \propto s^2_{\beta}c^2_{\beta} 
	(x_{\lambda}-x_\lambda^{-1})^2.
\end{equation}
The cross section is invariant under $t_{\beta} \leftrightarrow 1/t_{\beta}$
and under $x_{\lambda} \leftrightarrow 1/x_{\lambda}$.  
It reaches a maximum at $t_\beta = 1$, and vanishes at $x_{\lambda} = 1$.
Away from unity, 
it falls like $t_\beta^{-2}\ (t_\beta^{2})$ for large (small) $t_\beta$, 
and grows like $x_{\lambda}^2\  (x_{\lambda}^{-2})$ for large 
(small) $x_{\lambda}$.
The cross section is shown in Fig.~\ref{fig:TSU3} for $t_{\beta} = 3$
and various values of $x_{\lambda}$.
We see that the cross section is similar in size to that in the Littlest
Higgs model, depending on the parameter values in either model.

The dominant decay modes of $T$ in all little Higgs models
are $tH$, $tZ$ and $bW$.  The partial widths of $T$ to these 
final states are all controlled by the same coupling $\lambda_T$,
\begin{equation}
	\Gamma(T \to tH) = \Gamma(T \to tZ) = \frac{1}{2} \Gamma(T \to bW)
	= \frac{\lambda_T^2}{32 \pi} M_T
	= 9.9 \lambda_T^2 \left( \frac{M_T}{\rm TeV} \right) {\rm GeV},
	\label{eq:Tpartialwidths}
\end{equation}
where we neglect final-state masses compared to $M_T$.
If these are the only decays of $T$, then its total width is 
$40 \lambda_T^2(M_T/{\rm TeV})$ GeV.
The branching fractions of $T$ into these final states are then given by
\begin{equation}
	{\rm BR}(T \to tH) = {\rm BR}(T \to tZ) = 1/4,
	\qquad {\rm BR}(T \to bW) = 1/2.
	\label{eq:Tbrs}
\end{equation}
This simple relation between the branching fractions is easily understood 
in terms of the Goldstone boson
equivalence theorem: the decay modes at high energies (large $M_T$)
are just those into the four components of the SM Higgs doublet, i.e., the 
three Goldstone degrees of freedom and the physical Higgs boson.

Phenomenological studies of these $T$ decays have been performed at the level
of somewhat realistic detector simulations in Ref.~\cite{ATLAS}.
The $T$ mass can be reconstructed from each of these three channels; 
$T \to Zt \to \ell^+ \ell^- b \ell \not \! \! E_T$ 
provides the cleanest mass peak \cite{ATLAS}.

If the only significant decays of $T$ are into $tH$, $tZ$ and $bW$,
then the branching fractions of $T$ are predicted independent of any
model parameters by Eq.~(\ref{eq:Tbrs}).
A measurement of the rate for single $T$ production
with decays into any one of the three final states is sufficient to determine
the production cross section, and thus extract $\lambda_T$.
The measurement of the characteristic pattern of branching fractions 
also provides a test of the model (see Sec.~\ref{sec:top4thgen}).

In the SU(3) simple group model, $T$ has additional possible decay modes
due to the additional particles in the spectrum.  In particular, $T$
can also decay to $t \eta$, $t Y^0$, and $b X^+$ final states, depending 
on the relative masses of $T$, $\eta$, and $X,Y$.
In order to measure the single $T$ production cross section, and hence
$\lambda_T$, one needs to know the branching fraction(s) of the decay 
mode(s) in which $T$ is observed.  Assuming the SU(3) simple group model
structure, these can be predicted as follows.  
The $T$ mass can be reconstructed in, e.g., 
$T \to Zt \to \ell^+ \ell^- b \ell \not \! \! E_T$ as discussed above.
The $X,Y$ gauge boson masses are fixed in terms of $M_{Z^{\prime}}$, 
which will be easily measurable from its decays to dileptons 
(see Sec.~\ref{sec:gauge}).  The $T$ partial widths to $tY$ and $bX$ 
can then be calculated in terms of the gauge couplings in Table~\ref{table-Q}.
The $T$ partial width to $\eta$ can be calculated from the coupling
in Table~\ref{table-Q}
once the $\eta$ mass is measured, e.g., in decays of $\eta$ to dijets.
The partial widths to $tH$, $tZ$ and $bW$ are proportional to 
$\lambda_T^2$; thus the only remaining free parameter to be extracted
from the rate measurement in any given final state is $\lambda_T$.
Measurements of the pattern of branching fractions then provide a 
nontrivial test of the model.
Similarly, in the Littlest Higgs model with two U(1) groups gauged, $T$
can decay into $t A_H$.  Once the $A_H$ mass is measured, a similar analysis
can be applied.

\subsubsection{$Q$ production and decay}
\label{sec:Qprod}

The heavy quarks $Q$ in the SU(3) simple group model can be produced 
at the LHC via, e.g., $Wd \to U$, $Zu \to U$.
The production couplings are given in Table~\ref{table-Q}; for fixed
$M_Q$, the cross section depends on only one model parameter, $\delta_{\nu}$;
in particular the cross section is proportional to 
$\delta_{\nu}^2 = v^2 / 2 f^2 t_{\beta}^2$.  The single production cross
section for $U + \overline{U}$ is shown in Fig.~\ref{fig:USU3}, together
with the $U \overline{U}$ pair production cross section from QCD.

\FIGURE{
\resizebox{0.7\textwidth}{!}{\includegraphics{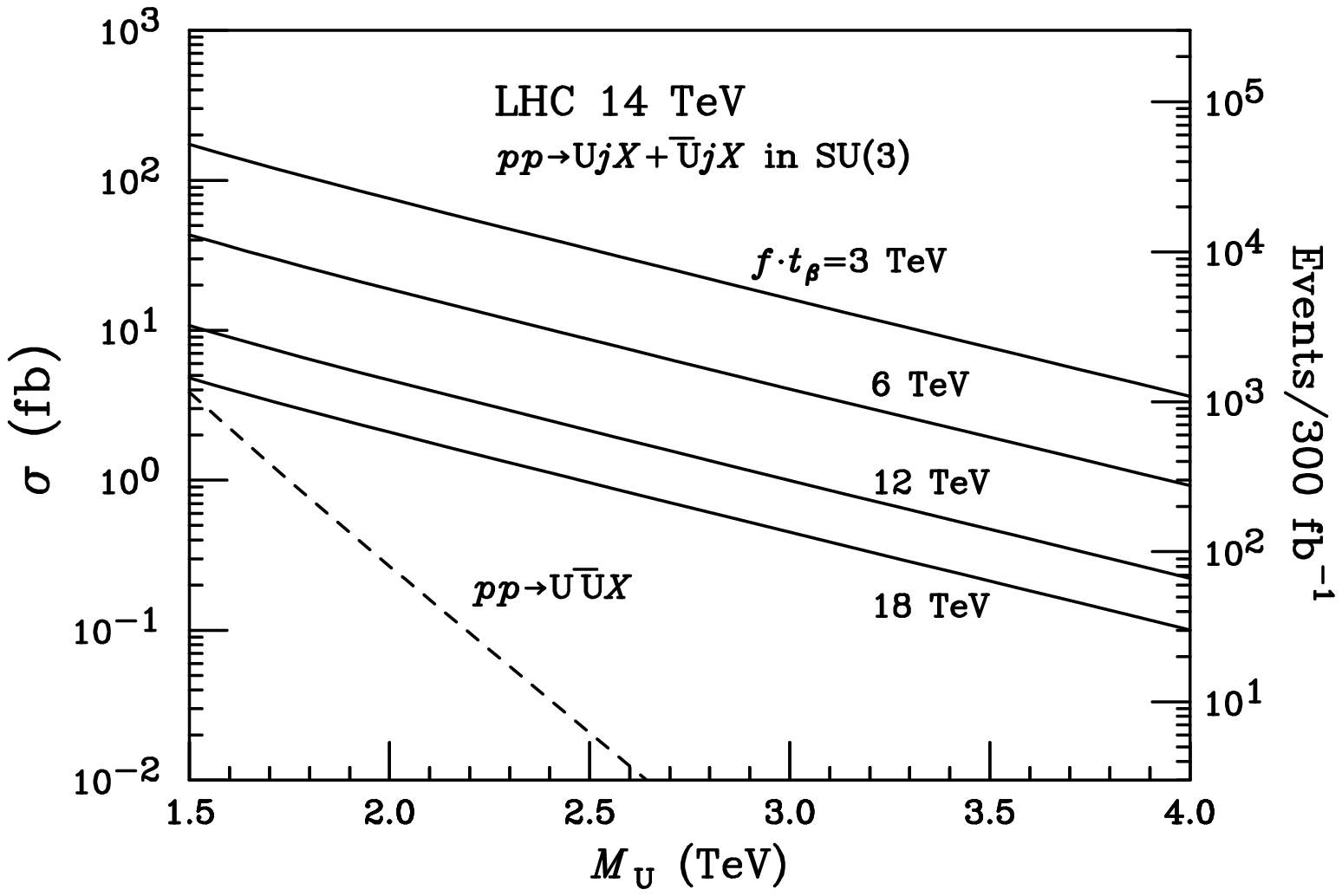}}
\caption{\label{fig:USU3}
Production cross sections for $U$ in the SU(3) simple group model. 
The single $U$ production cross section is shown for
various values of $f t_{\beta}$ (solid lines).}
}

The single $U$ production cross section is quite large compared to 
single production of $T$ 
at a comparable mass because $T$ production requires a $b$ quark 
in the initial state, while $U$ production proceeds from a valence
$u$ or $d$ quark.
By measuring both $M_U$ and the single $U$ production cross 
section, as well as $f$ from measurements in the gauge sector
(see Sec.~\ref{sec:gauge}), one can determine $\lambda_U$ and $t_{\beta}$
from Eqs.~(\ref{eq:MQ}) and (\ref{eq:deltaq}).
This measurement of $t_{\beta}$ is independent from that in the $T$ sector
and can be used as a nontrivial test of the model, as will be discussed
further in Sec.~\ref{sec:topdivtest}.

Production of the heavy quark partners of the first generation
offers an additional powerful handle on the SU(3) simple group model.
First, consider single $U$ production in the universal fermion embedding.  
This proceeds via the subprocesses
\begin{equation}
	d W^+ \to U, \qquad u Z \to U;
	\qquad \qquad \qquad
	\overline{d} W^- \to \overline{U}, 
	\qquad \overline{u} Z \to \overline{U}.
\end{equation}
At a proton-proton collider such as the LHC, we expect the cross section 
for $U$ production, from initial-state valence $u$ 
and $d$ quarks, will be much larger than that
for $\overline{U}$, from initial-state sea $\overline{u}$ and 
$\overline{d}$ antiquarks.  In fact, $\overline{U}$ production
constitutes less than 10\% of the total $U + \overline{U}$ cross section
shown in Fig.~\ref{fig:USU3}.
There will thus be a large asymmetry in the charge
of the final lepton in $U,\overline{U}$ decays to $W^{\pm}$, with many more
positively charged leptons.

In the anomaly-free embedding, single $D$ production 
proceeds via the subprocesses
\begin{equation}
	u W^- \to D, \qquad d Z \to D;
	\qquad \qquad \qquad
	\overline{u} W^+ \to \overline{D}, 
	\qquad \overline{d} Z \to \overline{D}.
\end{equation}
Because of the parton densities in the proton, the rate for $D$ production via 
charged current will be somewhat higher than for $U$, while the rate for $D$
production via neutral current will be somewhat lower than for $U$, resulting
in a comparable total cross section.
Again, there will be a large asymmetry in the charge of the final 
lepton in $D, \overline{D}$ decays to $W^{\mp}$, with many more negatively
charged leptons.  This allows a simple measurement of the dominant lepton 
charge in $Q \to q^{\prime} W(\to \ell \nu)$ decays to distinguish
the universal fermion embedding from the anomaly-free fermion embedding.
The fermion embedding must be known in order for the model parameters
to be extracted from the single-$Q$ production cross section because
the embedding determines which parton densities enter the production
cross section calculation.

Just as for $T$, the decay modes of $U$ in the SU(3) simple group model
depend on the spectrum of masses.  The $U$ quark decays into $uH$, $uZ$ and
$dW$ with partial widths 
\begin{equation}
	\Gamma(U \to uH) = \Gamma(U \to uZ) = \frac{1}{2} \Gamma(U \to dW)
	= 5.0 \left( \frac{\rm TeV}{f t_{\beta}} \right)^2
	\left( \frac{M_U}{\rm TeV} \right)^3 {\rm GeV}.
\end{equation}
$U$ can also decay into $u\eta$; however, the coupling at
leading order in $v/f$ is proportional to the up quark Yukawa coupling,
so this decay is extremely suppressed and can be neglected.
If $U$ is heavy enough,
it can also decay into $uY$ and $dX$ with partial widths that depend only
on the heavy gauge boson mass $M_{X,Y}$; the $UuY$ and $UdX$ couplings
are fixed in terms of the SM gauge coupling $g$.  The heavy gauge boson mass
$M_{X,Y}$ can be obtained from the $Z^{\prime}$ mass measurement 
(see Sec.~\ref{sec:gauge}).
The partial widths to $uH$, $uZ$ and $dW$ can then be extracted together
with $\delta_{\nu}$ from the rate measurement into any final state.
The above discussion applies equally to $D$ in the
anomaly-free fermion embedding.

\FIGURE{
\resizebox{0.4\textwidth}{!}{\includegraphics{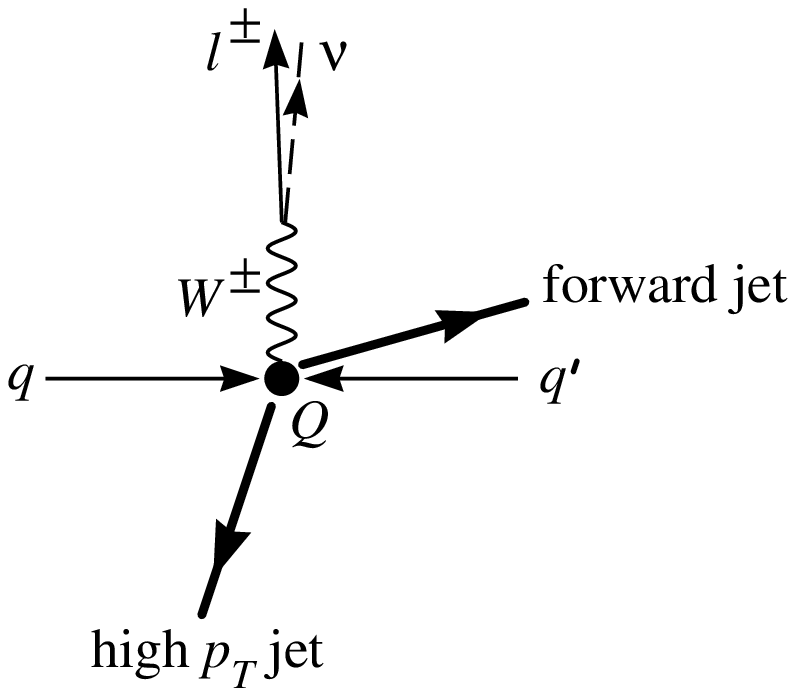}}
\caption{\label{fig:Qdecay}
Kinematics of $Q = U, D$ production and decay.  
}
}

The signal kinematics are as follows.  $U$ is produced via 
$dW^+$ or $uZ$ fusion, yielding a forward jet from which the 
$W$ or $Z$ was radiated.  $U$ then decays into a high-$p_T$ quark 
and a $W$ boson, with $W \to \ell \nu$.
The $W$ is highly boosted, with a momentum of roughly half the $U$ mass,
so that the momenta of the neutrino and charged lepton are almost parallel.
The decay kinematics are sketched in Fig.~\ref{fig:Qdecay}.

We can take advantage of the large boost of the $W$ boson in $U$ decay 
to reconstruct the $U$ mass.  Normally such a decay involving
a neutrino in the final state would
allow only the reconstruction of the $U$ transverse mass.
However, because $U$ is very heavy, we can neglect the $W$ mass relative
to its momentum and approximate the direction of the neutrino momentum to be
parallel to that of the charged lepton.  We can then reconstruct the
full neutrino momentum and combine it with that of the charged lepton
and the high-$p_T$ jet to reconstruct a mass peak for $U$.

We apply the following cuts to select $U$ production events over
the SM $W^+jj$ background.  We require
a positively-charged electron or muon with
\begin{equation}
	|\eta_{\ell}| < 3, \qquad \qquad
	p_{T \ell} > 20 \ {\rm GeV}.
\end{equation}
For the central high-$p_T$ jet we require
\begin{equation}
	|\eta_{j_1}| < 3, \qquad \qquad
	p_{T j_1} > 300 \ {\rm GeV}.
\end{equation}
We also require that the forward jet be tagged, with
\begin{equation}
	3 < |\eta_{j_2}| < 5, \qquad \qquad
	p_{T j_2} > 30 \ {\rm GeV}.
\end{equation}
Finally we require missing transverse momentum,
\begin{equation}
	\ptmiss > 30 \ {\rm GeV}.
\end{equation}
To simulate the detector effects, we smear the energies
for the charged lepton and the jets according to a Gaussian form,
$\Delta E/ E = a/ \sqrt{E/{\rm GeV}} \oplus b$,
with $a=5\%$, $b=1\%$ for a charged lepton and
$a=50\%$, $b=2\%$ for a jet.

The $p_T$ distribution of the highest-$p_T$ jet is shown in the left panel
of Fig.~\ref{fig:Umass}, together with the $W^+jj$ background.  
The signal distribution clearly exhibits a Jacobian peak near $M_U/2$.
The right panel of Fig.~\ref{fig:Umass} shows the $U$
transverse mass and the fully reconstructed $U$ mass.
The $U$ mass is reconstructed
from the momenta of $\ell^+$ and the highest-$p_T$ jet, as well as 
the missing momentum assumed to point along the 
direction of the $\ell^+$ momentum.
The reconstructed mass variable indeed leads to a sharper peak than the
transverse mass.

\FIGURE{
\resizebox{\textwidth}{!}{\includegraphics{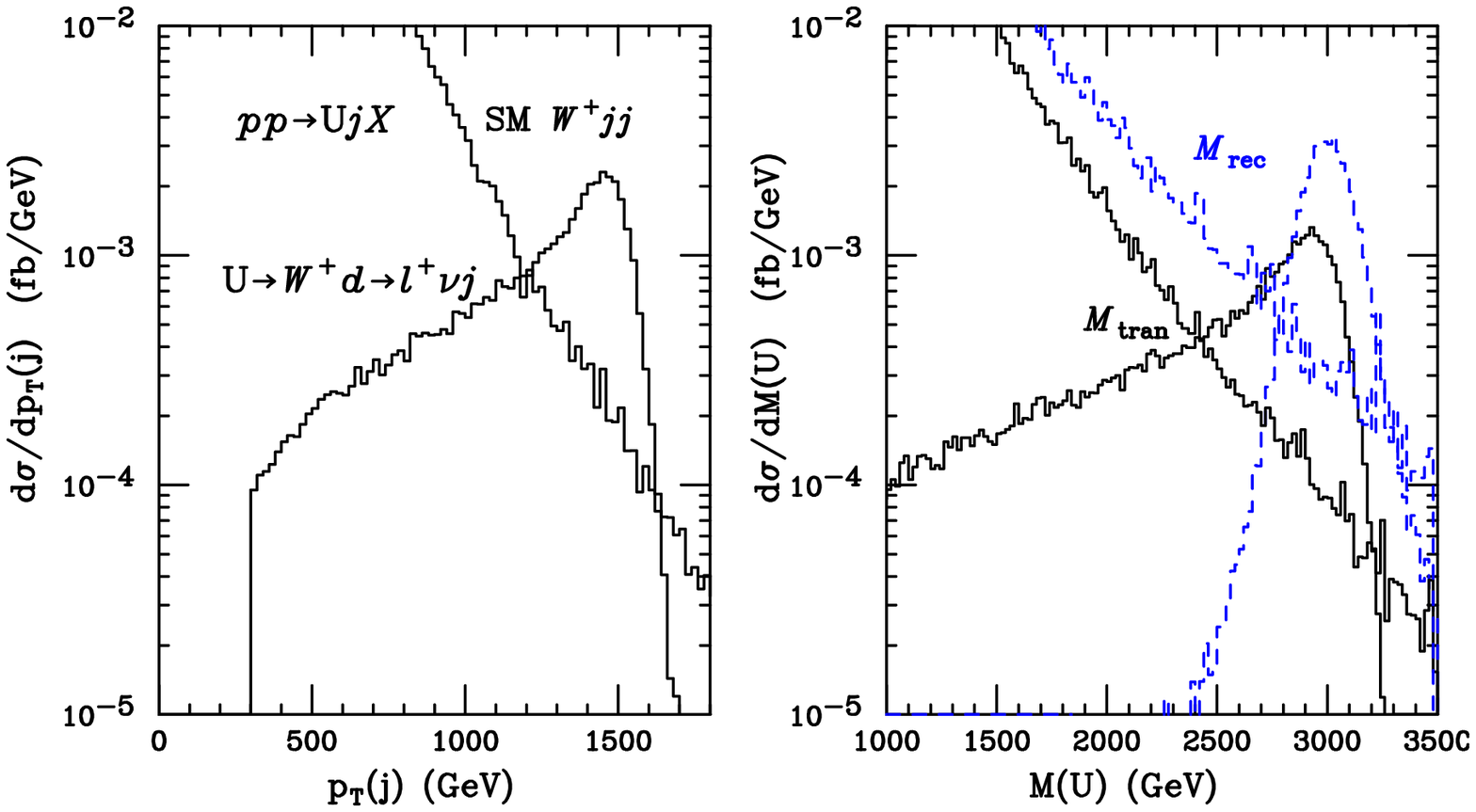}}
\caption{\label{fig:Umass}
Mass reconstruction of $U$ in $pp \to Uq \to \ell^+ \nu jj$, 
for $M_U = 3$ TeV and $f \, t_{\beta} = 3$ TeV.
(Left) $p_T$ of the highest-$p_T$ jet in the event.
(Right) Transverse mass $M_{\rm tran}$ (solid black histograms) 
and the full reconstructed mass $M_{\rm rec}$ (dashed blue histograms).
Also shown is the background from SM $W^+jj$.
}
}

In Fig.~\ref{fig:Umass} we have included only $U$ production (without the 
$\overline{U}$ contribution), and folded
in the branching fractions of $U \to W^+ u$ and $W^+ \to \ell^+ \nu$, with
$\ell^+ = e^+,\mu^+$.  The signal cross section after cuts for $M_U = 3$ TeV 
and $f \, t_{\beta} = 3$ TeV is about 0.66 fb, resulting in close to
200 signal events in 300 fb$^{-1}$ of LHC luminosity.  The background 
is well under control.
Additional statistics can be gained by considering the decay channels
$U \to uZ, uH$.

One can do a similar analysis for single $C$ ($S$) production, 
using $M_C$ ($M_S$) and the production cross section together with $f$ from 
the gauge sector measurements to determine $\lambda_C$ ($\lambda_S$) 
and make another independent measurement of $t_{\beta}$.  
However, because $C$ ($S$) is produced from inital-state
sea quarks $c$ and $s$, its production rate will be lower, only
10--20\% of that of $U$ ($D$).  Further, since the sea quark and antiquark 
distributions are equal, there will be no asymmetry in the charge
of the final lepton in $C$ ($S$) decays to $W^{\pm}$.  
This allows the $C$ ($S$) resonance to be experimentally distinguished 
from the $U$ ($D$) resonance, if enough events can be collected above
background.

\subsection{Testing the Higgs mass divergence cancellation in the top sector}
\label{sec:topdivtest}

The key experimental test of the little Higgs models is to verify
the cancellation of the Higgs mass quadratic divergence, embodied in 
the crucial relation of Eq.~(\ref{eq:topdivrel}).
Ideally, one could hope to measure the couplings $\lambda_T$ and 
$\lambda_T^{\prime}$ directly, without making any assumptions about the model
structure.
The coupling $\lambda_T$ controls the $T$ production cross 
section in $Wb$ fusion, where it can be extracted \cite{LHPheno,Peskin} 
by measuring
the single-$T$ production rate and the $T$ mass from signal kinematics.
The coupling $\lambda_T^{\prime}$ could in principle be extracted
from a measurement of the associated $TH$  production cross section.
However, a quick estimate \cite{THestimate} indicates that the cross 
section is too small to be observable at the LHC. 
Instead, the relation in Eq.~(\ref{eq:topdivrel}) for the Higgs mass
divergence cancellation must be checked within the context of the 
particular model.  Once the model is determined,
the relevant independent parameters that control
the top sector must be overconstrained to make a nontrivial test of the
model.

In the Littlest Higgs model, one can use the model relation 
$\lambda_T^{\prime} = \lambda_T M_T/f$ to write the divergence cancellation
condition in terms of the four observables $(\lambda_t, \lambda_T, M_T, f)$.
Note that only three of these are independent in the Littlest Higgs model;
$\lambda_T$ and $M_T$ can both be written in terms of $f$, $\lambda_t$ 
and $x_{\lambda}$.
Combining $T$-sector measurements of $M_T$ and $\lambda_T$ with a measurement 
of $f$ from the heavy gauge boson sector, one can overconstrain the
parameters and verify the cancellation of the quadratic divergence.

In the SU(3) simple group model the situation is
more complicated because of the ratio of the two vacuum condensates, 
$f_2/f_1 = t_{\beta}$, which appears in the fermion sector of the model.
Thus, in addition to the four parameters $(\lambda_t, \lambda_T, M_T, f)$
measurable in the $T$ and heavy gauge boson sectors, one needs a measurement
of $t_{\beta}$ in order to overconstrain the parameters and verify 
the relation in Eq.~(\ref{eq:topdivrel}).
Fortunately, $t_{\beta}$ can be extracted independently of the $\lambda_T$
and $M_T$ measurements by measuring the mass
and production cross section of the $U$ or $D$ quarks,
since their production couplings are proportional to $1/t_{\beta}$.

\subsection{Comparison with other models}

\subsubsection{A fourth generation sequential top-prime}
\label{sec:top4thgen}

The key feature that distinguishes $T$ from a fourth generation
sequential top-prime is the fact that it is an SU(2) singlet
before mixing with the top quark.
This feature allows for the presence of a vectorlike mass term for $T$
and flavor-changing $TtH$ and $TtZ$ couplings in the mass basis,
both of which are forbidden by electroweak symmetry in a fourth-generation
model.  As pointed out in Ref.~\cite{ATLAS}, detecting and measuring 
the flavor-changing neutral current decays $T \to Zt$ and
$T \to Ht$, with equal branching fractions, allows one to 
rule out the fourth-generation hypothesis and conclude that $T$ 
is an electroweak singlet, acquiring its coupling to the Higgs via a
gauge-invariant $TtH$ term.

\subsubsection{The top quark see-saw}

In the top quark see-saw model \cite{topseesaw,seesawlT}, EWSB
occurs via the condensation of the top quark in the presence of an extra
vectorlike SU(2)-singlet quark, forming a composite Higgs 
boson.  In order to reproduce the correct electroweak scale, the condensate 
mass must be large, of order 600 GeV.  The vectorlike singlet quark
joins the top in a see-saw, yielding the physical top mass (adjusted to 
the experimental value) and a multi-TeV mass for the vectorlike quark.
The little Higgs models thus generically contain an extended top sector
with the same electroweak quantum numbers as in the top see-saw model, i.e.,
a (multi-)TeV-scale isosinglet vectorlike quark $T$ 
with a small mixing with the SM top quark that gives rise to $TtZ$, 
$TtH$ and $TbW$ couplings.

The most important difference between the top see-saw model and the little
Higgs models is that the top see-saw model makes no prediction for the
dimension-5 $HHTT$ coupling $\lambda_T^{\prime}$, although this coupling
can be generated radiatively.  Thus, the top see-saw 
model does not in general satisfy the condition for cancellation of the 
Higgs mass quadratic divergence given in Eq.~(\ref{eq:topdivrel}).

In the top see-saw model, the $TtH$ coupling $\lambda_T$ is constrained
by the compositeness condition, which requires the wavefunction
renormalization of the composite Higgs field to vanish at the compositeness
scale $M_c$.
Ignoring the effect of EWSB, the effective Lagrangian of
the top see-saw model is \cite{seesawlT,seesawpheno}
\begin{equation}
	{\mathcal{L}}=Z_h |\mathcal{D}h|^2 
	+ \left[ 
	  \sqrt{2} y_t \bar{\psi}_L t_R \sqrt{Z_h} h + \sqrt{2} \lambda_T
	\bar{\psi}_L T_R \sqrt{Z_h} h - M_T \bar{T}_L T_R 
	+ {\rm h.c.} \right] + V_h,
\end{equation}
where $Z_h$ is the wavefunction renormalization of the composite
Higgs field $h$ and $V_h$ is the usual SM Higgs potential.
In the large-$N_c$ approximation, this implies \cite{seesawlT}
\begin{equation}
	\lambda_T^2 = \frac{4 \pi^2}{N_c \log (M_c/M_T)} - \frac{m_t^2}{v^2}.
\end{equation}
The compositeness scale $M_c$ should not be too far away from the scale of the 
heavy states.  For $M_c/M_T \sim 10$--100 and
$N_c = 3$, we obtain $\lambda_T \sim 5.2$--2.4; in particular, 
the compositeness condition generally requires a fairly large value
for $\lambda_T$.  In little Higgs models, on the other hand, $\lambda_T$
is typically of order one or smaller.  In the Littlest Higgs model, 
$\lambda_T = x_{\lambda} m_t / v \simeq x_{\lambda}/\sqrt{2}$,
which reaches the typical top quark see-saw values only for 
$x_{\lambda} \gtrsim 4$.  Large values of $x_{\lambda}$ in the Littlest
Higgs model tend to push up the $T$ mass, leading to greater fine tuning
in the electroweak scale.
In the SU(3) simple group model,
$\lambda_T = s_{\beta} c_{\beta} (x_{\lambda} - x_{\lambda}^{-1}) m_t/v$,
which is further suppressed by the $s_{\beta} c_{\beta} \leq 1/2$ factor 
in front.

\section{The gauge sector}
\label{sec:gauge}

Little Higgs models extend the electroweak gauge group at the TeV scale.
The structure of the extended electroweak gauge group 
determines crucial properties of the little Higgs model, which can be 
revealed by studying the new gauge bosons at the TeV scale.  Therefore,
we continue with a study of the heavy gauge boson sectors of little
Higgs models.

\subsection{Heavy gauge boson masses and parameters}

The extra gauge bosons get their masses from the $f$ condensate, which
breaks the extended gauge symmetry.
For our two prototype models, the gauge boson masses are given in terms
of the model parameters by
\begin{eqnarray}
	&& \left.
	\begin{array}{l}
	M_{W_H} = M_{Z_H} = gf/2sc = 0.65 f/\sin 2\theta \\
	M_{A_H} = g s_W f/2 \sqrt{5} c_W s^{\prime} c^{\prime}
	= 0.16 f/\sin 2 \theta^{\prime}
	\end{array}
	\right\} {\rm in \ the \ Littlest \ Higgs \ model,} 
	\nonumber \\
	&& \left.
	\begin{array}{l}
	M_{Z^{\prime}} = \sqrt{2} g f/\sqrt{3 - t_W^2} = 0.56 f \\
	M_X = M_Y = g f/\sqrt{2} = 0.46 f = 0.82 M_{Z^{\prime}}
	\end{array}
	\right\} {\rm in \ the \ SU(3) \ simple \ group \ model.}
	\label{eq:gaugemasses}
\end{eqnarray}
In the SU(3) simple group model the heavy gauge boson masses are determined by
only one free parameter, the scale $f = \sqrt{f_1^2 + f_2^2}$.
The Littlest Higgs model has two additional gauge sector parameters, 
$\tan\theta = s/c = g_2/g_1$ [in the SU(2)$^2\to$SU(2) breaking sector] and 
$\tan\theta^{\prime} = s^{\prime}/c^{\prime} = g_2^{\prime}/g_1^{\prime}$
[in the U(1)$^2\to$U(1) breaking sector].
If only one copy of U(1) is gauged \cite{GrahamEW2}, 
the $A_H$ state is not present
and the gauge sector of the Littlest Higgs model is controlled by only 
two free parameters, $f$ and $\tan\theta$.  Because the model with
only one copy of U(1) gauged is favored by the electroweak precision 
constraints, and since the U(1) sectors of the product
group models are quite model-dependent, we focus in what follows on 
the heavy SU(2) gauge bosons $W_H$ and $Z_H$.  The $W_H$ and $Z_H$ bosons 
capture the crucial features of the gauge sector of the Littlest Higgs 
model and their phenomenology can be applied directly to the other 
product group models.

\subsection{Heavy gauge boson interactions with SM particles}

The gauge couplings of the Higgs doublet take the general form
\begin{equation}
	\mathcal{L} = \left\{ \begin{array}{l}
	\left[ G_{HHVV} V V
	+ G_{HHV^{\prime}V^{\prime}} V^{\prime} V^{\prime}
	+ G_{HHVV^{\prime}} V V^{\prime} \right] H^2 \\
	\left[ G_{HHV^+V^-} V^+ V^-
	+ G_{HHV^{\prime +}V^{\prime -}} V^{\prime +} V^{\prime -} 
	+ G_{HHV^+V^{\prime -}} ( V^+ V^{\prime -} + V^- V^{\prime +} ) 
	\right] H^2,
	\end{array} \right.
\label{eq:GHHVV}
\end{equation}
where the top line is for $V$ neutral and the bottom line is for $V$ charged.
Here $V$ and $V^{\prime}$ stand for the SM and heavy gauge bosons, 
respectively.
This Lagrangian leads to two quadratically divergent
diagrams contributing to the Higgs mass: one involving a loop of $V$,
proportional to $G_{HHVV}$, and the other involving a loop of $V^{\prime}$,
proportional to $G_{HHV^{\prime}V^{\prime}}$.  The divergence cancellation in
the gauge sector can thus be written as
\begin{equation}
	\sum_i G_{HHV_iV_i} = 0,
	\label{eq:gaugecancellation}
\end{equation}
where the sum runs over all gauge bosons in the model.
The couplings in the models under consideration are given in 
Table~\ref{tab:gauge1}.
In the SU(3) simple group model, the quadratic divergence cancels
between the $Z$ and $Z^{\prime}$ loops and between the $W$ and $X$ loops.
In the Littlest Higgs model, the quadratic divergence cancels between
the $W$ and $W_H$ loops and there is a partial cancellation between the
$Z$ and $Z_H$ loops.  Including the $A_H$ loop leads to a complete
cancellation of the quadratic divergence from the $Z$ loop.
The key test of the little Higgs mechanism in the gauge sector is
the experimental verification of Eq.~(\ref{eq:gaugecancellation});
we discuss the prospects further in Sec.~\ref{sec:gaugedivtest}.

\TABLE{
\begin{tabular}{|c|c|c|}
\hline \hline
 & Littlest Higgs & SU(3) simple group \\
\hline
$G_{HHZZ}$ & $g^2/8c_W^2$ & $g^2/8c_W^2$ \\
$G_{HHW^+W^-}$ & $g^2/4$ & $g^2/4$ \\
\hline
$G_{HHV^{\prime}V^{\prime}}$
	& $G_{HHZ_HZ_H} = -g^2/8$ 
		& $G_{HHZ^{\prime}Z^{\prime}} = -g^2/8c_W^2$ \\
	& $G_{HHW_H^+W_H^-} = -g^2/4$
		& $G_{HHX^+X^-} = -g^2/4$ \\
\hline
$G_{HHVV^{\prime}}$
	& $G_{HHZZ_H} = -g^2 \cot 2\theta /4 c_W$
		& $G_{HHZZ^{\prime}} = g^2(1-t_W^2)/4 c_W \sqrt{3-t_W^2}$ \\
	& $G_{HHW^+W_H^-} = -g^2 \cot 2\theta / 4$
		& $G_{HHW^+X^-} = 0$ \\
\hline
$\delta_Z$ & $-\sin 4\theta \, v^2 / 8 c_W f^2$
	& $-(1-t_W^2)\sqrt{3-t_W^2} \, v^2 / 8 c_W f^2$ \\
$\delta_W$ & $c_W \delta_Z$ & $0$ \\
\hline
$g_{VVV^{\prime}}$
	& $g_{W^+W^-Z_H} = -g c_W \delta_Z$ 
		& $g_{W^+W^-Z^{\prime}} = g c_W \delta_Z$ \\
	& $g_{W^+W^-_HZ} = -g \delta_Z$
		& $g_{W^+X^-Z} = 0$ \\
\hline
$g_{VV^{\prime}V^{\prime}}$
	& $g_{W_H^+W_H^-Z} = -gc_W$
		& $g_{X^+X^-Z} = -g(1-2s_W^2)/2c_W$ \\
	&
		& $g_{Y^0 \overline{Y}^0 Z} = -g/2c_W$ \\
	& $g_{W^+W_H^-Z_H} = -g$
		& $g_{W^+X^-\overline{Y}^0} = g/\sqrt{2}$ \\
\hline
$g_{V^{\prime}V^{\prime}V^{\prime}}$ 
	& $g_{W_H^+W_H^-Z_H} = 2g \cot 2\theta$ 
		& $g_{X^+X^-Z^{\prime}} = g_{\overline{Y}^0 Y^0 Z^{\prime}} 
			= g/\sqrt{2}$ \\
\hline \hline
\end{tabular}
\caption{\label{tab:gauge1}
Heavy gauge boson parameters and couplings
in the Littlest Higgs model and the SU(3) simple group model.
The triple gauge coupling Feynman rule for 
$V_1^{\mu}(k_1) V_2^{\nu}(k_2) V_3^{\rho}(k_3)$
is given in the form 
	$- i g_{V_1V_2V_3} \left[ g^{\mu\nu} (k_1 - k_2)^{\rho}
	+ g^{\nu\rho} (k_2 - k_3)^{\mu} + g^{\rho\mu} (k_3 - k_1)^{\nu}
	\right]$, with the convention
$g_{W^+W^-Z} = -g c_W$.
}
}

After EWSB, the couplings of $H^2$ to one heavy
and one SM gauge boson induce mixing between the heavy and SM gauge bosons:
\begin{equation}
	V^{\prime} = V^{\prime}_0 - \delta_V V_0,  
	\qquad \qquad
	\delta_V = - v^2 G_{HHVV^{\prime}} / M_{V^{\prime}}^2,
\end{equation}
where $V^{\prime}_0$, $V_0$ stand for the states before EWSB.  The mixing
parameters $\delta_V$ are given in Table~\ref{tab:gauge1}.  This mixing 
gives rise to triple gauge couplings between one heavy and two SM gauge 
bosons, also shown in Table~\ref{tab:gauge1}.
In the SU(3) simple group model, EWSB also splits the $X$ and 
$Y$ gauge boson masses by a small amount,
\begin{equation}
	M_Y - M_X = \frac{gv^2}{4\sqrt{2} f}
	\simeq 3.9 \left( \frac{\rm TeV}{M_{Z^{\prime}}} \right) {\rm GeV}.
\end{equation}

In the Littlest Higgs model, the couplings of the heavy gauge bosons to
the SU(2)$_L$ fermion currents take the form
\begin{equation}
	Z_H^{\mu} \overline{f} f: \ \ i g \cot\theta T^3_f \gamma^{\mu} P_L,
	\qquad \qquad
	W_H^{+ \mu} \overline{u} d: \ \ -\frac{ig}{\sqrt{2}} \cot\theta 
		\gamma^{\mu} P_L,
\end{equation}
where $T^3_f = 1/2$ $(-1/2)$ for up (down) type fermions.
Below the TeV scale, exchange of $W_H$ and $Z_H$ gives rise to four-fermi 
operators, which are constrained by the electroweak precision data.  The 
experimental constraints are loosened by 
going to small values of $\cot\theta$, for which the couplings of the heavy
gauge bosons are suppressed.

In the SU(3) simple group model, the $Z^{\prime}$ couples to SM fermions
with gauge strength, while the $X,Y$ gauge bosons couple only via the mixing
between SM fermions and their TeV-scale partners.  The couplings are
given in Table~\ref{tab:gauge2}.

\TABLE{
\begin{tabular}{|c|c|}
\hline \hline
	& SU(3) simple group \\
\hline
$Z^{\prime} \overline t t:$ & $-\frac{ig}{c_W \sqrt{3-4s_W^2}}
		[(\frac{1}{2} - \frac{1}{3}s_W^2)P_L + \frac{2}{3}s_W^2P_R]$ \\
$Z^{\prime} \overline b b:$ & $-\frac{ig}{c_W \sqrt{3-4s_W^2}}
		[(\frac{1}{2} - \frac{1}{3}s_W^2)P_L - \frac{1}{3}s_W^2P_R]$ \\
\hline
$Z^{\prime} \overline u u:$ & $-\frac{ig}{c_W \sqrt{3-4s_W^2}}
		[(-\frac{1}{2} + \frac{2}{3}s_W^2)P_L + \frac{2}{3}s_W^2P_R]$
		(anomaly free) \\
	& $-\frac{ig}{c_W \sqrt{3-4s_W^2}}
		[(\frac{1}{2} - \frac{1}{3}s_W^2)P_L + \frac{2}{3}s_W^2P_R]$ 
		(universal) \\
\hline
$Z^{\prime} \overline d d:$ & $-\frac{ig}{c_W \sqrt{3-4s_W^2}}
		[(-\frac{1}{2} + \frac{2}{3}s_W^2)P_L - \frac{1}{3}s_W^2P_R]$
		(anomaly free) \\
	& $-\frac{ig}{c_W \sqrt{3-4s_W^2}}
		[(\frac{1}{2} - \frac{1}{3}s_W^2)P_L - \frac{1}{3}s_W^2P_R]$
		(universal) \\
\hline
$Z^{\prime} \overline e e:$ & $-\frac{ig}{c_W \sqrt{3-4s_W^2}}
		[(\frac{1}{2} - s_W^2)P_L  - s_W^2P_R]$ \\
$Z^{\prime} \overline \nu \nu:$ & $-\frac{ig}{c_W \sqrt{3-4s_W^2}}
		(\frac{1}{2} - s_W^2)P_L$ \\
\hline
$X^-_{\mu} \bar b t:$ & $\frac{g}{\sqrt{2}} \delta_t \gamma_{\mu} P_L$ \\
$X^-_{\mu} \bar d u:$ & $\frac{g}{\sqrt{2}} \delta_{\nu} \gamma_{\mu} P_L$ \\
$X^-_{\mu} \bar e \nu:$ & $\frac{g}{\sqrt{2}} \delta_{\nu} \gamma_{\mu} P_L$ \\
$Y^0_{\mu} \bar t t:$ & $\frac{g}{\sqrt{2}} \delta_t \gamma_{\mu} P_L$ \\
\hline
$Y^0_{\mu} \bar u u:$ & $0$ (anomaly free) \\
		      & $\frac{g}{\sqrt{2}} \delta_{\nu} \gamma_{\mu} P_L$ 
				(universal) \\
\hline
$Y^0_{\mu} \bar d d:$ & $\frac{g}{\sqrt{2}} \delta_{\nu} \gamma_{\mu} P_L$
				(anomaly free) \\
		      & $0$ (universal) \\
\hline
$Y^0_{\mu} \bar e e:$ & $0$ \\
$Y^0_{\mu} \bar \nu \nu:$ 
	& $\frac{g}{\sqrt{2}} \delta_{\nu} \gamma_{\mu} P_L$ \\
\hline
$Y^0_{\mu} H \eta:$ & $\frac{ig}{2\sqrt{2}} (p_{\eta} - p_H)_{\mu}$ \\
\hline \hline
\end{tabular}
\caption{\label{tab:gauge2}
Heavy gauge boson couplings 
in the SU(3) simple group model.  We neglect flavor misalignments.
The momenta $p_{\eta,H}$ of the scalars are outgoing.}
}

\subsection{Heavy gauge boson production and decay}
\label{sec:gaugeproddecay}

The best way to discover new heavy gauge bosons at the LHC is 
generally through Drell-Yan production.  This is certainly true
in the little Higgs models.

In the Littlest Higgs model, the heavy gauge bosons $Z_H,W_H$ couple
to pairs of SM fermions through the SU(2)$_L$ current, with coupling
strength scaled by $\cot\theta$ compared to the SM SU(2)$_L$ couplings.
They thus have large production cross sections, as shown in 
Fig.~\ref{fig:gauge}, controlled by one common free parameter, 
$\cot\theta$.\footnote{Note that the electroweak precision data tend
to favor small values of $\cot\theta$, which reduces the contribution
of $W_H,Z_H$ to four-Fermi operators at low energy.  Small $\cot\theta$
lowers the Drell-Yan cross section, reducing the LHC reach for $W_H,Z_H$
discovery.}
In addition, because $Z_H$ and $W_H$ form an SU(2) triplet, they are
degenerate in mass up to very small EWSB effects.
Thus, the measurement of the $Z_H$ mass in dileptons
predicts the transverse mass distribution of the $W_H$ in $W_H \to \ell \nu$,
and the measurement of the rate for $Z_H$ into dileptons predicts the
rate for $W_H$ into leptons, allowing a test of the SU(2) triplet 
nautre of $(W_H, Z_H)$.

\FIGURE{
  \resizebox{\textwidth}{!}{
    \includegraphics{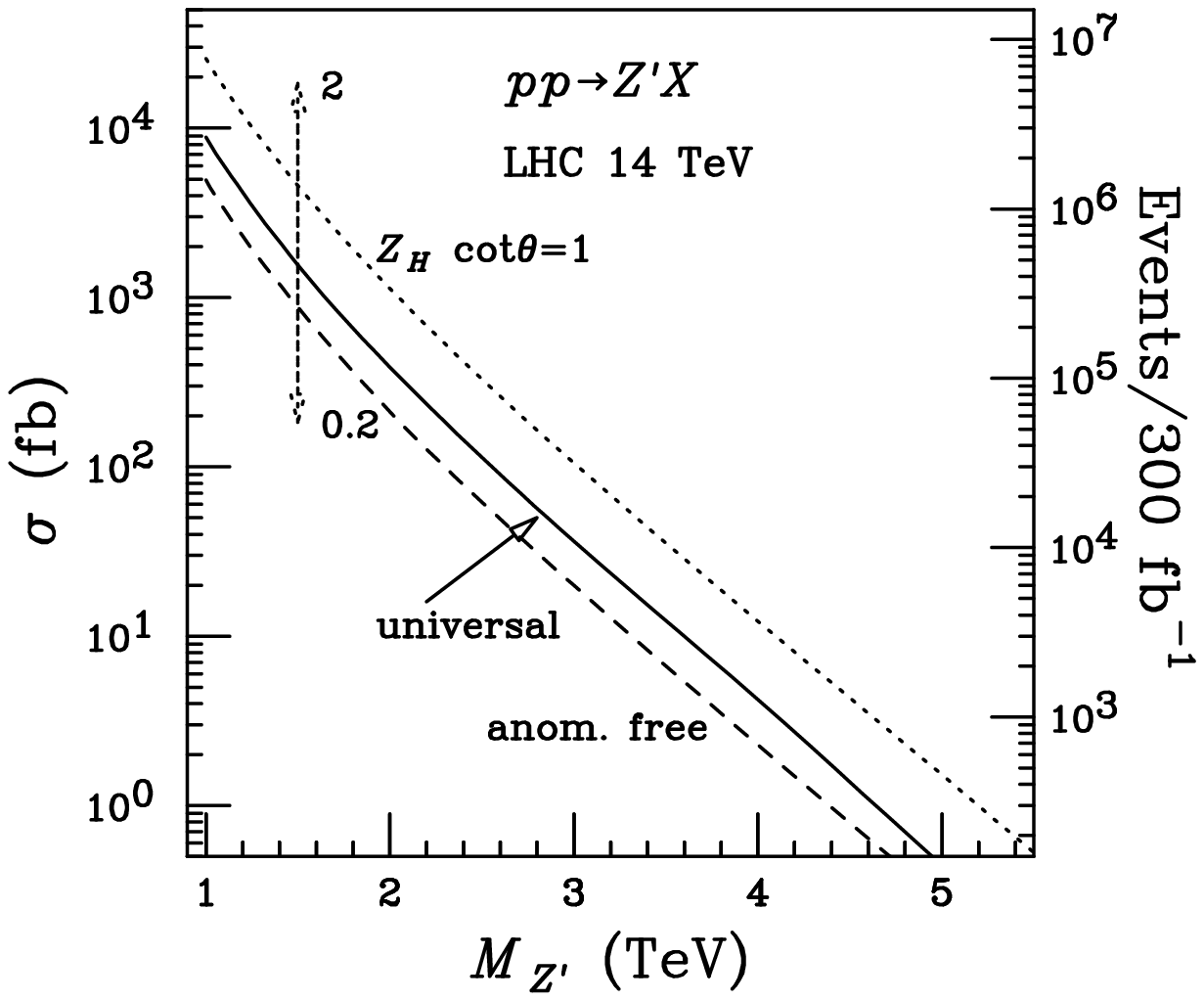}
    \includegraphics{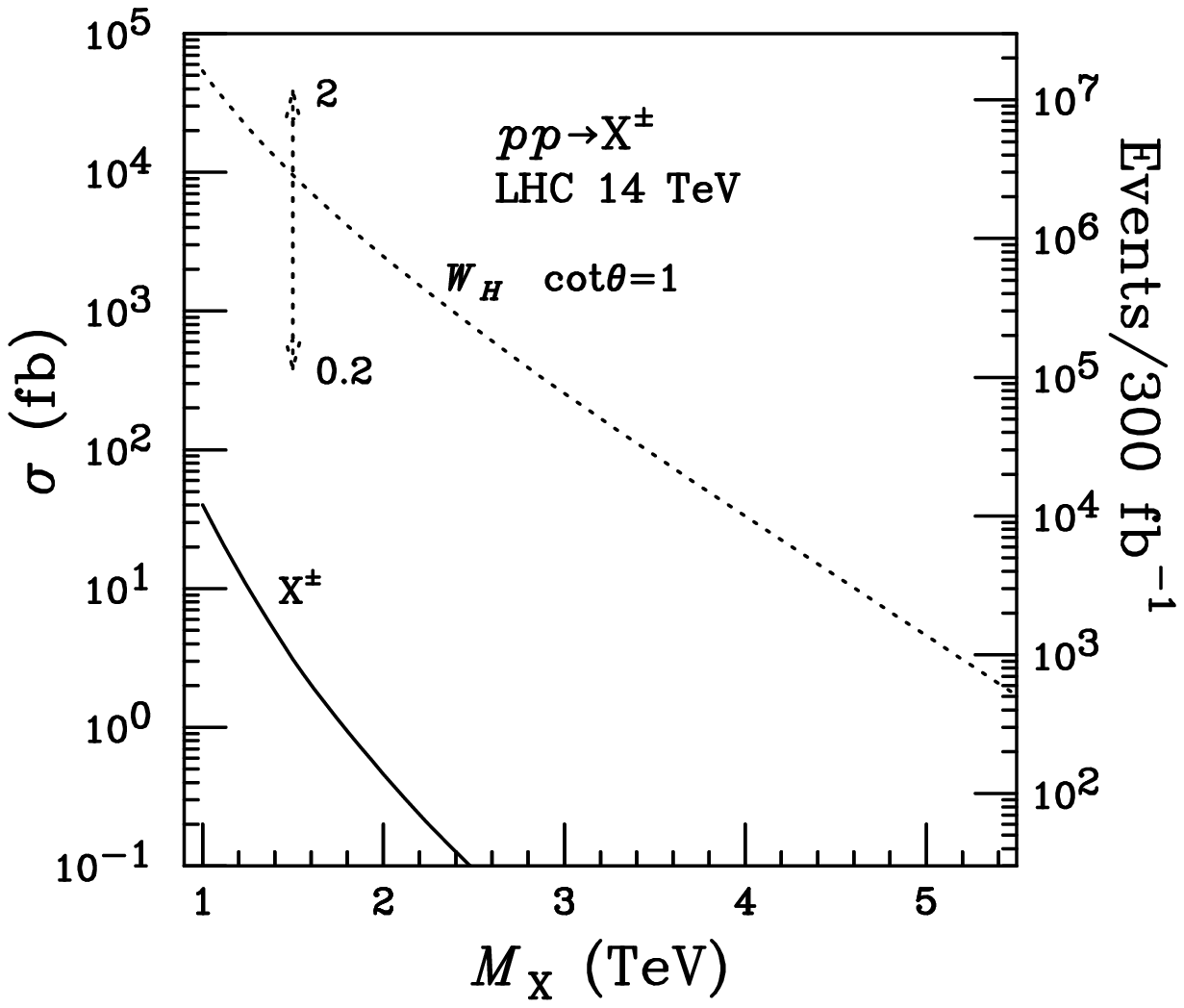}}
\caption{\label{fig:gauge}
Cross sections for neutral (left) and charged (right) 
heavy gauge boson production at the LHC, as a function of the mass
of the produced particle.
Dotted lines show $Z_H$ (left) and $W_H$ (right) production in the
Littlest Higgs model for $\cot\theta = 1$; 
the variation in cross section for $\cot\theta = 2$ and $0.2$ is 
shown by the dotted arrows.
For the SU(3) simple group model, $Z^{\prime}$ production is shown 
in the left panel for the universal (solid) and anomaly-free (dashed) 
fermion embeddings, and $X^{\pm}$ production is shown in the right 
panel for $t_{\beta} = 3$.  The $X^{\pm}$ cross section
is proportional to $1/t_{\beta}^2$.}
}

In the SU(3) simple group model, the heavy gauge boson $Z^{\prime}$
couples to pairs of SM fermions with couplings fixed in terms of the 
SM gauge couplings and depending only on the (discrete) choice of 
the fermion embedding, as shown in the left panel
of Fig.~\ref{fig:gauge}.  Unlike the $Z_H$ of the Littlest Higgs
model, there is no tunable parameter in the $Z^{\prime}$ cross
section.%
\footnote{This parameter independence is the most characteristic feature of
the $Z^{\prime}$ in simple group models with the
extended gauge group SU(3)$\times$U(1)$_X$ 
\cite{KS,Schmaltznote,SkibaTerning}.
Models with a larger extended gauge group, SU($N$)$\times$U(1)$_X$
with $N>3$, lose this parameter independence because they contain 
$N-2$ broken diagonal generators, which mix in general.  For 
example, the SU(4)$\times$U(1)$_X$ model of Ref.~\cite{KS}
contains two broken diagonal generators, $Z^{\prime}_1$ (which couples
to SM fermion pairs with fixed strength) and $Z^{\prime}_2$
(which does not couple to fermion pairs).  After mixing, the 
mass eigenstates $Z^{\prime},Z^{\prime\prime}$ share the fermion
couplings with the mixing angle as a free parameter.  If the fermion
couplings of both states can be measured, the parameter independence
reappears in the form of a coupling sum rule.}
The heavy gauge bosons $X,Y$ of the SU(3) simple group model have
a very different phenomenology, rooted in their identity as the SU(2)$_L$ 
doublet $(X^-,Y^0)$ of broken off-diagonal generators of SU(3).  
Because they couple to SM quark pairs only through $q-Q$ mixing
as given in Table~\ref{tab:gauge2}, their production cross sections
in Drell-Yan are suppressed by $\delta_{\nu}^2 \propto v^2/f^2$.
This is shown for $X$ in the right panel of Fig.~\ref{fig:gauge}.  
Because of this large cross section difference, $X^{\pm}$ cannot be 
mistaken for the charged members of an SU(2) triplet containing
$Z^{\prime}$, providing an easy distinction between simple group
and product group models.  The $\sim 20\%$ mass splitting between 
$X^{\pm}$ and $Z^{\prime}$ given in Eq.~(\ref{eq:gaugemasses}) also 
serves to distinguish $X^{\pm},Z^{\prime}$ from an SU(2) triplet.

An important feature of the product group models is the
couplings of $Z_H$, $W_H$ to dibosons, which gives rise to the decays 
$Z_H \to Z H$, $W^+W^-$ and $W_H \to W H$, $W Z$.  These couplings 
arise from a $W_H^a W^a h h^{\dagger}$ term in the Lagrangian \cite{Burdman}
and are proportional to $\cot 2\theta$ due to the
characteristic collective breaking structure of the gauge couplings
in the product group models.
The bosonic decay modes are dominated by the longitudinal components
of the final-state bosons; their partial widths can be shown 
by the Goldstone boson equivalence theorem
to obey the relation
$\Gamma(Z_H \to Z H) = \Gamma(Z_H \to W^+W^-) 
= \Gamma(W_H \to W H) = \Gamma(W_H \to W Z) \equiv \Gamma(V_H \to V H)$,
where we negect final-state masses and
\begin{equation}
	\Gamma(V_H \to V H) = \frac{g^2 \cot^2 2\theta}{192 \pi} M_{V_H}
	= 0.70 \cot^2 2\theta 
		\left( \frac{M_{V_H}}{\rm TeV} \right) {\rm GeV}.
\end{equation}
Here $M_{V_H}$ is the mass of $Z_H$ or $W_H$.
The measurement of $\cot\theta$
from $Z_H \to \ell^+\ell^-$ thus predicts the rates for decays of both
$Z_H$ and $W_H$ into dibosons.
The decay branching fractions of $Z_H$ and $W_H$ in the Littlest Higgs model
are shown as a function of $\cot\theta$ in Fig.~\ref{fig:ZHWHdecay}.
We neglect final-state masses and assume that no decays to 
$A_H$ are present (namely, $Z_H \to A_H H$ and $W_H \to A_H W$).

\FIGURE{
\resizebox{\textwidth}{!}{
	\includegraphics{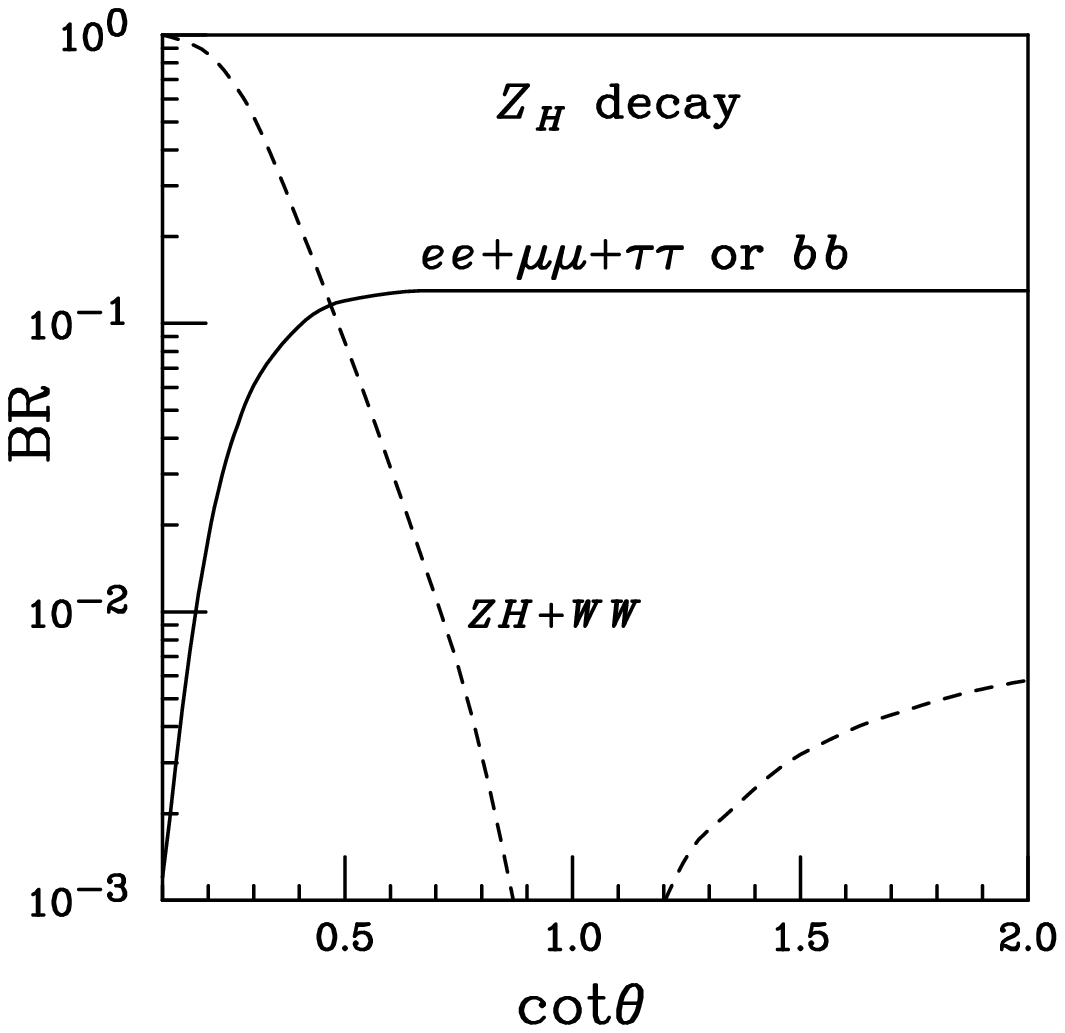}
    	\includegraphics{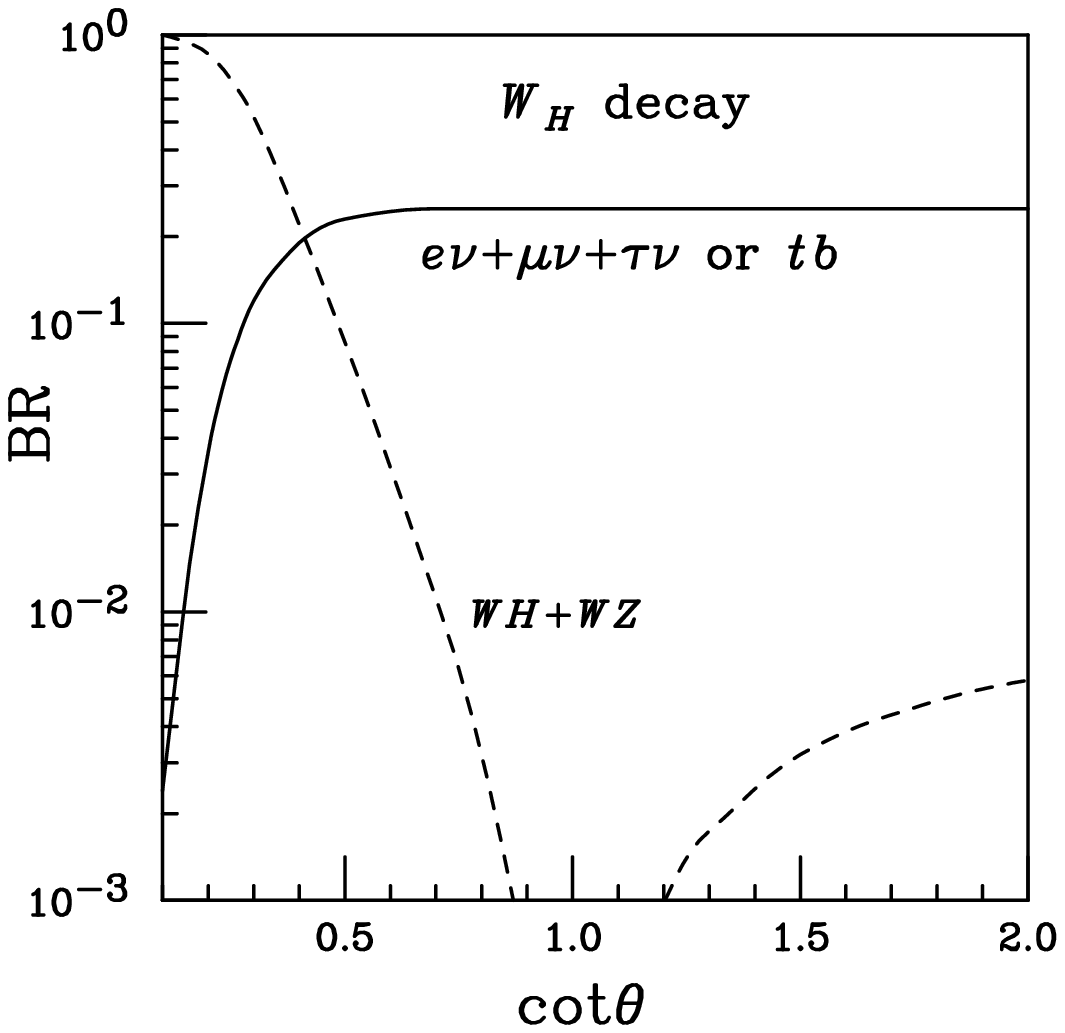}}
\caption{\label{fig:ZHWHdecay}
Decay branching fractions of $Z_H$ (left) and $W_H$ (right) 
in the Littlest Higgs model, as a function of $\cot\theta$.
Final-state masses are neglected.}
}

In the SU(3) simple group model, the decay partial widths of
$Z^{\prime}$ into pairs of SM bosons, $ZH$ and $W^+W^-$, are fixed
in terms of the $Z^{\prime}$ mass (neglecting final-state masses) to be
\begin{equation}
	\Gamma(Z^{\prime} \to Z H) = \Gamma(Z^{\prime} \to W^+W^-)
	= \frac{g^2 (1-t_W^2)^2}{192 \pi (3-t_W^2)} M_{Z^{\prime}}
	= 0.13 \left( \frac{M_{Z^{\prime}}}{\rm TeV} \right) {\rm GeV},
\end{equation}
and the decay partial widths into pairs of SM fermions
are fixed once the fermion embedding is chosen.  As discussed in
Sec.~\ref{sec:Qprod}, the fermion embedding can be determined at the LHC by
detecting the TeV-scale quark partner of the first generation, $U$ or $D$, 
decaying into $Wq$; the charge asymmetry of the final-state $W$ 
then determines the embedding.  Knowledge of the fermion embedding
from the fermion sector can be used to compute the 
$Z^{\prime}$ couplings uniquely and perform a cross-check the model.
If the TeV-scale fermion partners $T$ and/or $Q_m$ 
are not too heavy, they can be present 
in $Z^{\prime}$ boson decays.  If kinematically accessible, 
decays of $Z^{\prime}$ to pairs of TeV-scale fermion partners proceed
via gauge couplings.  This is in contrast to the product group models,
in which the TeV-scale top quark partner is mostly electroweak singlet and
couples to $Z_H$ only through its electroweak doublet admixture 
at order $v^2/f^2$.  The $Z^{\prime}$ can also decay to one SM fermion
and one TeV-scale fermion partner; however, the partial widths of these 
decays are suppressed by $\delta_t^2,\delta_{\nu}^2 \propto v^2/f^2$
and will be numerically unimportant.  Finally, the decay
$Z^{\prime} \to Y^0 \eta$ will be kinematically accessible if
$\eta$ is lighter than the $Z^{\prime}$--$Y^0$ mass splitting,
\begin{equation}
	M_{Z^{\prime}} - M_{Y} = 0.18 M_{Z^{\prime}}
	= 180 \left( \frac{M_{Z^{\prime}}}{\rm TeV} \right) {\rm GeV}.
	\label{eq:ZpYsplitting}
\end{equation}
The decay branching fractions of $Z^{\prime}$ in the SU(3) simple group
model are given in Table~\ref{tab:ZprimeBRs}, assuming that decays
to TeV-scale fermion-partner pairs or to $Y^0 \eta$ are kinematically 
forbidden and neglecting final-state masses.

\TABLE{
\begin{tabular}{|l|cccc|}
\hline \hline
Decay mode & \multicolumn{4}{c|}{Branching fraction} \\
                 & \multicolumn{2}{c}{SU(3) simple group}
                     & \multicolumn{2}{c|}{Littlest Higgs} \\
                 & universal & anomaly-free 
                     & $\cot\theta = 1$ & $\cot\theta = 0.2$ \\
\hline
$ee=\mu\mu=\tau\tau$ & 3.0\% & 3.7\% & 4.2\% & 0.60\% \\
$\sum_{i=1}^3 \nu_i \bar\nu_i$ & 5.2\% & 6.3\% & 12.5\% & 1.8\% \\
$t\bar t$ & 15\% & 18\% & 12.5\% & 1.8\% \\
$b\bar b$ & 13\% & 16\% & 12.5\% & 1.8\% \\
$u\bar u = c \bar c$ & 15\% & 13\% & 12.5\% & 1.8\% \\
$d \bar d = s \bar s$ & 13\% & 11\% & 12.5\% & 1.8\% \\
$ZH = WW$ & 0.87\% & 1.1\% & 0 & 43\% \\
\hline
Total width & 15 $\left( \frac{M_{Z^{\prime}}}{\rm TeV} \right)$ GeV 
		& 12 $\left( \frac{M_{Z^{\prime}}}{\rm TeV} \right)$ GeV 
     & 34 $\left( \frac{M_{Z_H}}{\rm TeV} \right)$ GeV
         & 9.5 $\left( \frac{M_{Z_H}}{\rm TeV} \right)$ GeV \\
\hline \hline
\end{tabular}
\caption{\label{tab:ZprimeBRs}
Decay branching fractions of $Z^{\prime}$ in the SU(3) simple group
model with universal and anomaly-free fermion embeddings, and of $Z_H$ in
the Littlest Higgs model for $\cot\theta = 1$ and 0.2.
Final-state masses are neglected.}
}

\subsection{Testing the Higgs mass divergence cancellation in the gauge sector}
\label{sec:gaugedivtest}

The defining feature of the little Higgs models is the cancellation of
the Higgs mass quadratic divergence at one-loop level.  Here we
investigate this cancellation in the gauge sector, as embodied in 
Eq.~(\ref{eq:gaugecancellation}).
Ideally, one could hope to measure directly the couplings 
$G_{HHV^{\prime}V^{\prime}}$ for each heavy gauge boson $V^{\prime}$ 
in the model.  This could be
done by measuring associated production of $H$ with a heavy gauge 
boson; e.g., $Z^{\prime}H$ associated production in the SU(3) 
simple group model.  This probes $G_{HHZ^{\prime}Z^{\prime}}$ through
the diagram involving $q \bar q \to Z^{\prime *} \to Z^{\prime} H$, where
one Higgs boson has been replaced by its vev in the interaction vertex.  
Ideally, one will want to measure both the magnitude and the sign of 
$G_{HHZ^{\prime}Z^{\prime}}$, perhaps through its interference with 
the similar diagram containing an $s$-channel $Z$.  A detailed study is
needed.

In addition to testing the divergence cancellation, the measurement
of the $HHV^{\prime}V^{\prime}$ couplings
also sheds light onto the structure of the model by revealing which
heavy gauge bosons are involved in the cancellation of each SM 
contribution to the Higgs mass quadratic divergence.
In the Littlest Higgs model, $Z_H$ cancels the divergence from the SM 
$W^3$ boson, $W_H^+$ and $W_H^-$ cancel the divergence from the SM $W^{\pm}$ 
bosons, and $A_H$ (if it is present) cancels the divergence from the 
SM hypercharge boson.
In contrast, in the SU(3) simple group model, $Z^{\prime}$ cancels the
divergences from the SM $W^3$ boson \emph{and} the hypercharge boson,
while $X$ (together with its isospin partner $Y$) cancels the divergence 
from the SM $W^{\pm}$ bosons.  Thus the 
$HHZ^{\prime}Z^{\prime}$ coupling strength 
that is characteristic of the little 
Higgs divergence cancellation mechanism can vary from model to model.  
In all product group models with SU(2)$^2\to$SU(2)$_L$ breaking
structure, the 
value of this coupling will be the same as in the Littlest Higgs model.  In 
simple group models the value of the coupling will be different, and 
may depend on the model.  For example, in the SU(4)$\times$U(1)$_X$ model
of Ref.~\cite{KS}, the two broken diagonal generators mix to form
mass eigenstates $Z^{\prime}$ and $Z^{\prime\prime}$, which both take 
part in the divergence cancellation; the sum rule then reads
\begin{equation}
	G_{HHZZ} + G_{HHZ^{\prime}Z^{\prime}} 
	+ G_{HHZ^{\prime\prime}Z^{\prime\prime}} = 0.
\end{equation}

A second approach to test the Higgs mass divergence cancellation,
first described in Ref.~\cite{Burdman}, 
is to measure the couplings of Higgs bosons to
one SM gauge boson and one new heavy gauge boson: e.g., 
$HHW^+W_H^-$, $HHZZ_H$ in the Littlest Higgs model \cite{Burdman}.
This approach works only for the product group models,
in which these couplings show a characteristic
$\cot 2\theta$ dependence which is fixed by the collective breaking
structure of the gauge couplings and the nonlinear transformation
of the SM Higgs doublet under the enlarged gauge symmetry.
A ``Big Higgs'' model, in which the
Higgs doublet transformed linearly under one of the two SU(2) gauge groups
as the fermion doublets do, would have a $HHZZ_H$ coupling 
proportional to $g \cot\theta$ [if $h$ transformed under SU(2)$_1$]
or $g \tan\theta$ [if $h$ transformed under SU(2)$_2$].
These couplings can be probed in the decays $Z_H \to Z H$ and 
$W_H \to W H$ \cite{Burdman} from $Z_H,W_H$ bosons produced on-shell, 
and will thus be more straightforward 
to measure than the $HHV^{\prime}V^{\prime}$ couplings discussed above.
The $\cot\theta$ dependence of the $Z_H$ production cross section
and decay to dileptons and the $\cot 2\theta$ dependence of the $Z_H$
decay to $ZH$ can be probed simultaneously by measuring the rate
into dileptons and the rate into $ZH$ \cite{Burdman}; these rates
will fall upon the curve shown in Fig.~\ref{fig:llZH} for the Littlest
Higgs model.

\FIGURE{
\resizebox{0.7\textwidth}{!}
{\rotatebox{270}{\includegraphics[50,50][555,590]{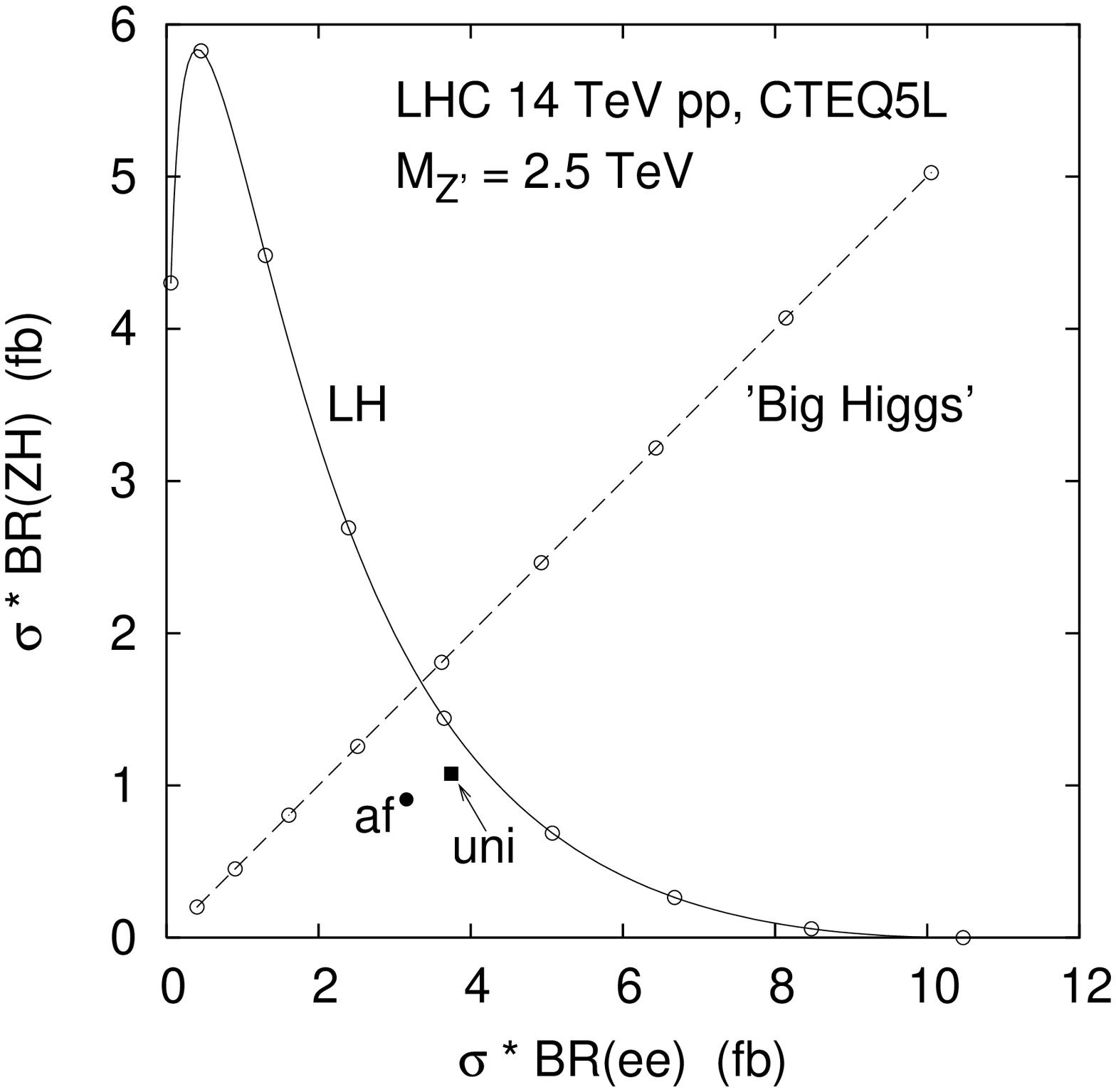}}}
\caption{\label{fig:llZH}
Cross section times branching ratio into $ee$ versus $ZH$
for a 2.5 TeV $Z^{\prime}$ boson in the Littlest Higgs model 
(`LH', solid line),
the SU(3) simple group model with anomaly-free (`af', filled circle)
and universal (`uni', filled square) fermion embeddings, and 
the ``Big Higgs'' model of Ref.~\cite{Burdman} (dashed line).
Open circles on the Littlest Higgs and Big Higgs lines indicate $\cot\theta$
values from 0.2 to 1 (left to right) in steps of 0.1.
Branching ratios are computed assuming that only decays into pairs of 
SM particles are present; we ignore, e.g., $Z_H \to A_H H$ and 
$Z^{\prime} \to Y \eta$.  We neglect all final-state particle
masses except that of the top quark.}
}

In simple group models, the $HHZZ^{\prime}$ coupling does \emph{not} 
provide a probe of the Higgs mass divergence cancellation because in these
models this coupling is not directly related to the crucial 
$HHZ^{\prime}Z^{\prime}$ vertex that takes part in the cancellation 
of the Higgs mass quadratic divergence in the gauge sector.
In fact, in the SU(3) simple group model, the $HHZZ^{\prime}$ coupling 
is fixed by the extended gauge structure and
would be the same in any model with the gauge group SU(3)$\times$U(1), 
whether or not the little Higgs mechanism were realized.
The rates of $Z^{\prime}$ into dileptons and into $ZH$ 
in the SU(3) simple group model are predicted uniquely
for the universal and anomaly-free fermion embeddings, as shown 
in Fig.~\ref{fig:llZH}.
In order to test the cancellation of the quadratic divergence
in simple group models, it is thus very important to uncover the gauge
structure and fermion embedding of the model. For this purpose, we now turn
to a discussion of the determination of the $Z^{\prime}$ 
properties in the simple group models.

\subsection{Identifying the $Z^{\prime}$}

In addition to testing the little Higgs \emph{mechanism} in the gauge
sector as described in the previous section, one must also identify
the \emph{model} to which a newly-discovered $Z^{\prime}$ boson
belongs.  This entails identifying the extended gauge structure and
determining how the SM fits into it.
We examine here some techniques that can be used at the LHC to shed 
light on the couplings of the $Z^{\prime}$.  We consider
the $Z_H$ of the Littlest Higgs model and the $Z^{\prime}$ of the SU(3)
simple group model, with both the universal and anomaly-free fermion 
embeddings.  
As examples of other new physics possibilities, we also
consider a sequential $Z^{\prime}$ with the same couplings to fermions
as the SM $Z$ boson,
the $Z^{\prime}_{\psi}$ and $Z^{\prime}_{\chi}$ bosons of the $E_6$ model
\cite{E6model},
and $Z_R$ of the left-right symmetric model \cite{LRsymmodel}.

\subsubsection{Rate in dileptons}

A $Z^{\prime}$ boson will most likely be first discovered in decays
to dileptons.  The dilepton rate then immediately
tells us the production cross section times the leptonic branching ratio,
and thus fixes a combination of the $Z^{\prime}$ couplings to up and 
down quarks (in the production cross section), the $Z^{\prime}$ coupling
to leptons (in the decay partial width), and the $Z^{\prime}$ total
width (which enters the branching ratio to leptons).
While the $Z^{\prime}$ couplings to up and down quarks enter the
production cross section together, multiplied by the appropriate 
parton densities, it may be possible to separate them experimentally
by fitting the shape of the $Z^{\prime}$ rapidity distribution to 
high-precision measurements of the up and down quark parton 
densities \cite{Dittmar}.

The SU(3) simple group model gives a definite prediction for the 
$Z^{\prime} \to \ell^+\ell^-$ rate in each of the fermion embeddings,
shown on the horizontal axis of Fig.~\ref{fig:llZH}.
If extra decay modes of $Z^{\prime}$ to the heavy fermion partners 
are kinematically allowed, they will increase the 
$Z^{\prime}$ total width and thus decrease the rate into dileptons.
Decays of $Z^{\prime}$ into one SM
and one heavy fermion are suppressed by the heavy-light mixing, 
$\sim v^2/f^2$.  Thus only decays into pairs of heavy fermions can 
contribute significantly; these are likely to be either kinematically 
inaccessible or heavily suppressed by phase space.
In the Littlest Higgs model, the rate of $Z_H$ into dileptons depends
on the free parameter $\cot\theta$.  Thus, in this channel, the
Littlest Higgs model can fake any other $Z^{\prime}$ model for an
appropriate value of $\cot\theta$. 

The rate in dileptons is uniquely predicted for 
the left-right symmetric model $Z_R$ and for
a sequential $Z^{\prime}$ (unless a tunable coupling is introduced by hand).
The $Z^{\prime}$ bosons in the $E_6$ model can mix, introducing a
free parameter in their cross sections; however, the cross section is
still constrained within a particular range for a $Z^{\prime}$ of given mass,
and the mixing angle can be extracted from the cross section.
A $Z^{\prime}$ from an extra U(1) gives a rate in dileptons tunable with
the U(1) coupling.  Therefore, while this rate measurement gives some
valuable information about the $Z^{\prime}$ couplings, it cannot uniquely
determine the model.

\subsubsection{Decay branching fractions to other fermion species}

In order to probe the $Z^{\prime}$ couplings to fermions in more
detail, one must look for $Z^{\prime}$ decays into additional 
fermion species.  This opens a window onto the relative couplings 
of the $Z^{\prime}$ to particles with different hypercharges.
Decays into neutrinos are only accessible through the $Z^{\prime}$ 
total width, which in little Higgs models is typically smaller 
than the detector dilepton mass resolution (see Table~\ref{tab:ZprimeBRs}).  
We thus consider decays into pairs of quarks.
This is a more difficult search than detecting the $Z^{\prime}$ in 
dileptons because of the large dijet background at the LHC.
However, it may be possible to detect the $Z^{\prime}$ decaying into 
top quark pairs, as a peak in the $t\bar t$ invariant mass spectrum, 
or into bottom quark pairs, as a peak in the $b$-tagged dijet invariant 
mass spectrum.

Measuring the rate of the $Z^{\prime}$ into top (bottom) 
quark pairs and taking 
the ratio with the rate to dileptons gives the ratio of partial widths
into top (bottom) versus electrons, shown in Table~\ref{tab:BRtBRe}.
\TABLE{
\begin{tabular}{|c|ccc|cccc|}
\hline \hline
 & $Z_H$ & $Z^{\prime}_{\rm uni}$ & $Z^{\prime}_{\rm af}$ &
	$Z^{\prime}_{\rm seq}$ & $Z^{\prime}_{\psi}$ & $Z^{\prime}_{\chi}$ &
		$Z_R$ \\
\hline
${\rm BR}(tt)/{\rm BR}(ee)$ &
	3.0 & 4.8 & 4.8 &
		3.4 & 3.0 & 0.6 & 4.4 \\
${\rm BR}(bb)/{\rm BR}(ee)$ &
	3.0 & 4.4 & 4.4 & 
		4.4 & 3.0 & 3.0 & 7.8 \\
\hline \hline
\end{tabular}
\caption{\label{tab:BRtBRe}
Ratio of branching fractions into $t \bar t$ ($b \bar b$) 
versus $e^+e^-$ for $Z^{\prime}$ bosons in various models.
From left to right: $Z_H$ (Littlest Higgs), $Z^{\prime}_{\rm uni}$
and $Z^{\prime}_{\rm af}$ (SU(3) simple group with universal and anomaly-free
fermion embeddings, respectively), $Z^{\prime}_{\rm seq}$ (sequential
$Z^{\prime}$), $Z^{\prime}_{\psi}$ and $Z^{\prime}_{\chi}$ ($E_6$ model),
and $Z_R$ (left-right symmetric model).
Final-state masses are neglected; the top mass dependence can be included
by multiplying ${\rm BR}(tt)/{\rm BR}(ee)$ by $(1-R)\sqrt{1-4R}$,
where $R = m_t^2/M_{Z^{\prime}}^2$.}
}
In the Littlest Higgs model, this ratio is fixed independent of
$\cot\theta$ because the $\cot\theta$ dependence enters
the couplings to all fermions in the same way.  Further, 
because $Z_H$ couples universally to all fermion doublets,
this ratio is just given by the number of color degrees of freedom, 
$N_c = 3$ (neglecting final-state masses).
This ratio is also fixed in the SU(3) simple group model;
it is different from the value in the Littlest Higgs model
because of the U(1)$_X$ content of the $Z^{\prime}$,
which introduces a dependence on the fermion hypercharge.
Note that the ratio of top (bottom) to electron partial widths is the 
same in the universal and the anomaly-free fermion embeddings, because
in both embeddings the leptons and the third generation of quarks all
transform as $\mathbf{3}$s of SU(3); the difference between the two 
embeddings appears only in the first two generations of quarks.

Similarly, these ratios are independent of model parameters for 
a sequential $Z^{\prime}$,
the $E_6$ $Z^{\prime}_{\psi}$ and $Z^{\prime}_{\chi}$, 
and the left-right symmetric $Z_R$.  The $E_6$ $Z^{\prime}_{\psi}$ and 
$Z^{\prime}_{\chi}$ mix in general, leading to intermediate 
values of the partial width raitos.  $Z^{\prime}_{\psi}$ has the
same ${\rm BR}(tt)/{\rm BR}(ee)$ and ${\rm BR}(bb)/{\rm BR}(ee)$ as
the Littlest Higgs $Z_H$,
and $Z^{\prime}_{\chi}$ has the same ${\rm BR}(bb)/{\rm BR}(ee)$,
as the Littlest Higgs $Z_H$.
Likewise, the sequential $Z^{\prime}$ has the same ${\rm BR}(bb)/{\rm BR}(ee)$ 
as the SU(3) simple group model $Z^{\prime}$; however, its
${\rm BR}(tt)/{\rm BR}(ee)$ is rather different.
Of course, the couplings of a $Z^{\prime}$ from an anomalous extra U(1) 
can be tuned to duplicate the predictions of any of these models.

\subsubsection{Forward-backward asymmetry}

The forward-backward asymmetry in 
$f_i \bar f_i \to Z^{\prime} \to f_f \bar f_f$
probes the chiral structure of the $Z^{\prime}$ couplings to the 
initial- and final-state fermions.  At the partonic level, this
asymmetry is defined as
\begin{equation}
	A_{FB}^{0,if} = \frac{N_F - N_B}{N_F + N_B}
	= \frac{3}{4} \mathcal{A}_i \mathcal{A}_f,
\end{equation}
where $N_F$ ($N_B$) is the number of events with the final-state fermion
momentum in the forward (backward) direction defined relative to the 
initial-state fermion.  The asymmetry $\mathcal{A}_f$ is defined in 
terms of the couplings $g_{L,R}^f$ as
\begin{equation}
	\mathcal{A}_f = \frac{(g_L^f)^2 - (g_R^f)^2}{(g_L^f)^2 + (g_R^f)^2}.
\end{equation}

Even though the LHC is a symmetric $pp$ collider, a forward-backward asymmetry
can be defined by taking advantage of the fact that the valence 
quarks in the proton tend to carry a higher momentum fraction $x$ 
than the sea (anti)quarks \cite{Rosner,Dittmar97}.
A ``hadronic'' forward-backward asymmetry can then be defined as
\begin{equation}
	A_{FB}^{\rm had} = \frac{N_F - N_B}{N_F + N_B},
\end{equation}
where now the forward direction for the final-state fermion is defined
relative to the boost direction of the $Z^{\prime}$ center-of-mass frame.
In the narrow-width approximation (neglecting interference between the 
$Z^{\prime}$ resonance and the continuum photon and $Z$ exchange),
$A_{FB}^{\rm had}$ is given in terms of the partonic asymmetries by
\begin{equation}
	A_{FB}^{\rm had} = 
	\frac{ \int dx_1 \sum_{q=u,d} A_{FB}^{0,qf} 
	\left( F_q(x_1)F_{\bar q}(x_2) - F_{\bar q}(x_1)F_q(x_2) \right) 
	{\rm sign}(x_1-x_2)}
	{ \int dx_1 \sum_{q=u,d,s,c} 
	\left( F_q(x_1)F_{\bar q}(x_2) + F_{\bar q}(x_1)F_q(x_2) \right) },
	\label{eq:Afbcalc}
\end{equation}
where $F_q(x_1)$ is the parton distribution function (PDF) 
for quark $q$ in the 
proton with momentum fraction $x_1$, evaluated at $Q^2 = M_{Z^{\prime}}^2$.
The momentum fraction $x_2$ is related to $x_1$ by the condition 
$x_1x_2 = M_{Z^{\prime}}^2/s$ in the narrow-width approximation.
Only $u$ and $d$ quarks contribute to the numerator since we explicitly
take the quark and antiquark PDFs to be identical for the sea quarks;
all flavors contribute to the denominator.

Here we consider $Z^{\prime}$ decays to $e^+e^-$ only, since it is 
much easier at LHC to determine the charge of a lepton than the charge
of a quark.  Decays to $\mu^+\mu^-$ can be added to double the statistics.
The relevant partonic asymmetries and $A_{FB}^{\rm had}$ 
are listed in Table~\ref{tab:asym} for the little Higgs models under 
consideration, as well as a number of other $Z^{\prime}$ models.
\TABLE{
\begin{tabular}{|c|ccc|cccc|}
\hline \hline
  & $Z_H$ & $Z^{\prime}_{\rm uni}$ & $Z^{\prime}_{\rm af}$
& $Z^{\prime}_{\rm seq}$ & $Z^{\prime}_{\psi}$ 
& $Z^{\prime}_{\chi}$ & $Z_R$ \\
\hline
$\mathcal{A}_e$ & 1
  & 0.15
  & 0.15
  & 0.15
  & 0
  & 0.8
  & $-0.28$
  \\
$\mathcal{A}_u$ & 1
  & 0.77
  & 0.67
  & 0.67
  & 0
  & 0
  & $-0.95$
  \\
$\mathcal{A}_d$ & 1
  & 0.94
  & 0.91
  & 0.94
  & 0
  & $-0.8$
  & $-0.97$
  \\
\hline
$A_{FB}^{0,ue}$ & 0.75
  & 0.087
  & 0.076
  & 0.076
  & 0
  & 0
  & 0.20
  \\
$A_{FB}^{0,de}$ & 0.75
  & 0.11
  & 0.10
  & 0.11
  & 0
  & $-0.48$
  & 0.20
  \\
\hline
$A_{FB}^{\rm had}$ & 0.44
  & 0.054
  & 0.049
  & 0.049
  & 0
  & $-0.077$
  & 0.12
  \\
\hline \hline
\end{tabular}
\caption{\label{tab:asym}
Coupling asymmetries before cuts for $Z^{\prime}$ bosons in the models 
listed in Table~\ref{tab:BRtBRe}.
$A_{FB}^{\rm had}$ is calculated for the LHC ($pp$ collisions at 14 TeV)
using CTEQ5L PDFs in the narrow width approximation,
with $M_{Z^{\prime}} = 2$ TeV.}
}
The hadronic forward-backward asymmetry
$A_{FB}^{\rm had}$ varies with $M_{Z^{\prime}}$ due to the shape of 
the PDFs.
The $Z^{\prime}$ mass dependence is shown in Fig.~\ref{fig:asym}
for the models included in Table~\ref{tab:asym}.  It is interesting
to note that the asymmetries of the $E_6$ $Z^{\prime}$ bosons are less than
or equal to zero, unlike the rest of the models.  The $E_6$ boson 
asymmetries remain negative definite for arbitrary mixing between
$Z^{\prime}_{\psi}$ and $Z^{\prime}_{\chi}$: $A_{FB}^{0,ue}$ is always
zero and $A_{FB}^{0,de}$ varies between $-0.75$ and 0 depending on the 
mixing angle.
\FIGURE{
\resizebox{0.7\textwidth}{!}
{\rotatebox{270}{\includegraphics[50,50][555,590]{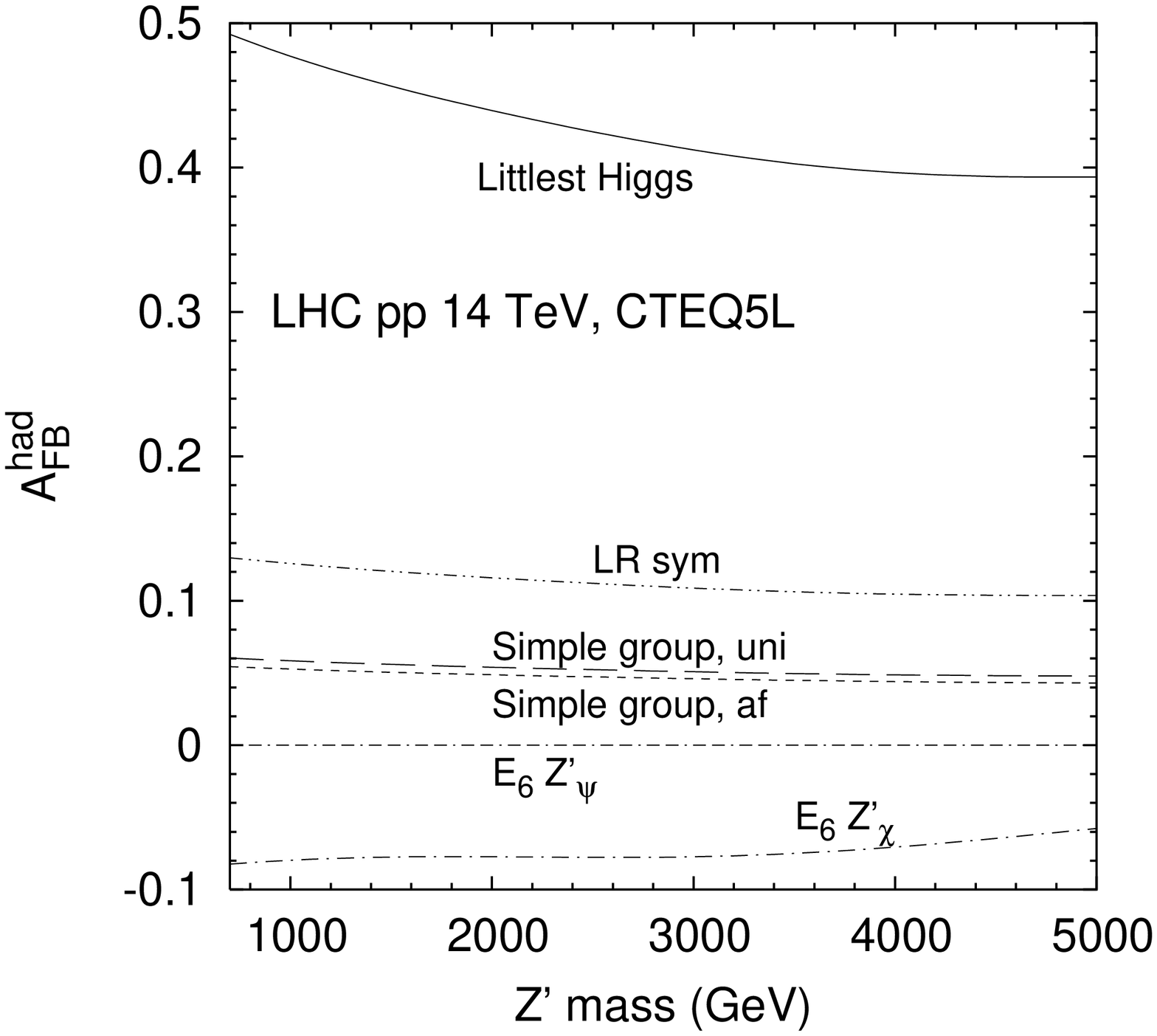}}}
\caption{\label{fig:asym}
The hadronic forward-backward asymmetry
$A_{FB}^{\rm had}$ as a function of $M_{Z^{\prime}}$ for the
models in Table~\ref{tab:asym}.  The curve for a sequential $Z^{\prime}$
is identical to the SU(3) simple group $Z^{\prime}$ with anomaly-free (af)
fermion embedding.}
}
In Eq.~(\ref{eq:Afbcalc}) we have expressed $A_{FB}^{\rm had}$ as a 
single number, integrated over rapidity, which depends on both 
$A_{FB}^{0,ue}$ and $A_{FB}^{0,de}$.  It may be possible to 
extract these two 
quantities separately by fitting the asymmetry as a function
of the $Z^{\prime}$ rapidity to high-precision measurements of the
up and down quark parton densities \cite{Dittmar};
however, this would require a huge amount of luminosity.

In the Littlest Higgs model, a measurement of $A_{FB}^{\rm had}$ 
would provide a spectacular test of the model because it would confirm
that $\mathcal{A}_u = \mathcal{A}_d = \mathcal{A}_e = \pm 1$;
that is,
that the $Z_H$ couplings to fermions are either purely left-handed
or purely right-handed.  The sign ambiguity is due to the fact that
$A_{FB}^{0,if}$ depends on the product $\mathcal{A}_i \mathcal{A}_f$.
Together with measurements of ${\rm BR}(tt)/{\rm BR}(ee)$ and/or
${\rm BR}(bb)/{\rm BR}(ee)$, which would demonstrate the universality
of the $Z_H$ couplings to fermions, and the discovery of the
$W_H^{\pm}$ degenerate in mass and with a related production rate,
this measurement would confirm $Z_H$ as a member of an SU(2) triplet
of gauge bosons.  In such a case we learn that the SM SU(2)$_L$ 
gauge symmetry arises from the diagonal breaking of [SU(2)]$^2$,
with the SM fermion doublets transforming under one of the two SU(2)
gauge groups.
A measurement of $A_{FB}^{\rm had}$ will also provide a test of
the SU(3) simple group model and the other $Z^{\prime}$ models
considered, since it probes another independent combination of
the $Z^{\prime}$ couplings to fermions.

\subsubsection{Bosonic decay modes}

Measuring the bosonic decay modes $Z^{\prime} \to ZH$ and 
$Z^{\prime} \to W^+W^-$
probes the transformation properties of the Higgs doublet under the
extended gauge symmetry and the mixing of $Z$ and $Z^{\prime}$ induced
by electroweak symmetry breaking.
As described in detail in Sec.~\ref{sec:gaugedivtest},
this can shed light on the little Higgs mechanism in the gauge sector,
but it also provides useful information about the model structure.
Also of interest are bosonic decay modes of the $Z^{\prime}$ involving 
non-SM bosons in the final state, such as $Z^{\prime} \to Y \eta$ in the
SU(3) simple group model or $Z_H \to A_H H$ in the Littlest Higgs model.
Detecting and measuring the branching 
fractions of these decay modes provides additional information on the structure
of the extended gauge group and the mixings among the new gauge bosons.

\section{Other phenomenological features of the SU(3) simple group model}
\label{sec:other}

In this section we collect some additional features of the SU(3) simple group
model not directly relevant to the simple group/product group classification
and the identification of the little Higgs mechanism.

\subsection{The heavy leptons}

In the SU(3) simple group model, the three lepton doublets of the SM are 
enlarged into triplets.  The model thus contains three heavy neutral 
states $N_m$.
The scalar interactions of the leptons can be written as
\begin{equation}
	\mathcal{L}_Y = i \lambda_{N_m} N_m^c \Phi_2^{\dagger} L_m
	+ \frac{i \lambda_e^{mn}}{\Lambda} e^c_m \epsilon_{ijk} 
	\Phi_1^i \Phi_2^j L_n^k + {\rm h.c.},
	\label{eq:LYSU3lept}
\end{equation}
where $m,n = 1,2,3$ are generation indices, $i,j,k = 1,2,3$ are SU(3) indices,
$L_m = (\nu, e, iN)_m^T$ are the lepton triplets, and $N_m^c$ are 
right-handed neutral leptons that marry the $N_m$ and get masses of order 
$f \sim$ TeV.  We neglect neutrino masses; a nice extension of
the SU(3) simple group model including neutrino masses was presented
in Ref.~\cite{LHseesaw}.

Equation~(\ref{eq:LYSU3lept}) generates masses for $N_m$,
\begin{equation}
	M_{N_m} = \lambda_{N_m} s_{\beta} f.
	\label{eq:MNm}
\end{equation}
The Lagrangian also contains a term
\begin{equation}
	\mathcal{L}_Y \supset - \frac{\lambda_{N_m} c_{\beta}}{\sqrt{2}} 
		H N_m^c \nu + {\rm h.c.}
	= - \frac{M_{N_m}}{\sqrt{2} f t_{\beta}} H N_m^c \nu + {\rm h.c.}
\end{equation}
for each generation, leading to mixing between the $N_m$ and the SM
neutrinos given by $N = N_0 - \delta_{\nu} \nu_0$, where $N_0,\nu_0$
denote the electroweak eigenstates of each generation and $\delta_{\nu}$
was given in Eq.~(\ref{eq:deltaq}).
This mixing gives rise to
the couplings of $N$ to $eW$ and $\nu Z$ with Feynman rules
\begin{equation}
	W_{\mu}^+ \overline{N} e: \ \frac{i g \delta_{\nu}}{\sqrt{2}}
	\gamma_{\mu} P_L,
	\qquad \qquad
	Z_{\mu} \overline{N} \nu: \ 
	\frac{i g \delta_{\nu}}{2 c_W} \gamma_{\mu} P_L.
\end{equation}
Because the $N_m$ carry lepton number, their production at the LHC 
requires an additional lepton in the final state and can thus proceed
only through $s$-channel gauge boson exchange, e.g., 
$q \bar q^{\prime} \to W^{+*} \to N e^+$.
Their decays, into $\nu H$, $e W$ and $\nu Z$, along with
$e X$, $\nu Y$ and $\nu \eta$ if kinematically accessible,
will be spectacular.
The $N_m$ could also be produced at a linear collider of sufficient 
energy through $t$-channel $W$ exchange, 
$e^+e^- \to \bar \nu N$.

\subsection{The $X$ and $Y$ gauge bosons}

The heavy gauge bosons $X^-,Y^0$ correspond to the off-diagonal broken
generators of SU(3) and thus communicate between the
SU(2)$_L$ doublet fermions and the SU(2)$_L$ singlets, with couplings
of gauge strength of the form $XQq^{\prime}$ and $YQq$ as summarized
in Table~\ref{table-Q}.  
These couplings can play a role in $T$ or $Q$ decay
if the corresponding final states are kinematically accessible.  They 
will not play a significant role in single $T$ or $Q$ production because the 
initial-state couplings of $X^-,Y^0$ to pairs of SM fermions are 
suppressed by $v/f$.
While $X^-,Y^0$ could be produced in association with $T$ or $Q$, e.g., 
$b \to T X^-$, these processes have two TeV-mass particles in the final state
and will be limited by phase space.

The production cross sections of the $X$ and $Y$ gauge bosons in Drell-Yan
are very small.  We thus consider other ways of producing these particles.
If they are light enough, $X$ and $Y$
can be produced in the decays of the TeV-scale quark partners:
\begin{equation}
	T \to X^+ b, \overline{Y}^0 t,
	\qquad \qquad
	U_j \to X^+ d_j, \overline{Y}^0 u_j
	\qquad {\rm or} \qquad
	D_j \to X^- u_j, Y^0 d_j.
	\label{eq:XYfprod}
\end{equation}
For example, taking $M_T = 1$ TeV, $M_Y = 0.9$ TeV and $\lambda_T = 1$,
we find ($T \to t \overline{Y}^0$ is kinematically forbidden for these masses),
\begin{equation}
	{\rm BR}(T \to b X^+) \simeq 0.55\%.
\end{equation}
Similarly, $X$ and $Y$ can be produced through the decays of the
heavy lepton partners, $N \to X^+ \ell^-, \overline{Y}^0 \nu$.
The $X$ and $Y$ bosons can also be pair produced by electroweak interactions
via the triple gauge couplings in Table~\ref{tab:gauge1};
however, pair production of these TeV-scale particles will suffer 
from reduced phase space and off-shell $s$-channel propagators compared
to Drell-Yan production of the $Z^{\prime}$.

If they are heavy enough, $X$ and $Y$ can decay to one SM fermion and one
TeV-scale fermion partner,
\begin{eqnarray}
	&&X^+ \to T \overline{b}, U_j \overline{d}_j, N_i \ell^+,
	\qquad \qquad
	Y^0 \to t \overline{T}, u_j \overline{U}_j, \nu_i \overline{N}_i
	\qquad \qquad {\rm (universal)}
	\nonumber \\
	&& X^+ \to T \overline{b}, u_j \overline{D}_j, N_i \ell^+,
	\qquad \qquad
	Y^0 \to t \overline{T}, d_j \overline{D}_j, \nu_i \overline{N}_i
	\qquad \qquad {\rm (anomaly \ free)}.
\end{eqnarray}
Neglecting the SM fermion mass, the partial widths for these decays
are given by
\begin{equation}
	\Gamma(V \to F \bar f) = \frac{N_c g^2}{32 \pi} 
		\beta^2 \left[ 1 - \frac{\beta}{3} \right] M_V
	= 4.2 \ N_c \beta^2 \left[ 1 - \frac{\beta}{3} \right]
	\left( \frac{M_V}{\rm TeV} \right) {\rm GeV},
\end{equation}
where $N_c=1$ or 3 is the number of colors and $\beta = (1 - M_F^2/M_V^2)$.
This decay mode and the production in Eq.~(\ref{eq:XYfprod}) are 
mutually exclusive, depending on the relative masses
of $X,Y$ and the TeV-scale fermion partners.

If the decay to one SM fermion and one TeV-scale fermion partner is
kinematically inaccessible, $X$ and $Y$ can decay to pairs of SM fermions 
through their mixings with the TeV-scale fermion partners,
with partial widths proportional to 
$\delta_t^2, \delta_{\nu}^2 \propto v^2/f^2$.
The decays of $X$ are independent of the fermion embedding,
\begin{equation}
	X^- \to b \bar t, d_j \bar u_j, \ell^- \bar \nu,
\end{equation}
while the decays of $Y$ depend on the fermion embedding, since
$Y$ can decay only to fermions that mix with a heavy partner:
\begin{equation}
	Y^0 \to t \bar t, u_j \bar u_j, \nu \bar \nu
	\qquad {\rm (universal)},
	\qquad \qquad
	Y^0 \to t \bar t, d_j \bar d_j, \nu \bar \nu
	\qquad {\rm (anomaly \ free)}.
\end{equation}
Unfortunately, there are no decays of $Y$ to charged dileptons because
$N_i$ mix only with the neutrinos.
The decays $Y^0 \to t \bar t$, $X^- \to b \bar t$ are controlled 
by $\delta_t$, while the decays to the first two quark generations 
and to the leptons are controlled by the smaller $\delta_{\nu}$.  
Thus, decays to third generation quarks will have a somewhat larger 
partial width.  Neglecting final-state masses, the relevant partial
widths are
\begin{eqnarray}
	\Gamma(X^- \to b \bar t) = \Gamma(Y^0 \to t \bar t)
	&=& \frac{3 g^2}{48 \pi} \delta_t^2 M_Y
	= 0.51 \lambda_T^2 \left( \frac{\rm TeV}{M_T} \right)^2
		\left( \frac{M_Y}{\rm TeV} \right) {\rm GeV},
	\nonumber \\
	\Gamma(X^- \to jj) = \Gamma(Y^0 \to jj)
	&=& 2 \frac{3 g^2}{48 \pi} \delta_{\nu}^2 M_Y
	= \frac{0.11}{t_{\beta}^2} 
		\left( \frac{\rm TeV}{M_Y} \right) {\rm GeV},
	\nonumber \\
	\Gamma(X^- \to \ell \bar\nu) = \Gamma(Y^0 \to \nu\bar\nu)
	&=& 3 \frac{g^2}{48 \pi} \delta_{\nu}^2 M_Y
	= \frac{0.054}{t_{\beta}^2} 
		\left( \frac{\rm TeV}{M_Y} \right) {\rm GeV},
\end{eqnarray}
where $jj$ denote jets from quarks of the first two generations and 
the decays to leptons are summed over all three generations.
Finally, $Y$ can decay to $H \eta$ via the coupling in 
the last row of Table~\ref{tab:gauge2},
\begin{equation}
	\Gamma(Y^0 \to H \eta) = \Gamma(\overline{Y}^0 \to H \eta)
	= \frac{g^2 M_Y}{384 \pi}
	= 0.35 \left( \frac{M_Y}{\rm TeV} \right) {\rm GeV}.
\end{equation}

\subsection{The singlet pseudoscalar $\eta$}
\label{sec:eta}

The scalar sectors of little Higgs models are very model-dependent.
For completeness, however, we briefly sketch here the decay modes of
the singlet (pseudo-)scalar $\eta$ in the SU(3) simple group model.  
A more detailed analysis of 
the $\eta$ phenomenology can be found in Ref.~\cite{KRR}.
The singlet scalar $\eta$, which naturally gets a mass of a couple hundred
GeV, can decay to pairs of SM fermions with couplings
that depend on the SM fermion masses.  These couplings receive contributions
from the usual fermion Yukawa couplings, via the expansion of the nonlinear
sigma model fields, and from the couplings of $\eta$ to a SM fermion and its
TeV-scale partner combined with the $F$--$f$ mixing.  
These couplings are all of order $m_f/f$, that is, suppressed by $v/f$ 
relative to the usual fermion Yukawa couplings.  
The $\eta$ can also decay into a Higgs boson and an off-shell $Y$, which
then decays to a pair of SM fermions with couplings suppressed by the
$F$--$f$ mixing.  We expect the decays
of $\eta$ into pairs of fermions to dominate, with branching fractions
proportional to the fermion masses up to order-one factors related to the
contribution from the $F$--$f$ mixing.  The total width of $\eta$ will be
suppressed by $v^2/f^2$ compared to that of a ``bosophobic'' Higgs of the
same mass; however, this width will be too narrow to measure directly and
too wide to give rise to displaced vertices, and thus can only be probed
through production cross sections.

\section{Conclusions}
\label{sec:conclusions}

The little Higgs models represent a new approach to electroweak
symmetry breaking that will be accessible at future high-energy
colliders.  These models stabilize the 
hierarchy between a relatively low cutoff scale $\sim 10$ TeV and 
the electroweak scale by making the Higgs a pseudo-Goldstone boson
of a spontaneously broken approximate global symmetry.
Implementing such a global symmetry requires enlarging the gauge,
fermion and scalar sectors of the SM.  Little Higgs models therefore
predict new gauge bosons, fermions and scalars at or below the TeV 
scale, which offer exciting possibilities for beyond-the-SM collider 
phenomenology at the LHC.

However, many models of physics beyond the SM contain new gauge bosons,
fermions, and/or scalars at or below the TeV scale.  If such particles
are discovered, one will want to know whether they implement the 
little Higgs mechanism by canceling the one-loop quadratic divergence
in the Higgs mass due to the SM gauge bosons, top quark, and Higgs 
quartic coupling.  

We categorized the many little Higgs models into two classes based on
the structure of the extended electroweak gauge group: 
\begin{itemize}
\item[(a)] product
group models, in which the SM SU(2)$_L$ gauge group arises from the
diagonal breaking of two or more gauge groups, and 
\item[(b)] simple group
models, in which the SM SU(2)$_L$ gauge group arises from the breaking
of a single larger gauge group down to an SU(2) subgroup.
\end{itemize}
As prototypes of each class, we studied the experimental signatures
of the Littlest Higgs model and the SU(3) simple group model, 
respectively.

The ``smoking guns'' for the little Higgs mechanism -- the cancellation
of the Higgs mass quadratic divergences between loops of SM particles
and loops of the new particles -- are quite straightforward and allow
one to distinguish models that implement the little Higgs mechanism 
from other models that have a similar superficial phenomenology.
In the top sector, the little Higgs mechanism appears as a sum rule 
involving the top quark Yukawa coupling, the $TtH$ or $TbW$ coupling 
$\lambda_T$, and the dimension-five $TTHH$ coupling $\lambda_T^{\prime}$.  
In product group models, the simple structure of the
top mass generation mechanism ensures that $\lambda_T^{\prime}$ can
be expressed in terms of $\lambda_T$, $M_T$ and the top Yukawa coupling.
The little Higgs mechanism can then be checked by measuring 
$\lambda_T$ and $M_T$,
computing the condensate $f$, and comparing with $f$ from the gauge sector.
In simple group models, on the other hand, the top mass generation mechanism
is slightly more complicated and involves two (or more) TeV-scale condensates.
This introduces an extra free parameter into the top sector (which can 
be chosen as the ratio of the two condensates, $f_2/f_1 \equiv t_{\beta}$),
so that all three parameters $\lambda_T$, 
$\lambda_T^{\prime}$, and $M_T$ must be 
measured in the top sector.  We have not found a way to measure 
$\lambda_T^{\prime}$ directly at the LHC.  Instead, the required 
third parameter
can be measured from the production rate of the TeV-scale quarks associated 
with the first two generations in the simple group models.
These measurements of the extended top sector and the 
TeV-scale quark partners of the first two generations, if present, thus 
allow one to test the little Higgs mechanism in the top sector,
distinguish the structure of the top quark mass generation 
mechanism, and extract the model parameters that control the fermion sector.
We showed explicitly how these measurements allow one to distinguish the
top sector of a little Higgs model from a fourth-generation top-prime
and from a top see-saw model.

In the gauge sector, the little Higgs mechanism appears as a sum rule
involving the Higgs boson coupling to pairs of SM vector bosons and
to pairs of the new TeV-scale vector bosons.  The couplings involved in the
sum rule can be directly measured via 
$q \bar q \to V^{\prime *} \to V^{\prime} H$ 
associated production.  Measurement of these couplings allows one to 
test which new particles are responsible for canceling each of the SM 
contributions to the Higgs mass-squared quadratic divergence.
In product group models, the test of the little Higgs mechanism is 
particularly simple because of the collective breaking structure of 
the Higgs couplings to gauge bosons: it is enough to measure the
$Z_H Z H$ ($W_H W H$) couplings, which are accessible through
$Z_H \to Z H$ ($W_H \to W H$) decays.  The simple group models
have a different collective breaking structure in the gauge sector, however, 
so that a direct measurement of the $V^{\prime} V^{\prime} H$ 
couplings is necessary.
Additional measurements in the gauge sector will
shed light on the structure of the extended electroweak gauge group.
We showed explicitly how measurements of the properties of a $Z^{\prime}$
allow one to distinguish the $Z^{\prime}$ states present in little Higgs
models from the $Z^{\prime}$s in the $E_6$ and left-right symmetric
models and from a sequential $Z^{\prime}$.

The scalar sector is very model dependent.  It depends on the global
symmetry structure; therefore the classification of models into 
product group and simple group does not give a useful classification
of the scalar sector phenomenology.


\acknowledgments
We thank B.~Dobrescu, C.~Hill, 
B.~McElrath, J.~Terning, D.~Rainwater, M.~Schmaltz,
T.~Tait, and W.~Skiba for useful conversations.
HEL and LTW thank the Aspen Center for Physics for hospitality while
this work was initiated.
This work was supported in part by the U.S.~Department of Energy
under grant DE-FG02-95ER40896
and in part by the Wisconsin Alumni Research Foundation.
TH was also supported in part
by the National Natural Science Foundation of China.
LW was also supported in part by U.S.~Department of Energy
under grant DE-FG02-91ER40654.

\appendix
\section{Survey of little Higgs models}
\label{survey}

\subsection{Product group models}
\label{survey-product}

The majority of little Higgs models are product group models.
In addition to the Littlest Higgs, these include
the theory space models (the Big Moose \cite{LHidea} and the Minimal
Moose \cite{MinMoose}), the SU(6)/Sp(6) model of Ref.~\cite{SU6Sp6},
and two extensions of the Littlest Higgs with built-in custodial
SU(2) symmetry \cite{ChangWacker,ChangSU2}.  There are also product
group models with $T$-parity in the literature 
\cite{Tparity1,Tparity12,Tparity2,Tparity-pheno}; however, we 
do not address them here in any detail.  In general, the phenomenology
of models with $T$-parity is quite different from that discussed here;
however, the top partner is typically $T$-parity even so that its 
phenomenology can be taken over directly from the Littlest Higgs case.

We start with the theory space models.  The Minimal Moose \cite{MinMoose}
consists of two sites (where the gauge groups live) connected by four link
fields (scalar fields transforming under the gauge groups at either end 
of the link).  The electroweak gauge symmetry at one site is 
SU(2)$\times$U(1), while at the other it is SU(3) [or alternatively, a 
second copy of SU(2)$\times$U(1); electroweak precision constraints 
\cite{MinMooseEW} favor this second possibility].
The diagonal breaking of the gauge symmetry down to SU(2)$_L \times$U(1)$_Y$ 
leaves a set of
SU(3) gauge bosons [alternatively the broken SU(2)$\times$U(1) gauge bosons]
at the TeV scale.  The top quark mass is generated
by an interaction of the same form as Eq.~(\ref{eq:LYLH}), leaving a heavy
charge 2/3 electroweak singlet quark at the TeV scale.  The scalar spectrum
consists of two Higgs doublets, a complex triplet and a complex singlet
at the weak scale, with an additional Higgs doublet, triplet, and 
singlet at the TeV scale.  
The Big Moose \cite{LHidea} is an extended version of this structure,
with a longer chain of gauge groups connected by link fields that break
down to the diagonal SU(2)$\times$U(1), leaving a larger number of broken
gauge generators at the TeV scale.
Many different theory space structures yield the little Higgs mechanism,
with only mild topological constraints on the shape of the theory space
\cite{MooseTopology}.  In particular, the theory space can be chosen 
such that the low-energy theory contains only two Higgs doublets, giving
the extra light scalars of the Minimal Moose masses at the TeV scale 
\cite{MooseTopology}.  Theory space models always contain at least two
light Higgs doublets.

The SU(6)/Sp(6) model \cite{SU6Sp6} is similar to the Littlest Higgs, 
but starting with a global SU(6) symmetry broken down to Sp(6) at the
TeV scale by an antisymmetric condensate.  A subgroup 
[SU(2)$\times$U(1)]$^2$ of the global symmetry is gauged; the gauge
symmetry is broken down to SU(2)$_L \times$U(1)$_Y$ by the condensate,
leaving a set of SU(2)$\times$U(1) gauge bosons at the TeV scale.
The top quark mass is generated in exact analogy to Eq.~(\ref{eq:LYLH}),
leaving a heavy charge 2/3 electroweak singlet quark at the TeV scale.
The scalar spectrum consists of two light Higgs doublets, plus a complex
singlet at the TeV scale.

The extensions of the Littlest Higgs with built-in custodial
SU(2) symmetry \cite{ChangWacker,ChangSU2} were constructed in order to
avoid some of the electroweak precision constraints on the Littlest Higgs model
\cite{GrahamEW1,JoAnneEW,GrahamEW2}.  The first such extension is a hybrid
of the Littlest Higgs and the Minimal Moose with an 
SO(5)$\times$[SU(2)$\times$U(1)] gauge symmetry \cite{ChangWacker}.
It contains two light Higgs doublets, plus additional scalars at the TeV
scale due to the enlarged global symmetry.  It also contains extra TeV-scale
gauge bosons from the enlarged gauge symmetry.
The second such extension expands the global symmetry group to SO(9),
spontaneously broken down to SO(5)$\times$SO(4) \cite{ChangSU2}.
This model contains only a single light Higgs doublet, with three
scalar triplets and a singlet at the TeV scale.  The gauge symmetry
is [SU(2)$_L \times$SU(2)$_R$]$\times$[SU(2)$\times$U(1)], broken down
to the SM electroweak gauge group by the symmetry breaking condensate.
The model thus contains extra TeV-scale gauge bosons compared to the
Littlest Higgs.  The top sectors of both extensions are identical to that
of the Littlest Higgs.

The product group models all share two features.  First, the models
all contain a set of SU(2) gauge bosons at the TeV scale, obtained from
the diagonal breaking of two gauge groups down to SU(2)$_L$.  Some models
contain additional TeV-scale gauge bosons as well, from the breaking of
more than two SU(2) gauge groups or from the breaking of gauge groups
larger than SU(2).
Second, the models all generate the top quark mass from a Lagrangian
involving two terms, only one of which couples to the scalar sector
of the model.  This results in an extended top quark sector of the
same form as in the Littlest Higgs model.
These two features distinguish the product group
models from the simple group models, which we consider next.

\subsection{Simple group models}

In addition to the SU(3) simple group model, there are two other simple
group models in the
literature to date: the SU(4) simple group model \cite{KS} and the 
SU(9)/SU(8) model of Ref.~\cite{SkibaTerning}.  These two models depart
from the SU(3) simple group model in different directions.

The SU(4) simple group model \cite{KS} is a straightforward extension of
the SU(3) model to the electroweak gauge group SU(4)$\times$U(1)$_X$.  
It was introduced because the simplest version of the SU(3) model 
generates a Higgs quartic coupling only at one-loop level through
the Coleman-Weinberg potential, leading to a too-light
Higgs boson \cite{KS}.  This problem can be fixed by adding an extra 
term to the scalar Lagrangian \cite{Schmaltznote}, which explicitly
breaks a global U(1) symmetry in the model (and 
has the added benefit of giving mass to the $\eta$ pseudoscalar, which would 
otherwise be a Nambu-Goldstone boson).  The SU(4) model, on the other 
hand, generates a Higgs quartic coupling at tree-level, so the Higgs 
mass is easily large enough.

In the SU(4) simple group model the isospin doublets of the SM are all 
extended to quadruplets under SU(4).
A total of four scalar quadruplets are needed to break SU(4)$\times$U(1)$_X$
down to SU(2)$_L \times$U(1)$_Y$, which leads to extra light scalars
so that the low-energy theory contains two light Higgs doublets and 
two real singlets, plus three 
complex singlets which get masses of order $f \sim$ TeV.
The potential generated for the two Higgs doublets is not the most
general possible, yielding interesting relations among the Higgs masses 
and couplings; in fact, the potential for the two Higgs doublets is of 
the same form as the one in the SU(6)/Sp(6) product group model.
There are now four symmetry breaking vevs, $f_{1,\ldots,4}$.
The fermion sector contains two heavy quark-partners and two heavy 
lepton-partners for each generation.  
Only one of the heavy quark-partners in each generation
mixes with the corresponding SM quark.
Like in the SU(3) model, the fermions can be embedded
in a universal (but anomalous) way into SU(4) or in an anomaly-free way
\cite{Kong}.  Again, the anomaly-free embedding only works if the number
of fermion generations is a multiple of three.
The heavy gauge sector contains 
the broken generators of SU(4)$\to$SU(2), namely two neutral gauge bosons
$Z^{\prime}$ and $Z^{\prime\prime}$ (which
mix in general), two complex SU(2) doublets $(Y^0,X^-)$, 
$(Y^{0 \prime},X^{- \prime})$, and a complex SU(2) singlet 
$Y^{0 \prime\prime}$.  The phenomenology of the first $Z^{\prime}$ and the 
first doublet $(Y^0,X^-)$ are similar to those of the SU(3) model.

The SU(9)/SU(8) model of Ref.~\cite{SkibaTerning} contains exactly the
same gauge group and fermion sector as the SU(3) simple group model.
Thus the gauge and fermion sectors contain the same particle content 
and interactions as in the SU(3) simple group model.
The only difference is the global symmetry structure, which leads to
a different scalar sector.
The global symmetry group is SU(9), broken down to SU(8) by a vacuum 
condensate with two independent vevs, $f_{1,2}$.
The Higgs quartic coupling in this model is generated at tree level
by Lagrangian terms that explicitly break the SU(9) global symmetry.
The scalar sector contains two light Higgs doublets, plus two complex
singlets that get masses of order $f \sim$ TeV.
As in the SU(4) model, the potential generated for the two Higgs 
doublets is far from the most general possible, yielding interesting 
relations among the Higgs masses and couplings.

The simple group models share two features which distinguish them from
the product group models.  First, the models all contain an
SU($N$)$\times$U(1) gauge symmetry that is broken down to 
SU(2)$_L \times$U(1)$_Y$, yielding the TeV-scale gauge bosons.
The gauge couplings of the expanded SU($N$)$\times$U(1) symmetry are 
thus fixed in terms of the known SM gauge couplings.  The gauge 
structure also forbids mixing between the SM $W^{\pm}$ bosons and the 
TeV-scale gauge bosons, in contrast to the product group models.
Second, the top quark mass is generated from a Lagrangian involving 
two terms, which couple the top quark to two different nonlinear sigma
model fields.  This structure introduces an additional parameter into
the top sector, which complicates the phenomenology and allows the heavy
top-partner to be made lighter relative to the TeV-scale gauge bosons
than in the product group models, thereby reducing the fine-tuning.


\section{The SU(3) simple group model}
\label{appendixB}

In this Appendix we collect some technical details of the SU(3) 
simple group model of Refs.~\cite{KS,Schmaltznote} and derive the 
interaction Lagrangian in the mass basis.

The SU(3) simple group model \cite{KS,Schmaltznote} is constructed by
enlarging the SM SU(2)$_L \times$U(1)$_Y$ gauge group to 
SU(3)$\times$U(1)$_X$.  This requires enlarging the SU(2) doublets of 
the SM to SU(3) triplets and adding the additional SU(3) gauge bosons.
The SU(3)$\times$U(1)$_X$ gauge symmetry is broken down to the SM 
electroweak gauge group by two complex scalar fields
$\Phi_{1,2}$, which are triplets under 
the SU(3) with aligned vevs $f_{1,2}$, both of order a TeV.
We start with a scalar potential for $\Phi_{1,2}$ which has a
[SU(3)$\times$U(1)]$^2$ global symmetry.  After $\Phi_{1,2}$ acquire
vevs, the global symmetry is spontaneously broken down to
[SU(2)$\times$U(1)]$^2$.  At the same time,
the global symmetry is broken explicitly down to its diagonal 
SU(3)$\times$U(1) subgroup by the gauge interactions.
The scalar fields are parameterized as a nonlinear sigma model with
\begin{equation}
	\Theta = \frac{1}{f} \left[ 
		\left( \begin{array}{cc}
		\begin{array}{cc} 0 & 0 \\ 0 & 0 \end{array}
			& h \\ 
		h^{\dagger} & 0 \end{array} \right)
		+ \frac{\eta}{\sqrt{2}} 
		\left( \begin{array}{ccr}
		1 & 0 & 0 \\
		0 & 1 & 0 \\
		0 & 0 & 1 \end{array} \right) \right],
	\qquad \qquad 
	h = \left( \begin{array}{c}
		h^0 \\  h^-  \end{array} \right),
\end{equation}
and
\begin{eqnarray}
	\Phi_1 &=& e^{i \Theta f_2/f_1} \left( \begin{array}{c}
			0 \\ 0 \\ f_1 \end{array} \right)
		= f c_{\beta} \left[ 
		\left( \begin{array}{c}
			0 \\ 0 \\ 1 \end{array} \right)
		+ \frac{i t_{\beta}}{f} \left( \begin{array}{c}
			h \\ \eta/\sqrt{2} \end{array} \right)
		- \frac{t_{\beta}^2}{2 f^2} \left( \begin{array}{c}
			\sqrt{2} \eta h \\ h^{\dagger}h + \eta^2/2 
			\end{array} \right)
		+ \cdots \right],
	\\
	\Phi_2 &=& e^{-i \Theta f_1/f_2} \left( \begin{array}{c}
			0 \\ 0 \\ f_2 \end{array} \right)
		= f s_{\beta} \left[
		\left( \begin{array}{c}
			0 \\ 0 \\ 1 \end{array} \right)
		- \frac{i}{t_{\beta} f} \left( \begin{array}{c}
			h \\ \eta/\sqrt{2} \end{array} \right)
		- \frac{1}{2 t_{\beta}^2 f^2} \left( \begin{array}{c}
			\sqrt{2} \eta h \\ h^{\dagger}h + \eta^2/2
			\end{array} \right)
		+ \cdots \right].  \nonumber
\end{eqnarray}
We define $f^2 \equiv f_1^2 + f_2^2$ and 
$t_{\beta} \equiv \tan\beta = f_2/f_1$.
Under the SU(2)$_L$ SM gauge group, $h$ transforms as a doublet and will 
be identified as the SM Higgs doublet with a vev 
$v \equiv \sqrt{2}\langle h^0 \rangle = 246$ GeV,
while $\eta$ is a real singlet which also remains light.
We have chosen $\eta$ proportional to the unit matrix because this state
remains unmixed with the unphysical (eaten) Goldstone bosons after 
EWSB.\footnote{We thank Dave Rainwater for 
enlightening discussions on this point.}
We do not write down the Goldstone bosons that are eaten by the broken 
gauge generators.  

The SU(3) gauge bosons can be written in matrix form as
\begin{equation}
        A^a T^a = \frac{A^3}{2} \left( \begin{array}{ccc}
                1 & \quad & \quad \\
                \quad & -1 & \quad \\
                \quad & \quad & 0 \end{array} \right)
        + \frac{A^8}{2\sqrt{3}} \left( \begin{array}{ccc}
                1 & \quad & \quad \\
                \quad & 1 & \quad \\
                \quad & \quad & -2 \end{array} \right)
        + \frac{1}{\sqrt{2}} \left( \begin{array}{ccc}
                \quad & W^+ & Y^0 \\
                W^- & \quad & X^- \\
                \overline{Y}^0 & X^+ & \quad \end{array} \right).
	\label{eq:SU3AaTa}
\end{equation}
The $\Phi$ vevs break the SU(3)$\times$U(1)$_X$ gauge symmetry down
to the SM SU(2)$_L \times$U(1)$_Y$ via the covariant derivative term
\begin{equation}
	\mathcal{L}_{\Phi} = \left| \left( \partial_{\mu}
	+ i g A_{\mu}^a T^a - \frac{i g_x}{3} B_{\mu}^x \right) 
	\Phi_i \right|^2,
	\label{eq:SU3LPhi}
\end{equation}
where the SU(3) gauge coupling $g$ is equal to the SM SU(2)$_L$ gauge coupling
and the U(1)$_X$ gauge coupling $g_x$ is fixed in terms of $g$ and 
the weak mixing angle $t_W \equiv \tan\theta_W$ by
\begin{equation}
	g_x = \frac{g t_W}{\sqrt{1 - t_W^2/3}}.
\end{equation}
The broken gauge generators get masses of order $f \sim$ TeV and consist
of a $Z^{\prime}$ boson (a linear combination of $A^8$ and $B^x$) and a 
complex SU(2)$_L$ doublet $(Y^0,X^-)$.

\subsection{Gauge and Higgs sectors}

Before EWSB, the $X$ and $Y$ gauge bosons
and a linear combination $Z^{\prime}$ of the $A^8$ and $B^x$
gauge bosons get masses from the $f$ vevs.
The linear combination $Z^{\prime}$ that becomes massive is
\begin{equation}
        Z^{\prime}_0 = \frac{\sqrt{3} g A^8 + g_x B^x}{\sqrt{3g^2 + g_x^2}}
	= \frac{1}{\sqrt{3}} 
	\left( \sqrt{3 - t_W^2} A^8 + t_W B^x \right).
\end{equation}
We denote states and masses before EWSB with the subscript zero.
The orthogonal combination of $A^8$ and $B^x$
becomes the hypercharge gauge boson $B$,
\begin{equation}
        B = \frac{-g_x A^8 + \sqrt{3} g B^x}{\sqrt{3g^2 + g_x^2}}
	= \frac{1}{\sqrt{3}}
	\left( - t_W A^8 + \sqrt{3 - t_W^2} B^x \right).
\end{equation}
Hypercharge is given by
\begin{equation}
        Y = -\frac{1}{\sqrt{3}} T^8 + Q_x, \qquad \qquad
        T^8 = \frac{1}{2\sqrt{3}} {\rm diag}(1,1,-2),
\end{equation}
where $Q_x = -1/3$ for the scalar fields $\Phi_i$.
We also have the relations
\begin{eqnarray}
	A^3 &=& c_W Z_0 + s_W A,
	\qquad \qquad
	A^8 = \sqrt{1 - t_W^2/3} \, Z^{\prime}_0 
		+ \frac{s_W^2}{\sqrt{3} c_W} Z_0
		- \frac{s_W}{\sqrt{3}} A
	\nonumber \\
	B^x &=& \frac{t_W}{\sqrt{3}} Z^{\prime}_0
		- s_W \sqrt{1-t_W^2/3} \, Z_0
		+ c_W \sqrt{1-t_W^2/3} \, A,
\end{eqnarray}
where $A$ is the photon.

For use in precision corrections, we give the $W$ and $Z$
boson masses and their couplings to the Higgs at next-to-leading order
in $v^2/f^2$ in Table~\ref{tab:WZHnlo}.  The $WWH$ and $ZZH$ couplings
can be written in the form
\begin{equation}
	\mathcal{L} = 2 \frac{M^2_W}{v} y_W W^+ W^- H
	+ \frac{M_Z^2}{v} y_Z ZZH,
\end{equation}
with coefficients $y_{W,Z}$ given in Table~\ref{tab:WZHnlo}.

\TABLE{
\begin{tabular}{|c|c|}
\hline\hline
$M_W$ & $\frac{gv}{2} \left[ 1 - \frac{v^2}{12 f^2} \left( 
	\frac{s_{\beta}^4}{c_{\beta}^2} + \frac{c_{\beta}^4}{s_{\beta}^2} 
	\right) \right]$ \\
$M_Z$ & $\frac{gv}{2c_W} \left[ 1 - \frac{v^2}{12 f^2} \left( 
	\frac{s_{\beta}^4}{c_{\beta}^2} + \frac{c_{\beta}^4}{s_{\beta}^2} 
	\right) + \frac{v^2}{16 f^2} (1-t_W^2)^2 \right]$ \\
\hline
$W^+_{\mu} W^-_{\nu} H$: & $\frac{ig^2v}{2} \left[ 1 - 
	\frac{v^2}{3f^2} \left( 
	\frac{s_{\beta}^4}{c_{\beta}^2} + \frac{c_{\beta}^4}{s_{\beta}^2} 
	\right) \right] g_{\mu\nu}$ \\
$W^+_{\mu} W^-_{\nu} H H$: & $\frac{ig^2}{2} \left[ 1 - 
	\frac{v^2}{f^2} \left( 
	\frac{s_{\beta}^4}{c_{\beta}^2} + \frac{c_{\beta}^4}{s_{\beta}^2} 
	\right) \right] g_{\mu\nu}$ \\
$Z_{\mu} Z_{\nu} H$: & $\frac{ig^2v}{2c_W^2} \left[ 1 - 
	\frac{v^2}{3f^2} \left( 
	\frac{s_{\beta}^4}{c_{\beta}^2} + \frac{c_{\beta}^4}{s_{\beta}^2} 
	\right) - \frac{v^2}{4f^2} (1-t_W^2)^2 \right] g_{\mu\nu}$ \\
$Z_{\mu} Z_{\nu} H H$: & $\frac{ig^2}{2c_W^2} \left[ 1 - 
	\frac{v^2}{f^2} \left( 
	\frac{s_{\beta}^4}{c_{\beta}^2} + \frac{c_{\beta}^4}{s_{\beta}^2} 
	\right) - \frac{v^2}{4f^2} (1-t_W^2)^2 \right] g_{\mu\nu}$ \\
\hline
$y_W$ & $1 + \frac{v^2}{f^2} \left[ -\frac{1}{6} 
		\left( \frac{s_{\beta}^4}{c_{\beta}^2}
		+ \frac{c_{\beta}^4}{s_{\beta}^2} \right) \right]$ \\
$y_Z$ & $1 + \frac{v^2}{f^2} \left[ -\frac{1}{6} 
		\left( \frac{s_{\beta}^4}{c_{\beta}^2}
		+ \frac{c_{\beta}^4}{s_{\beta}^2} \right)
	- \frac{3}{8} (1 - t_W^2)^2 \right]$ \\
\hline \hline
\end{tabular}
\caption{\label{tab:WZHnlo}
$W$ and $Z$ boson masses and their couplings to the Higgs at 
next-to-leading order in $v^2/f^2$ in the SU(3) simple group model.}
}

\subsection{Fermion sector}
\label{sec:fermionappendix}

Because the model contains a gauged SU(3),
SM fermions that are doublets under SU(2) must
be expanded into triplets under the SU(3).
In addition, new SU(3)-singlet fermions must be
introduced to cancel the hypercharge anomalies and to marry and
give mass to the new third components of the SU(3)-triplet fermions.

The most straightforward way to construct a fermion sector for the 
SU(3) simple group model is to expand all the SU(2) doublets of the 
SM into SU(3)
triplets, adding additional SU(3)-singlet right-handed fermions as needed,
as was done in Ref.~\cite{KS}.  We call this embedding ``universal'',
since the three generations have identical quantum numbers.
The quarks and leptons of each generation are put into ${\mathbf{3}}$ 
representations of SU(3):
\begin{eqnarray}
	Q^T_m = (u,d,iU)_m, &\qquad \qquad&  iu^c_m, id^c_m, iU^c_m
	\qquad \qquad {\rm (universal)}
	\nonumber \\
        L^T_m = (\nu, e, iN)_m, &\qquad \qquad& ie^c_m, iN^c_m,
\end{eqnarray}
where $m$ is the generation index.
We do not include a right-handed neutrino at this stage, leaving the
neutrinos massless.  Neutrino masses could be incorporated, e.g., through
a see-saw mechanism in the UV completion of the little Higgs model \cite{KS}
or within the little Higgs theory itself \cite{LHseesaw};
however, this is beyond the scope of our current work.
The $Q_x$ charges of the fermions are given in Table~\ref{tab:Qxf}.

\TABLE{
\begin{tabular}{|c||c|c|c|c||c|c|c|}
\hline \hline
 & \multicolumn{7}{c|}{Universal embedding} \\
\hline
fermion & $Q_{1,2}$ & $Q_3$ & $u_m^c$, $T^c$, $U_m^c$ & $d_m^c$ 
	& $L_m$ & $N_m^c$ & $e_m^c$ \\
\hline
$Q_x$ charge & $1/3$ & $1/3$ & $-2/3$ & $1/3$ 
	& $-1/3$ & $0$ & 1 \\
\hline
SU(3) rep & {\bf 3} & {\bf 3} & {\bf 1} & {\bf 1} 
	& {\bf 3} & {\bf 1} & {\bf 1} \\
\hline \hline
 & \multicolumn{7}{c|}{Anomaly-free embedding} \\
\hline
fermion & $Q_{1,2}$ & $Q_3$ & $u^c_m, T^c$ & $d^c_m, D^c, S^c$ & $L_m$ 
	& $N^c_m$ & $e^c_m$ \\
\hline
$Q_x$ charge & $0$ & $1/3$ & $-2/3$ & $1/3$ & $-1/3$ & $0$ & 1 \\
\hline
SU(3) rep & {$\mathbf{\bar 3}$} & {\bf 3} & {\bf 1} & {\bf 1} & {\bf 3} 
	& {\bf 1} & {\bf 1} \\
\hline \hline
\end{tabular}
\caption{\label{tab:Qxf}
The $Q_x$ charges and SU(3) representations of the fermions
in the universal and anomaly-free embeddings.}
}

It was pointed out by Kong \cite{Kong} that such a universal 
fermion sector leads to SU(3) and U(1)$_x$ gauge anomalies,
although the SM SU(2) and U(1)$_Y$ gauge
groups remain anomaly-free.  These anomalies are not necessarily a problem
because the little Higgs model is only an effective theory valid
up to an energy scale $\Lambda \sim 4 \pi f$.
Additional fermions can be added at the scale $\Lambda$ to cancel the 
SU(3) and U(1)$_x$ gauge anomalies without affecting the phenomenology
at the $f$ scale.
Alternatively, one can construct a fermion sector that is anomaly-free
already at the $f$ scale and yet contains no more degrees of freedom than
the universal embedding, as proposed by Kong~\cite{Kong}.
This can be done by
putting the first two generations of quarks in {$\mathbf{\bar 3}$} 
representations of SU(3), while the third quark generation and all three 
lepton generations are in {$\mathbf{3}$}s of SU(3).  
We call this embedding ``anomaly-free''.
It is fascinating to note that with this fermion content,
the anomalies do not cancel within a single generation, as in the 
SM, but rather three generations (or a multiple thereof) are required
to cancel the anomalies. 
The anomaly cancellation pattern of this fermion content has been previously
pointed out in 3-3-1 models \cite{331} outside of the little Higgs context.

The quarks of the third generation and three generations of leptons
are put into {$\mathbf{3}$} representations of SU(3), exactly as in the
universal embedding.
The first two generations of quarks are put into {$\mathbf{\bar 3}$}
representations of SU(3):
\begin{eqnarray}
	Q^T_1 = (d, -u, iD), &\qquad \qquad&  id^c, iu^c, iD^c
	\qquad \qquad {\rm (anomaly \ free)}
	\nonumber \\
	Q^T_2 = (s, -c, iS), &\qquad \qquad&  is^c, ic^c, iS^c,
\end{eqnarray}
where the minus signs in front of $u$ and $c$ are there because
the $\mathbf{\bar 2}$ of SU(2) is $(d, -u)$
[which is equivalent to the $\mathbf{2}$, $(u,d)$].  
Notice that the heavy vector-like quarks of the first two generations
have electric charge $-1/3$, in contrast to the charge $+2/3$ heavy 
quark of the third generation.
The $Q_x$ charges of the fermions are given in Table~\ref{tab:Qxf}.

\subsubsection{Lepton masses and mixing}
\label{sec:leptons}

The lepton sector is identical in both the universal and anomaly-free 
embeddings.  The lepton masses are generated by the Lagrangian in 
Eq.~(\ref{eq:LYSU3lept}), where we have chosen the flavor basis
to correspond to the mass basis for the heavy neutrino partners $N_m$.  
The $N_m$ masses are then given by Eq.~(\ref{eq:MNm}).
The dimension-5 operator in Eq.~(\ref{eq:LYSU3lept}) normalized
by the cutoff scale $\Lambda$ gives masses to the charged leptons
via the $3\times 3$ Yukawa matrix $\lambda_e^{mn}$, which also generates
a CKM-like mixing matrix $V_{im}^{\ell}$ between the charged lepton
mass eigenstates $e_i$ and the heavy neutrino partners $N_m$.
This mixing matrix appears in the $X^- \bar e_i N_m$ couplings,
\begin{equation}
	\mathcal{L} \supset -\frac{g}{\sqrt{2}} V_{im}^{\ell}
	X^-_{\mu} \bar e_i \gamma^{\mu} P_L N_m.
\end{equation}
These couplings can lead to lepton flavor violating processes, such
as $\mu \to e \gamma$, via loops of $N_m$ and $X^-$.  As in the quark
sector of the SM, this lepton flavor violation will be GIM-suppressed
and will vanish in the limit that $V_{im}^{\ell}$ is diagonal, so that
the $N_m$ mass eigenstates are aligned
with the charged lepton mass eigenstates.  The lepton flavor violation
will also vanish in the limit that the $N_m$ are degenerate.  The 
experimental limits on lepton flavor violation therefore put stringent
constraints on the $\lambda_{N_m}$ couplings and/or on the structure
of the $\lambda_e^{mn}$ matrix.

After EWSB, the $h$ vev induces mixing between $N_{m 0}$ and the 
corresponding neutrino $\nu_{m 0}$ at order $v/f$, where as usual
we use a subscript 0 to denote the SU(3) eigenstates and no subscript
to denote the mass eigenstates after EWSB.
Because of the structure of the $N_m$ mass term in 
Eq.~(\ref{eq:LYSU3lept}), $N_m$ 
mixes only with the neutrino in the same SU(3) triplet, with 
a mixing angle $\delta_{\nu}$ given in Eq.~(\ref{eq:deltaq})
that is the same for all three generations.
Note that $t_{\beta} > 1$ suppresses $\delta_{\nu}$.
The SU(3) eigenstates $N_{m 0}$ and $\nu_{i0}$ are given in
terms of the mass eigenstates $N_m$ and the SM neutrinos
in the charged lepton mass basis ($\nu_i = \nu_e,\nu_{\mu},\nu_{\tau}$) by
\begin{equation}
	N_{m 0} = N_m + \delta_{\nu} V_{mi}^{\ell \dagger} \nu_i,
	\qquad \qquad
	\nu_{i 0} = \left( 1 - \frac{1}{2} \delta_{\nu}^2 \right) \nu_i
	- \delta_{\nu} V_{im}^{\ell} N_m,
\end{equation}
where we have kept the $\delta_{\nu}^2$ term in the neutrino mixing
because it will modify the well-measured couplings of neutrinos to the
$W$ and $Z$ bosons at order $v^2/f^2$.
In particular, the Fermi constant $G_F$ is measured in muon decay.
The four-Fermi effective interaction Lagrangian is
\begin{equation}
	\mathcal{L} = - 2 \sqrt{2} G_F J^{+ \mu} J^-_{\mu}
	= -\frac{g^2}{2 M_W^2} J^{+ \mu} J^-_{\mu} 
		\left( 1 - \delta_{\nu}^2 \right).
\end{equation}
Plugging in $M_W^2$ (from Table~\ref{tab:WZHnlo}) and $\delta_{\nu}$, we have,
\begin{equation}
	\frac{1}{G_F} = \sqrt{2} v^2 \left\{ 1 + \frac{v^2}{f^2}
	\left[ - \frac{1}{6} \left( \frac{s_{\beta}^4}{c_{\beta}^2}
		+ \frac{c_{\beta}^4}{s_{\beta}^2} \right)
		+ \frac{1}{2 t_{\beta}^2} \right] \right\}.
\end{equation}

\TABLE{
\begin{tabular}{|c|c|}
\hline\hline
$H \overline{e_i} e_i$: & $-\frac{im_{e_i}}{v} y_{\ell}$ \\
$H \overline{N_m} \nu_i$: & $-\frac{ic_{\beta}}{\sqrt{2}} \lambda_{N_m}
	V_{mi}^{\ell \dagger} P_L$ \\
$H \overline{N_m} N_m$: & $\mathcal{O}(v^2/f^2)$ \\
$\eta \overline{e_i} e_i$: & $\frac{\sqrt{2}m_{e_i}}{f} \cot 2\beta 
	\gamma_5$ \\
$\eta \overline{N_m} N_m$: & $-\frac{c_{\beta}}{\sqrt{2}} \lambda_{N_m}
	\gamma_5$ \\
\hline
$y_{\ell}$ & $1 - \frac{v^2}{6f^2} \left(3 + \frac{s_{\beta}^4}{c_{\beta}^2}
	+ \frac{c_{\beta}^4}{s_{\beta}^2} \right)$ \\
\hline \hline
\end{tabular}
\caption{\label{tab:Hll}
Couplings of $H$ and $\eta$ to lepton pairs.}
}

The couplings of the scalars $H$ and $\eta$ to lepton pairs are given
in Table~\ref{tab:Hll}.  The couplings of charged leptons to $H$ get a 
multiplicative correction factor $y_{\ell}$ relative to the SM Yukawa 
couplings in terms of the lepton mass due to the nonlinear 
sigma model expansion.

\subsubsection{Lepton couplings to gauge bosons}

The fermion couplings to gauge bosons are given by
the fermion kinetic term,
\begin{equation}
        \mathcal{L} = \bar \psi i \mathcal{D}_{\mu} \gamma^{\mu} \psi,
	\qquad
	\mathcal{D} = \partial + i g A^a T^a + i g_x Q_x B^x,
\end{equation}
with the $Q_x$ charges given in Table~\ref{tab:Qxf}.
The generators $T^a$ of the fundamental $\mathbf{3}$ representation of SU(3)
are given in Eq.~(\ref{eq:SU3AaTa}).

The couplings of the $Z^{\prime}$ to lepton pairs were given in 
Table~\ref{tab:gauge2}.
The couplings of the heavy off-diagonal gauge bosons $X^{\mp}$,
$Y^0$ and $\overline{Y}^0$ to leptons were given in Table~\ref{table-Q},
neglecting flavor misalignment between the charged leptons and the
$N_m$.  Allowing for the possibility of flavor misalignment, we have
\begin{eqnarray}
	\mathcal{L}_{X,Y} &=& - \frac{g}{\sqrt{2}} \left[
	i X^-_{\mu} \bar e_i \gamma^{\mu} \left( V^{\ell}_{im} N_m
	+ \delta_{\nu} \nu_i \right)
	+ i Y^0_{\mu} \bar \nu_i \gamma^{\mu} \left( V^{\ell}_{im} N_m 
	+ \delta_{\nu} \nu_i \right)
	+ {\rm h.c.} \right],
\end{eqnarray}
where all fermion fields are left-handed and we have taken the neutrinos 
in the charged lepton mass basis, $\nu_i = \nu_e,\nu_{\mu},\nu_{\tau}$; 
$N_m$ are the heavy neutral leptons in their mass basis.
The couplings of $W^{\pm}$ to lepton pairs, keeping
terms of order $v^2/f^2$ in interactions involving only SM particles
and terms of order $v/f$ in interactions involving one or more heavy
particles, are
\begin{eqnarray}
	\mathcal{L}_W &=& - \frac{g W^+_{\mu}}{\sqrt{2}} \left[
	\left( 1 - \frac{1}{2} \delta_{\nu}^2 \right)
		\overline{\nu}_i \gamma^{\mu} e_i 
	- \delta_{\nu} V_{mi}^{\ell \dagger} 
		\overline{N}_m \gamma^{\mu} e_i 
	+ {\rm h.c.} \right].
\end{eqnarray}
The couplings of the $Z$ boson to leptons, including the corrections
from mixing between $Z$ and $Z^{\prime}$ and mixing between the heavy
neutral leptons and the SM neutrinos, are
\begin{eqnarray}
	&\mathcal{L}_Z& = - Z_{\mu} \frac{g}{c_W} \left\{
	\left( J_3^{\mu} - s^2_W J_Q^{\mu} \right)
	- \frac{1}{2} \delta_{\nu}^2 \overline{\nu}_i \gamma^{\mu} \nu_i
	- \frac{1}{2} \left[\delta_{\nu} V_{im}^{\ell *} 
		\overline{N}_m \gamma^{\mu} \nu_i
		+ {\rm h.c.} \right]
	\right.  \\ &+& \left.
	\frac{\delta_Z}{\sqrt{3 - 4 s_W^2}}
	\left[\left( \frac{1}{2} - s_W^2 \right)
		\left( \overline{\nu}_i \gamma^{\mu} \nu_i 
		+ \overline{e}_i \gamma^{\mu} e_i \right)
	+ s_W^2 \overline{e}_i^c \gamma^{\mu} e_i^c
	+ \left( -1 + s_W^2 \right) \overline{N}_i \gamma^{\mu} N_i
	\right]
	\right\}, \nonumber
\end{eqnarray}
where the leading-order coupling is given in terms of the standard fermion 
currents
\begin{equation}
	J_3^{\mu} = \overline{f} \gamma^{\mu} T^3 f,
	\qquad \qquad
	J_Q^{\mu} = \overline{f} \gamma^{\mu} Q_f f 
		- \overline{f}^c \gamma^{\mu} Q_{f^c} f^c.
\label{eq:currents}
\end{equation}
The couplings of the photon to fermions are given by the 
electromagnetic current as usual, $\mathcal{L}_A = - A_{\mu} e J_Q^{\mu}$.

\subsubsection{Quark masses and mixing: anomaly-free embedding}
\label{AFmq}

The quark sector is more complicated than the lepton sector
because of the anomaly-free embedding structure.
The relevant Lagrangian terms for the third generation and for the first two 
generations are
\begin{eqnarray}
	\mathcal{L}_3 &=& \lambda_1^t iu_1^c 
		\Phi_1^{\dagger} Q_3
		+ \lambda_2^t iu_2^c \Phi_2^{\dagger} Q_3 
		+ \frac{\lambda_b^m}{\Lambda} id_m^c
		\epsilon_{ijk} \Phi_1^i \Phi_2^j Q_3^k + {\rm h.c.}
	\nonumber \\
	\mathcal{L}_{1,2} &=& \lambda_1^{d n} id^{nc}_1 Q_n^T \Phi_1
		+ \lambda_2^{d n} id^{nc}_2 Q_n^T \Phi_2
		+ \frac{\lambda_u^{mn}}{\Lambda}
		iu^c_m \epsilon_{ijk} \Phi_1^{*i} \Phi_2^{*j} Q_n^k 
		+ {\rm h.c.},
\label{eq:L312}
\end{eqnarray}
where $n = 1,2$; $i,j,k = 1,2,3$ are SU(3) indexes;
$u_1^c$ and $u_2^c$ are linear combinations of $t^c$ and $T^c$ 
[see Eqs.~(\ref{eq:Tc}) and (\ref{eq:tc}) below]; 
$b_m^c$ runs over all the down-type 
conjugate quarks ($d^c, s^c, b^c, D^c, S^c$);
$d^{nc}_1$ and $d^{nc}_2$ are linear combinations of $d^c$ and $D^c$
for $n=1$ and of $s^c$ and $S^c$ for $n=2$
[see Eqs.~(\ref{eq:Dc}) and (\ref{eq:dc}) below];
and $u^c_m$ runs over all the up-type conjugate quarks ($u^c, c^c, t^c, T^c$).

The $f$ vevs generate mass terms for three heavy quarks.
The state
\begin{equation}
	T^c = \frac{\lambda_1^t c_{\beta} u_1^c + \lambda_2^t s_{\beta} u_2^c}
	{\sqrt{\lambda_1^{t 2} c_{\beta}^2 + \lambda_2^{t 2} s_{\beta}^2}}
\label{eq:Tc}
\end{equation}
marries $T$, giving it a mass of
\begin{equation}
	M_T = f \sqrt{\lambda_1^{t 2} c_{\beta}^2 
		+ \lambda_2^{t 2} s_{\beta}^2}
\end{equation}
and leaving the orthogonal combination of $u_1^c$ and $u_2^c$ massless:
\begin{equation}
	t^c = \frac{ -\lambda_2^t s_{\beta} u_1^c 
			+ \lambda_1^t c_{\beta} u_2^c}
	{\sqrt{\lambda_1^{t 2} c_{\beta}^2 + \lambda_2^{t 2} s_{\beta}^2}}.
\label{eq:tc}
\end{equation}
The states (here we denote $\lambda_{1,2}^{dn}$ by $\lambda_{1,2}^d$ for 
$n=1$ and by $\lambda_{1,2}^s$ for $n=2$)
\begin{equation}
	D^c = \frac{\lambda_1^d c_{\beta} d_1^{1c} 
			+ \lambda_2^d s_{\beta} d_2^{1c}}
	{\sqrt{\lambda_1^{d 2} c_{\beta}^2 + \lambda_2^{d 2} s_{\beta}^2}},
	\qquad \qquad
	S^c = \frac{\lambda_1^s c_{\beta} d_1^{2c} 
			+ \lambda_2^s s_{\beta} d_2^{2c}}
	{\sqrt{\lambda_1^{s 2} c_{\beta}^2 + \lambda_2^{s 2} s_{\beta}^2}}
\label{eq:Dc}
\end{equation}
marry $D$ and $S$, respectively, giving them masses of
\begin{equation}
	M_D = f \sqrt{\lambda_1^{d 2} c_{\beta}^2 
		+ \lambda_2^{d 2} s_{\beta}^2},
	\qquad \qquad
	M_S = f \sqrt{\lambda_1^{s 2} c_{\beta}^2 
		+ \lambda_2^{s 2} s_{\beta}^2},
\end{equation}
and leaving the orthogonal combinations massless:
\begin{equation}
	d^c = \frac{-\lambda_2^d s_{\beta} d_1^{1c} 
			+ \lambda_1^d c_{\beta} d_2^{1c}}
	{\sqrt{\lambda_1^{d 2} c_{\beta}^2 + \lambda_2^{d 2} s_{\beta}^2}},
	\qquad \qquad
	s^c = \frac{-\lambda_2^s s_{\beta} d_1^{2c} 
			+ \lambda_1^s c_{\beta} d_2^{2c}}
	{\sqrt{\lambda_1^{s 2} c_{\beta}^2 + \lambda_2^{s 2} s_{\beta}^2}}.
\label{eq:dc}
\end{equation}
After EWSB, the quark mass terms are
\begin{eqnarray}
	\mathcal{L}_{\rm up \ mass} &=&
	- M_T T^c T
	+ \frac{v}{\sqrt{2}} 
	\frac{s_{\beta} c_{\beta} (\lambda_1^{t 2} - \lambda_2^{t 2})}
	{\sqrt{\lambda_1^{t 2} c_{\beta}^2 + \lambda_2^{t 2} s_{\beta}^2}}
	T^c t
	- \frac{v}{\sqrt{2}} 
	\frac{\lambda_1^t \lambda_2^t}
	{\sqrt{\lambda_1^{t 2} c_{\beta}^2 + \lambda_2^{t 2} s_{\beta}^2}} 
	t^c t
	\nonumber \\ &&
	+ \frac{v}{\sqrt{2}} \frac{f}{\Lambda} \lambda_u^{mn} u_m^c u_n
	+ {\rm h.c.}
\label{eq:Lup}
	\\
	\mathcal{L}_{\rm down \ mass} &=&
	- M_D D^c D
	- \frac{v}{\sqrt{2}} 
	\frac{s_{\beta} c_{\beta} (\lambda_1^{d 2} - \lambda_2^{d 2})}
	{\sqrt{\lambda_1^{d 2} c_{\beta}^2 + \lambda_2^{d 2} s_{\beta}^2}} 
	D^c d
	+ \frac{v}{\sqrt{2}} 
	\frac{\lambda_1^d \lambda_2^d}
	{\sqrt{\lambda_1^{d 2} c_{\beta}^2 + \lambda_2^{d 2} s_{\beta}^2}} 
	d^c d
	\nonumber \\ &&
	- M_S S^c S
	- \frac{v}{\sqrt{2}} 
	\frac{s_{\beta} c_{\beta} (\lambda_1^{s 2} - \lambda_2^{s 2})}
	{\sqrt{\lambda_1^{s 2} c_{\beta}^2 + \lambda_2^{s 2} s_{\beta}^2}} 
	S^c s
	+ \frac{v}{\sqrt{2}} 
	\frac{\lambda_1^s \lambda_2^s}
	{\sqrt{\lambda_1^{s 2} c_{\beta}^2 + \lambda_2^{s 2} s_{\beta}^2}} 
	s^c s
	\nonumber \\ &&
	+ \frac{v}{\sqrt{2}} \frac{f}{\Lambda} \lambda_b^m d^c_m b
	+ {\rm h.c.}
\label{eq:Ldown}
\end{eqnarray}
where $u_n = u, c$; $u_m^c = u^c, c^c, t^c, T^c$; and
$d^c_m = d^c, s^c, b^c, D^c, S^c$.

The couplings $\lambda_u^{mn}$ and $\lambda_b^m$ cause a misalignment
between the mass eigenstates in the up and down sectors, leading to
the CKM matrix.  They also cause an analogous misalignment between the
SM quark mass eigenstates and the heavy quarks $D$, $S$, and $T$, leading
to an analogous matrix.  We choose the ``flavor basis'' to be the mass 
basis for $D,S,T$.
Two unitary matrices are needed to rotate the left-handed
up- and down-type quarks from the flavor basis (primed fields) into the 
mass basis (unprimed fields):
\begin{equation}
	V^u \left( \begin{array}{c}
		u^{\prime} \\
		c^{\prime} \\
		t^{\prime} \end{array} \right)
	= \left( \begin{array}{c}
		u \\
		c \\
		t \end{array} \right),
	\qquad \qquad
	V^d \left( \begin{array}{c}
		d^{\prime} \\
		s^{\prime} \\
		b^{\prime} \end{array} \right)
	= \left( \begin{array}{c}
		d \\
		s \\
		b \end{array} \right).
\end{equation}
The CKM matrix is then given by
\begin{equation}
	V^{\rm CKM} = V^u V^{d \dagger}.
\label{eq:CKM}
\end{equation}
These matrices appear in the quark gauge couplings;
see Sec.~\ref{sec:fermion-gauge} for details. 
Note that, in contrast to the SM, there are \emph{two} physically 
meaningful mixing matrices.

Electroweak symmetry breaking also induces mixing 
between the heavy left-handed quarks $D,S,T$ and the SM quarks.
In the up-quark sector,
the terms in Eq.~(\ref{eq:Lup}) involving $T^c$ lead to mixing between
$T$ and $u,c,t$ that violates the SU(3) symmetry.  
As usual we use the subscript $0$ to denote SU(3) states; 
fields with no subscript denote the mass eigenstates after the mixing
induced by EWSB.
We can rewrite the SU(3) state $T_0$ in terms of the mass eigenstate $T$ 
and the SM fermions in the interaction basis (primed fields) as
\begin{equation}
	T_0 = T + \delta_{u_i} u_i^{\prime},
\end{equation}
with $i = 1,2,3$, where
\begin{equation}
	\delta_u = \frac{v}{\sqrt{2} \Lambda} 
	\frac{\lambda_u^{T^c u}}
	{\sqrt{\lambda_1^{t 2} c_{\beta}^2 + \lambda_2^{t 2} s_{\beta}^2}},
	\qquad
	\delta_c = \frac{v}{\sqrt{2} \Lambda} 
	\frac{\lambda_u^{T^c c}}
	{\sqrt{\lambda_1^{t 2} c_{\beta}^2 + \lambda_2^{t 2} s_{\beta}^2}},
	\qquad
	\delta_t = \frac{v}{\sqrt{2} f}
	\frac{s_{\beta} c_{\beta} (\lambda_1^{t 2} - \lambda_2^{t 2})}
	{(\lambda_1^{t 2} c_{\beta}^2 + \lambda_2^{t 2} s_{\beta}^2)}.
\end{equation}
One can choose the couplings $\lambda_u^{T^c u}$ and $\lambda_u^{T^c c}$
to be small in order to suppress the mixing effects in the first and 
second generations.
In the mass basis (unprimed fields) this becomes
\begin{equation}
	T_0 = T + \Delta_{u_i} u_i, \qquad \qquad
	\Delta_{u_i} = V^{u *}_{ij} \delta_{u_j}
	\simeq V^{u *}_{i3} \delta_t,
\end{equation}
where in the last approximate equality we neglect $\lambda_u^{T^c u}$ 
and $\lambda_u^{T^c c}$.
After mixing, the up quarks in the mass basis become
\begin{equation}
	u_{i 0} = \left( 1 - \frac{1}{2} |\Delta_{u_i}|^2 \right) u_i
		- \Delta_{u_i} T,
\end{equation}
where we have kept the $|\Delta_{u_i}|^2$ term 
(which is of order $v^2/f^2$) because it will modify the well-measured 
couplings of quarks to the $W$ boson.

Similarly, in the down-quark sector, 
the terms in Eq.~(\ref{eq:Ldown}) involving $D^c$ ($S^c$) lead to mixing
between $D$ and $d,b$ ($S$ and $s,b$).
As in the up sector, we can rewrite the SU(3) states $D_0$ and $S_0$
in terms of the mass eigenstates $D$ and $S$ and the SM fermions
in the interaction basis (primed fields) as
\begin{equation}
	D_0 = D + \delta_{D d_i} d_i^{\prime},
	\qquad \qquad
	S_0 = S + \delta_{S d_i} d_i^{\prime},
\end{equation}
with $i = 1,2,3$, where
\begin{eqnarray}
	\delta_{Dd} &=& \frac{-v}{\sqrt{2} f}
	\frac{s_{\beta} c_{\beta} (\lambda_1^{d 2} - \lambda_2^{d 2})}
	{(\lambda_1^{d 2} c_{\beta}^2 + \lambda_2^{d 2} s_{\beta}^2)},
	\qquad
	\delta_{Ds} = 0,
	\qquad
	\delta_{Db} = \frac{v}{\sqrt{2} \Lambda}
	\frac{\lambda_b^{D^c}}
	{\sqrt{\lambda_1^{d 2} c_{\beta}^2 + \lambda_2^{d 2} s_{\beta}^2}},
	\nonumber \\
	\delta_{Sd} &=& 0,
	\qquad
	\delta_{Ss} = \frac{-v}{\sqrt{2} f}
	\frac{s_{\beta} c_{\beta} (\lambda_1^{s 2} - \lambda_2^{s 2})}
	{(\lambda_1^{s 2} c_{\beta}^2 + \lambda_2^{s 2} s_{\beta}^2)},
	\qquad
	\delta_{Sb} = \frac{v}{\sqrt{2} \Lambda}
	\frac{\lambda_b^{S^c}}
	{\sqrt{\lambda_1^{s 2} c_{\beta}^2 + \lambda_2^{s 2} s_{\beta}^2}}.
\end{eqnarray}
The zero mixings, $\delta_{Ds} = \delta_{Sd} = 0$, are a consequence
of the collective breaking mass generation for $d$ and $s$ in the 
$D,S$ mass basis.  
One can choose $\lambda_b^{D^c}$ and $\lambda_b^{S^c}$ to be small
in order to suppress the mixing effects in the $b$ quark sector.
From Eq.~(\ref{eq:Ldown}), the small mass of the $d$ ($s$) quark
requires one of the couplings $\lambda_{1,2}^d$ ($\lambda_{1,2}^s$) 
to be very small.  We choose the small coupling to be $\lambda_1^d$
($\lambda_1^s$) so that the mixing effects in the down-quark sector are 
suppressed in the same $t_{\beta} > 1$ limit as the mixing effects 
in the neutrino sector.  We then have,
\begin{equation}
	\delta_{Dd} \simeq \delta_{Ss} 
	\simeq \frac{v}{\sqrt{2} t_{\beta} f}
	= -\delta_{\nu}.
\end{equation}

In the mass basis (unprimed fields), the $D$ and $S$ states become
\begin{eqnarray}
	&D_0 = D + \Delta_{Dd_i} d_i, \qquad \qquad 
	&\Delta_{Dd_i} = V^{d *}_{ij} \delta_{Dd_j}
	\simeq -V^{d *}_{i1} \delta_{\nu},
	\nonumber \\
	&S_0 = S + \Delta_{Sd_i} d_i, \qquad \qquad
	&\Delta_{Sd_i} = V^{d *}_{ij} \delta_{Sd_j}
	\simeq -V^{d *}_{i2} \delta_{\nu},
\end{eqnarray}
where in the last approximate equalities we neglect
$\lambda_b^{D^c}$ and $\lambda_b^{S^c}$.
After mixing, the down quarks in the mass basis become
\begin{equation}
	d_{i 0} = \left( 1 - \frac{1}{2} |\Delta_{Dd_i}|^2 
		- \frac{1}{2} |\Delta_{Sd_i}|^2 \right) d_i
	- \Delta_{Dd_i} D - \Delta_{Sd_i} S,
\end{equation}
where we again have kept the $|\Delta_{Dd_i}|^2$ and $|\Delta_{Sd_i}|^2$ terms,
which are of order $v^2/f^2$.

We now write the couplings of the scalars, $H$ and $\eta$, to quark pairs,
taking into account corrections from the expansion of the nonlinear sigma
model and the mixing between the SM quarks and the heavy quarks.
The different treatment of the third quark generation in the anomaly-free
fermion embedding [Eq.~(\ref{eq:L312})] 
leads to flavor-changing couplings of quarks
to $H$ (at order $v^2/f^2$) and to $\eta$ (at order $v/f$). 
The full parameter dependence of the flavor changing couplings
depends on the exact form of the up and down quark mass matrices, which
determine the quark mixing in the left- and right-handed sectors.
A detailed exploration of the quark mass matrices is beyond the scope of
this work.  Instead, we write down the scalar couplings ignoring
the mixing of the right-handed top quark $t^c$ with the first two 
generations.

We begin with the couplings of $T$ quark pairs.
$T$ couples to $\eta$ with a coupling of order one and to $H$ with
a coupling of order $v/f$:
\begin{eqnarray}
	\mathcal{L}_{T^cT} &\simeq& 
	(H T^c T) \frac{v}{f} \left[
	(\lambda_1^{t2} s_{\beta}^2 + \lambda_2^{t2} c_{\beta}^2)
	\frac{f}{2 M_T}
	- s_{\beta}^2 c_{\beta}^2 (\lambda_1^{t2} - \lambda_2^{t2})^2
	\frac{f^3}{2 M_T^3}
	\right]
	\nonumber \\ && 
	+ (i \eta T^c T)
	s_{\beta} c_{\beta} ( \lambda_1^{t 2} - \lambda_2^{t 2} )
	\frac{f}{\sqrt{2} M_T} + {\rm h.c.},
\label{eq:LTT}
\end{eqnarray}
where we have neglected terms involving $\lambda_u^{T^c u}$
and $\lambda_u^{T^c c}$.
Similarly, $D$ and $S$ quark pairs couple to $\eta$ with a coupling of 
order one:
\begin{equation}
	\mathcal{L}_{D^c_m D_m} \simeq 
	\frac{c_{\beta}}{\sqrt{2}} \lambda_2^d (i \eta D^c D)
	+ \frac{c_{\beta}}{\sqrt{2}} \lambda_2^s (i \eta S^c S)
	+ {\rm h.c.},
\end{equation}
where we have neglected terms involving $\lambda_b^{D^c}$ and 
$\lambda_b^{S^c}$ and taken $\lambda_1^{d,s} \ll \lambda_2^{d,s}$
[if the top quark mass were neglected, Eq.~(\ref{eq:LTT}) would also
reduce to this simple form].
One would naively expect an $H D^c_m D_m$ coupling
at order $v/f$ coming from replacing one Higgs field by its vev in 
the nonlinear sigma model expansion term
$HH D^c_m D_m$; however, this term is exactly canceled by the contribution
from $H D^c_m d_m$ after $d-D$ mixing if the down and strange quark masses
are neglected.

The leading-order couplings of scalars to one $T$ quark and one SM 
up-type quark are
\begin{equation}
	\mathcal{L} \simeq (H T^c u_i) \left[
	s_{\beta} c_{\beta} ( \lambda_1^{t 2} - \lambda_2^{t 2} )
	\frac{f}{\sqrt{2} M_T} V^{u *}_{i3}
	\right]
	- (i \eta t^c T) \left[
	\frac{\lambda_1^t \lambda_2^t f}{\sqrt{2} M_T} \right]
	+ {\rm h.c.},
\label{eq:LTu}
\end{equation}
where we again neglect terms involving $\lambda_u^{T^c u}$
and $\lambda_u^{T^c c}$ and in the last term ignore the mixing of the 
right-handed top quark $t^c$ with the first two generations.
The last term can be written in terms of SM quark masses and 
mixing angles via the relation (again ignoring right-handed quark mixing)
\begin{equation}
	\frac{\lambda_1^t \lambda_2^t f}{\sqrt{2} M_T} =
	\sum_j \frac{m_{u_j}}{v} V^{u *}_{j3}
	\simeq \frac{m_t}{v} V_{33}^{u *},
\end{equation}
where we have used $m_u, m_c \ll m_t$.
The couplings in Eq.~(\ref{eq:LTu}) will lead to the decays
$T \to t H$ and $T \to t \eta$.

The couplings of scalars to $D,S$ and one SM down-type quark are
\begin{equation}
	\mathcal{L} \simeq
	\frac{c_{\beta}}{\sqrt{2}} \lambda_2^d V_{i1}^{d *}
	(H D^c d_i)
	+ \frac{c_{\beta}}{\sqrt{2}} \lambda_2^s V_{i2}^{d *}
	(H S^c d_i)
	+ {\rm h.c.},
\end{equation}
where we have neglected terms involving $\lambda_b^{D^c}$
and $\lambda_b^{S^c}$ and ignored couplings of $\eta$ proportional 
to the down or strange quark masses.
These couplings will lead to the decays $D,S \to d_i H$.

The couplings of scalars to a pair of SM up-type quarks 
(again ignoring right-handed quark mixing) are
\begin{eqnarray}
	\mathcal{L} &=& (H u_i^c u_j) \left\{
	\delta_{ij} \frac{-m_{u_i}}{v} \left[ 1 - \frac{v^2}{6 f^2}
	\left( 3 + \frac{s_{\beta}^4}{c_{\beta}^2} 
		+ \frac{c_{\beta}^4}{s_{\beta}^2} \right)
	\right]
	+ \delta_{i3} \frac{m_t}{2 \sqrt{2} f} V^{u *}_{33} \Delta_{u_j}
	\left( \frac{c_{\beta}^2 - s_{\beta}^2}{s_{\beta} c_{\beta}} \right)
	\right. \nonumber \\ && \left. \qquad \qquad
	- \delta_{i3} V^{u *}_{j3} \frac{v m_t}{2f^2} V^{u *}_{33}
	+ \delta_{u^c_i} \Delta_{u_j} \frac{M_T}{v}
	\right\}
	\nonumber \\ &&
	+ (i\eta u^c_i u_j) \left[
	\delta_{ij} \frac{m_{u_i}}{\sqrt{2} f}
	\left( \frac{s_{\beta}^2 - c_{\beta}^2}{s_{\beta} c_{\beta}} \right)
	+ \delta_{i3} \Delta_{u_j}
	\frac{m_t}{v} V^{u*}_{33}
	\right]  + {\rm h.c.}
\end{eqnarray}
Note the flavor-changing couplings involving $t^c$ from terms containing
a $\delta_{i3}$.
Here we have introduced the notation $\delta_{u_i^c}$ for the mixings
between $u_i^c$ and $T^c$, which occur at order $v^2/f^2$.  
They are given explicitly by
\begin{eqnarray}
	\delta_{u^c} &=& -\frac{v}{M_T} \frac{f}{\sqrt{2} \Lambda} 
	\left( \delta_u \lambda_u^{u^c u^{\prime}}
		+ \delta_c \lambda_u^{u^c c^{\prime}} \right),
	\qquad
	\delta_{c^c} = -\frac{v}{M_T} \frac{f}{\sqrt{2} \Lambda} 
	\left( \delta_u \lambda_u^{c^c u^{\prime}}
		+ \delta_c \lambda_u^{c^c c^{\prime}} \right),
	\nonumber \\
	\delta_{t^c} &=& \frac{v}{M_T} \left\{ 
	\frac{m_t}{v} V^{u *}_{33}
	\left[ \Delta_t 
	+ \frac{v}{2 \sqrt{2} f} 
	\left( \frac{c_{\beta}^2 - s_{\beta}^2}{s_{\beta} c_{\beta}} \right)
	\right]
	- \frac{f}{\sqrt{2} \Lambda} \left(
	\delta_u \lambda_u^{t^c u^{\prime}}
	+ \delta_c \lambda_u^{t^c c^{\prime}}
	\right) \right\},
\end{eqnarray}
where $T^c = T^c_0 - \delta_{u_i^c} u_{i0}^c$.

The couplings of scalars to a pair of SM down-type quarks
(again ignoring right-handed quark mixing) are
\begin{eqnarray}
	\mathcal{L} &=& (H d_i^c d_j) \left\{
	\delta_{ij} \frac{-m_{d_i}}{v} \left[ 1 - \frac{v^2}{6f^2}
	\left( 3 + \frac{s_{\beta}^4}{c_{\beta}^2} 
		+ \frac{c_{\beta}^4}{s_{\beta}^2} \right) 
	\right]
	+ \frac{v^2}{2 f^2} 
	\left[ -\delta_{i1} \frac{m_d}{v} V^{d*}_{11} V^{d*}_{j1}
	  - \delta_{i2} \frac{m_s}{v} V^{d*}_{22} V^{d*}_{j2} \right]
	\right. \nonumber \\ && \left.
	+ \frac{v}{2 \sqrt{2} f} 
	\left( \frac{c_{\beta}^2 - s_{\beta}^2}{s_{\beta} c_{\beta}} \right)
	\left[ -\delta_{i1} \Delta_{Dd_j} \frac{m_d}{v} V^{d*}_{11} 
	- \delta_{i2} \Delta_{Sd_j} \frac{m_s}{v} V^{d*}_{22}
	\right]
	\right. \nonumber \\ && \left.
	+ \delta_{Dd_i^c} \Delta_{Dd_j} \frac{M_D}{v} 
	+ \delta_{Sd_i^c} \Delta_{Sd_j} \frac{M_S}{v}
	\right\}
	\nonumber \\ &&
	+ (i\eta d_i^c d_j) \left[
	\delta_{ij} \frac{-m_{d_i}}{\sqrt{2} f}
	\left( \frac{s_{\beta}^2 - c_{\beta}^2}{s_{\beta} c_{\beta}} \right)
	+ \delta_{i1} \Delta_{Dd_j} \frac{m_d}{v} V^{d*}_{11}
	+ \delta_{i2} \Delta_{Sd_j} \frac{m_s}{v} V^{d*}_{22}
	\right] + {\rm h.c.},
\end{eqnarray}
where we have used (neglecting right-handed quark mixing)
\begin{equation}
	\frac{\lambda_1^d \lambda_2^d f}{\sqrt{2} M_D}
	= -\frac{m_d}{v} V^{d *}_{11},
	\qquad \qquad
	\frac{\lambda_1^s \lambda_2^s f}{\sqrt{2} M_S}
	= -\frac{m_s}{v} V^{d *}_{22}.
\end{equation}
Note the flavor-changing couplings involving $d^c$ ($s^c$) from 
terms containing a $\delta_{i1}$ ($\delta_{i2}$).
We also introduce the notation $\delta_{Dd_i^c}$, $\delta_{Sd_i^c}$
for the mixings between $d_i^c$ and $D^c$, $S^c$, respectively, which
occur at order $v^2/f^2$.  They are given explicitly by
\begin{eqnarray}
	\delta_{Dd^c} &=& -\frac{v}{M_D} \left\{ 
	-\frac{m_d}{v} V^{d*}_{11}
	\left[ \delta_{Dd} - \frac{v}{2 \sqrt{2} f} 
	\left( \frac{c_{\beta}^2 - s_{\beta}^2}{s_{\beta} c_{\beta}} \right) 
		\right]
	+ \frac{f}{\sqrt{2} \Lambda} \delta_{Db} \lambda_b^{d^c}
	\right\},
	\nonumber \\
	\delta_{Ds^c} &=& -\frac{v}{M_D} \frac{f}{\sqrt{2} \Lambda}
	\delta_{Db} \lambda_b^{s^c},
	\qquad
	\delta_{Db^c} = -\frac{v}{M_D} \frac{f}{\sqrt{2} \Lambda}
	\delta_{Db} \lambda_b^{b^c},
	\nonumber \\
	\delta_{Ss^c} &=& -\frac{v}{M_S} \left\{
	-\frac{m_s}{v} V^d_{22}
	\left[ \delta_{Ss} - \frac{v}{2 \sqrt{2} f} 
	\left( \frac{c_{\beta}^2 - s_{\beta}^2}{s_{\beta} c_{\beta}} \right) 
		\right]
	+ \frac{f}{\sqrt{2} \Lambda} \delta_{Sb} \lambda_b^{s^c}
	\right\},
	\nonumber \\
	\delta_{Sd^c} &=& -\frac{v}{M_S} \frac{f}{\sqrt{2} \Lambda}
	\delta_{Sb} \lambda_b^{d^c},
	\qquad
	\delta_{Sb^c} = -\frac{v}{M_S} \frac{f}{\sqrt{2} \Lambda}
	\delta_{Sb} \lambda_b^{b^c},
\end{eqnarray}
where $D^c = D^c_0 - \delta_{Dd_i^c} d_{i0}^c$
and $S^c = S^c_0 - \delta_{Sd_i^c} d_{i0}^c$.

\subsubsection{Quark couplings to gauge bosons: anomaly-free embedding}
\label{sec:fermion-gauge}

The couplings of the heavy off-diagonal gauge bosons $X^{\mp}$,
$Y^0$ and $\overline{Y}^0$ to quarks in the anomaly-free embedding were
given in Table~\ref{tab:gauge2}, neglecting
flavor misalignment and CKM mixing.  Allowing for the flavor 
misalignment, we have\footnote{The SU(3) generators for the 
quarks of the first two generations, in
the antifundamental $\mathbf{\bar 3}$ representation,
are given by $-T^{a*}$.}
\begin{eqnarray}
	\mathcal{L}_{X,Y} &=& - \frac{g}{\sqrt{2}} \left\{
	i X^-_{\mu} \bar d_i \gamma^{\mu} \left[ V^d_{i3} T
		+ \left( \Delta_{u_j} V_{i3}^d
		+ \Delta^*_{Dd_i} V_{j1}^{u *}
		+ \Delta^*_{Sd_i} V_{j2}^{u *} \right) u_j \right]
	\right. \nonumber \\ && \left.
	+ i X^+_{\mu} \bar u_i \gamma^{\mu} V^u_{ij} D_j
	+ i Y^0_{\mu} \bar u_i \gamma^{\mu} \left( V^u_{i3} T 
		+ \Delta_{u_k} V^u_{i3} u_k \right)
	\right. \nonumber \\ && \left.
	+ i \overline{Y}^0_{\mu} \bar d_i \gamma^{\mu} \left[ V^d_{ij} D_j
		+ \left( \Delta_{Dd_k} V^d_{i1} + \Delta_{Sd_k} V^d_{i2}
			\right) d_k \right]
	+ {\rm h.c.} \right\}.
\end{eqnarray}
The couplings of $W^{\pm}$ to quark pairs, keeping
terms of order $v^2/f^2$ in interactions involving only SM particles
and terms of order $v/f$ in interactions involving one or more heavy
particles, are
\begin{eqnarray}
	\mathcal{L}_{W} &=& - \frac{g W^+_{\mu}}{\sqrt{2}} \left[
	\left( 1 - \frac{1}{2} |\Delta_{u_i}|^2 - \frac{1}{2} |\Delta_{Dd_j}|^2
		- \frac{1}{2} |\Delta_{Sd_j}|^2 \right) V_{ij}^{\rm CKM}
		\bar u_i \gamma^{\mu} d_j
	\right. \nonumber \\ && \left.
	- V_{ij}^{\rm CKM} \Delta_{u_i}^* \overline{T} \gamma^{\mu} d_j 
	- V_{ij}^{\rm CKM} \Delta_{Dd_j} \bar u_i \gamma^{\mu} D 
	- V_{ij}^{\rm CKM} \Delta_{Sd_j} \bar u_i \gamma^{\mu} S
	+ {\rm h.c.} \right].
\end{eqnarray}
The couplings of the $Z^{\prime}$ boson to quarks
were also given in Table~\ref{tab:gauge2}, neglecting
flavor misalignment and CKM mixing.  Allowing for the flavor 
misalignment, we find flavor-changing couplings for the left-handed
quarks involving $V_{i3}^u V_{3j}^{u \dagger}$ in the up sector
and $V_{i3}^d V_{3j}^{d \dagger}$ in the down sector:
\begin{eqnarray}
	\mathcal{L}_{Z^{\prime}} &\supset&
	- \frac{g}{c_W} \frac{Z^{\prime}_{\mu}}{\sqrt{3 - 4 s_W^2}} 
	\left[ 
	\left( -\frac{1}{2} + \frac{2}{3} s_W^2 \right)
		\left( \bar u_i \gamma^{\mu} u_i 
		+ \bar d_i \gamma^{\mu} d_i \right)
	\right. \nonumber \\ && \left.
	+ \left( 1 - s_W^2 \right)
	\left( V_{i3}^u V_{3j}^{u \dagger} \bar u_i \gamma^{\mu} u_j
		+ V_{i3}^d V_{3j}^{d \dagger} \bar d_i \gamma^{\mu} d_j \right)
	\right].
\label{eq:afZprime}
\end{eqnarray}
The couplings of the $Z$ boson to quarks, including the corrections
from mixing between $Z$ and $Z^{\prime}$ and mixing between the TeV-scale
quarks and their SM partners, are
\begin{eqnarray}
	\mathcal{L}_Z &=& - Z_{\mu} \frac{g}{c_W} \left\{
	\left( J_3^{\mu} - s^2_W J_Q^{\mu} \right)
	+ \frac{1}{2} \left[ - |\Delta_{u_i}|^2 \bar u_i \gamma^{\mu} u_i 
		+ \left( |\Delta_{Dd_i}|^2 + |\Delta_{Sd_i}|^2 \right)
		\bar d_i \gamma^{\mu} d_i 
		\right]
	\right. \nonumber \\ && \left.
	+ \frac{\delta_Z}{\sqrt{3 - 4 s_W^2}}
	\left[ \left( -\frac{1}{2} + \frac{2}{3} s_W^2 \right) 
		\left( \bar u_i \gamma^{\mu} u_i 
		+ \bar d_i \gamma^{\mu} d_i \right)
	- \frac{2}{3} s_W^2 \bar u_i^c \gamma^{\mu} u_i^c
	+ \frac{1}{3} s_W^2 \bar d_i^c \gamma^{\mu} d_i^c
	\right. \right. \nonumber \\ && \left. \left.
	+ \left( 1 - s_W^2 \right)
		\left( V^u_{i3} V^{u \dagger}_{3j} \bar u_i \gamma^{\mu} u_j
		+ V^d_{i3} V^{d \dagger}_{3j} \bar d_i \gamma^{\mu} d_j \right)
	\right]
	\right. \nonumber \\ && \left.
	+ \frac{1}{2} \left[ - \Delta_{u_i} \overline{T} \gamma^{\mu} u_i
		+ \Delta_{Dd_i} \overline{D} \gamma^{\mu} d_i 
		+ \Delta_{Sd_i} \overline{S} \gamma^{\mu} d_i 
		+ {\rm h.c.} \right] 
	\right\},
\label{eq:afZ}
\end{eqnarray}
where the leading-order coupling is given in terms of the standard fermion 
currents defined in Eq.~(\ref{eq:currents}).
The $Z$ boson couples to pairs of heavy quarks at order one through 
the electromagnetic current $J_Q$.
Note the flavor-changing couplings induced by $Z-Z^{\prime}$ mixing.
The couplings of photons to fermions are given by the electromagnetic 
current as usual.

\subsubsection{Constraints from flavor physics: anomaly-free embedding}

The flavor-changing couplings of $Z^{\prime}$ to quark pairs
can feed into low-energy observables, leading to 
potentially large flavor-changing neutral currents.  The contributions
of the anomaly-free fermion embedding to mixing in the neutral
$K$, $D$, $B$, and $B_s$ systems and the rare decays $B_{d,s} \to \mu^+\mu^-$
and $B \to K \mu^+\mu^-$ were summarized in Ref.~\cite{Sher} in the context
of 3-3-1 models without the little Higgs mechanism.
If the quark mixing matrices take a Fritzsch-like structure
\cite{GomezDumm:1994tz}, $V^{u,d}_{ij} = \sqrt{m_j/m_i}$ ($i\geq j$), 
then the strongest bound on the $Z^{\prime}$ mass comes from 
$B$--$\bar B$ mixing \cite{Long} and requires $M_{Z^{\prime}} > 10.5$ TeV
\cite{Sher}.  The next-most-stringent constraint comes from 
$B_s$--$\bar B_s$ mixing \cite{Sher} and requires $M_{Z^{\prime}} > 5.0$ TeV.
Clearly, the down quark mixing matrix must be more diagonal than
the Fritzsch-like structure, in order to suppress flavor-changing
effects in the down quark sector.  In fact, one can choose
$V^d_{i3} = \delta_{i3}$, so that the $d$ couplings are flavor-diagonal;
this \emph{eliminates} flavor-changing effects in the down quark sector.
The flavor-changing effects are then pushed into the up sector.
The $u$ and $d$ couplings to $Z^{\prime}$ can never both be flavor-diagonal
because they are related by the CKM matrix [Eq.~(\ref{eq:CKM})].

\subsubsection{Quark masses and mixing: universal embedding}
\label{UNmq}

In the universal embedding, the quark Yukawa Lagrangian is given 
for all three generations by
\begin{equation}
	\mathcal{L} = \lambda_1^{un} iu_1^{nc} \Phi_1^{\dagger} Q_n
	+ \lambda_2^{un} iu_2^{nc} \Phi_2^{\dagger} Q_n
	+ \frac{\lambda_d^{mn}}{\Lambda} id_m^c \epsilon_{ijk}
	\Phi_1^i \Phi_2^j Q_n^k
	+ {\rm h.c.},
\label{eq:universalYuk}
\end{equation}
where $m,n = 1,2,3$ are generation indexes; $i,j,k = 1,2,3$ are SU(3)
indexes; $d_m^c$ runs over all the down-type conjugate quarks
($d^c,s^c,b^c$); and $u_{1,2}^{nc}$ are linear combinations of 
the up-type conjugate quarks as given in Eqs.~(\ref{eq:Ucdef})
and (\ref{eq:ucdef}) below.

The physics of the down quark sector in the universal embedding is 
exactly analogous to that of the charged leptons.  The down quark 
Higgs couplings are given by
\begin{equation}
	\mathcal{L} = - \frac{m_{d_i}}{v} y_d (H d_i^c d_i)
	+ {\rm h.c.},
	\qquad \qquad
	y_d = 1 - \frac{v^2}{6 f^2} \left( 3 + \frac{c_{\beta}^4}{s_{\beta}^2}
		+ \frac{s_{\beta}^4}{c_{\beta}^2} \right),
\end{equation}
and their couplings to $\eta$ are given by
\begin{equation}
	\mathcal{L} = - \frac{m_{d_i}}{v} \frac{v}{4 \sqrt{2} f}
	\left( \frac{c_{\beta}^2 - s_{\beta}^2}{s_{\beta} c_{\beta}} \right)
	(i \eta d_i^c d_i) + {\rm h.c.}
\end{equation}
In the up sector,
the $f$ vevs generate mass terms for the three heavy quarks with
charge $+2/3$.
The three states 
\begin{equation}
	U^c_n = \frac{\lambda_1^{un} c_{\beta} u_1^{nc} 
	+ \lambda_2^{un} s_{\beta} u_2^{nc}}
	{\sqrt{(\lambda_1^{un})^2 c_{\beta}^2 
	+ (\lambda_2^{un})^2 s_{\beta}^2}},
\label{eq:Ucdef}
\end{equation}
marry the three $U_n$ states, giving them masses of
\begin{equation}
	M_{U_n} = f \sqrt{(\lambda_1^{un})^2 c_{\beta}^2 
	+ (\lambda_2^{un})^2 s_{\beta}^2}
\end{equation}
and leaving the orthogonal combinations of $u_1^{nc}$ and $u_2^{nc}$
massless:
\begin{equation}
	u^c_n = \frac{-\lambda_2^{un} s_{\beta} u_1^{nc}
	+ \lambda_1^{un} c_{\beta} u_2^{nc}}
	{\sqrt{(\lambda_1^{un})^2 c_{\beta}^2 
	+ (\lambda_2^{un})^2 s_{\beta}^2}}.
\label{eq:ucdef}
\end{equation}
Note that the Yukawa Lagrangian in Eq.~(\ref{eq:universalYuk}) does 
\emph{not} generate a misalignment between the SM up quark mass eigenstates
and the heavy quarks.  Such a misalignment could be generated by adding an
additional dimension-5 operator,
\begin{equation}
	\frac{\lambda_u^{mn}}{\Lambda} iu_m^c \epsilon_{ijk}
	\Phi_1^{* i} \Phi_2^{* j} Q_n^k + {\rm h.c.},
\end{equation}
to generate off-diagonal entries in the up quark mass matrix.  We ignore
this possibility here.  The usual CKM matrix is generated by the off-diagonal
entries in the down quark mass matrix, controlled by $\lambda_d^{mn}$.

After EWSB, the up quark mass terms are
\begin{eqnarray}
	\mathcal{L}_{\rm up \ mass} &=&
	- M_{U_n} U_n^c U_n
	+ \frac{v}{\sqrt{2}} 
	\frac{s_{\beta} c_{\beta} [(\lambda_1^{un})^2 - (\lambda_2^{un})^2]}
	{\sqrt{(\lambda_1^{un})^2 c_{\beta}^2 
		+ (\lambda_2^{un})^2 s_{\beta}^2}}
	U_n^c u_n
	\nonumber \\ &&
	- \frac{v}{\sqrt{2}} 
	\frac{\lambda_1^{un} \lambda_2^{un}}
	{\sqrt{(\lambda_1^{un})^2 c_{\beta}^2 
		+ (\lambda_2^{un})^2 s_{\beta}^2}} 
	u_n^c u_n
	+ {\rm h.c.}
\end{eqnarray}
These terms lead to mixing between the heavy quarks and their corresponding
SM quark partners.  As usual, we use the subscript $0$ to denote SU(3)
states; fields with no subscript denote the mass eigenstates after the
mixing induced by EWSB.  We can rewrite the SU(3) state $U_{m 0}$ in terms 
of the mass eigenstate $U_m$ and the SM fermion $u_m$ as
\begin{equation}
	U_{m 0} = U_m + \delta_{u_m} u_m,
	\qquad \qquad
	u_{m 0} = 
	\left( 1 - \frac{1}{2} \delta_{u_m}^2 \right) u_m
	- \delta_{u_m} U_m,
\end{equation}
where
\begin{equation}
	\delta_{u_m} = \frac{v}{\sqrt{2} f}
	\frac{s_{\beta} c_{\beta} [(\lambda_1^{um})^2 - (\lambda_2^{um})^2]}
	{[(\lambda_1^{um})^2 c_{\beta}^2 + (\lambda_2^{um})^2 s_{\beta}^2]}.
\end{equation}
The masses of the SM up-type quarks are given to leading order by
\begin{equation}
	\frac{\lambda_1^{um} \lambda_2^{um} f}{\sqrt{2} M_{U_m}}
	= \frac{m_{u_m}}{v}.
\end{equation}
The small mass of the $u$ ($c$) quark requires one of the couplings
$\lambda_{1,2}^{u1}$ ($\lambda_{1,2}^{u2}$) to be very small.  We choose
the small coupling to be $\lambda_1^{u1}$ ($\lambda_1^{u2}$) so that the
mixing effects in the up-quark sector are suppressed in the same
$t_{\beta} > 1$ limit as the mixing effects in the neutrino sector.
We then have,
\begin{equation}
	M_U = f \lambda_U s_{\beta},
	\qquad \qquad
	M_C = f \lambda_C s_{\beta},
	\qquad \qquad
	M_T = f \sqrt{\lambda_1^2 c_{\beta}^2 + \lambda_2^2 s_{\beta}^2},
\end{equation}
where we define $\lambda_U = \lambda_2^{u1}$, $\lambda_C = \lambda_2^{u2}$,
$\lambda_1 = \lambda_1^{u3}$, and $\lambda_2 = \lambda_2^{u3}$.
For the mixing angles we also have
\begin{equation}
	\delta_u = \delta_c = \frac{-v}{\sqrt{2} t_{\beta} f} = \delta_{\nu},
	\qquad \qquad
	\delta_t = \frac{v f}{\sqrt{2} M_T^2} 
	s_{\beta} c_{\beta} (\lambda_1^2 - \lambda_2^2).
\end{equation}

We now write the up quark couplings to scalars.  The
couplings of heavy quark-partner pairs are given by
\begin{eqnarray}
	\mathcal{L} &=& 
	-(i \eta U^c U) \frac{c_{\beta}}{\sqrt{2}} \lambda_U
	- (i \eta C^c C) \frac{c_{\beta}}{\sqrt{2}} \lambda_C
	+ (i \eta T^c T)
	s_{\beta} c_{\beta} ( \lambda_1^2 - \lambda_2^2 )
	\frac{f}{\sqrt{2} M_T}
	\nonumber \\ &&
	+ (H T^c T) \frac{v}{f} \left[
	(\lambda_1^2 s_{\beta}^2 + \lambda_2^2 c_{\beta}^2)
	\frac{f}{2 M_T}
	- s_{\beta}^2 c_{\beta}^2 (\lambda_1^2 - \lambda_2^2)^2
	\frac{f^3}{2 M_T^3}
	\right] + {\rm h.c.}
\end{eqnarray}
One would naively expect an $H U^c_m U_m$ coupling
for the first two generations
at order $v/f$ coming from replacing one Higgs field by its vev in 
the $HH U^c_m U_m$ term that is generated by the expansion of the nonlinear
sigma model; however, this term is exactly canceled by the contribution
from $H U^c_m u_m$ after $u-U$ mixing in the first two generations if the
up and charm quark masses are neglected.

The leading-order couplings of the scalars to one heavy quark partner
and one SM up-type quark are
\begin{equation}
	\mathcal{L} = 
	- (H U^c u) \frac{c_{\beta} \lambda_U}{\sqrt{2}}
	- (H C^c c) \frac{c_{\beta} \lambda_C}{\sqrt{2}}
	+ (H T^c t) ( \lambda_1^2 - \lambda_2^2 )
	\frac{s_{\beta} c_{\beta} f}{\sqrt{2} M_T}
	- (i \eta t^c T) \frac{m_t}{v}
	+ {\rm h.c.},
\end{equation}
where in the $\eta$ couplings 
we neglect $m_u$ and $m_c$ in the couplings of the 
first two generations and neglect the $v/f$ suppressed coupling of the
third generation.  These couplings will lead to the decays
$U_m \to u_m H$ and $T \to t \eta$.

The couplings of scalars to a pair of SM up-type quarks are
\begin{eqnarray}
	\mathcal{L} &=&
	(H u_i^c u_i) \left\{ \frac{-m_{u_i}}{v} \left[
	1 - \frac{v^2}{6 f^2} \left( \frac{s_{\beta}^4}{c_{\beta}^2}
		+ \frac{c_{\beta}^4}{s_{\beta}^2} \right)
	- \delta_{u_i} \frac{v}{2\sqrt{2}f} 
	\left( \frac{c_{\beta}^2 - s_{\beta}^2}{s_{\beta}c_{\beta}} \right)
	\right]
	+ \frac{M_{U_i}}{v} \delta_{u_i} \delta_{u_i^c} \right\}
	\nonumber \\ &&
	+ (i\eta u^c_i u_i) \frac{m_{u_i}}{v}
	\left[ \frac{v}{\sqrt{2}f} 
	\left( \frac{s_{\beta}^2-c_{\beta}^2}{s_{\beta}c_{\beta}} \right)
	+ \delta_{u_i} \right]  + {\rm h.c.},
\end{eqnarray}
where the mixing between $u_i^c$ and $U_i^c$ at order $v^2/f^2$ is given by
$U_i^c = U_{i0}^c - \delta_{u_i^c} u_{i0}^c$, with
\begin{equation}
        \delta_{u_i^c} = \frac{m_{u_i}}{M_{U_i}}
	\left[ \delta_{u_i} + \frac{v}{2\sqrt{2}f}
	  \left( \frac{c_{\beta}^2 - s_{\beta}^2}{s_{\beta}c_{\beta}} \right)
	\right].
\end{equation}

\subsubsection{Quark couplings to gauge bosons: universal embedding}

The couplings of the $Z^{\prime}$ boson to quarks in the universal
embedding were given in Table~\ref{tab:gauge2}.
These couplings are purely flavor-diagonal in the universal fermion
embedding.
The couplings of the heavy off-diagonal gauge bosons $X^-$ and
$Y^0$ to quarks in the universal embedding were also
given in Table~\ref{tab:gauge2}, neglecting
CKM mixing.  Keeping the full CKM dependence, we have
\begin{equation}
	\mathcal{L}_{X,Y} =
	- \frac{g}{\sqrt{2}} \left[
	i X^-_{\mu} \bar d_i \gamma^{\mu} \left( V_{ji}^{\rm CKM *} U_j 
		+ \delta_{u_j} V_{ji}^{\rm CKM *} u_j \right)
	+ i Y^0_{\mu} \bar u_i \gamma^{\mu} \left( U_i 
		+ \delta_{u_i} u_i \right)
	+ \rm {h.c.} \right].
\end{equation}
The couplings of $W^{\pm}$ to quark pairs, keeping
terms of order $v^2/f^2$ in interactions involving only SM particles
and terms of order $v/f$ in interactions involving one or more heavy
particles, are
\begin{equation}
	\mathcal{L}_W =
	- \frac{g W^+_{\mu}}{\sqrt{2}} \left[
	\left( 1 - \frac{1}{2} \delta_{u_i}^2 \right)
		V_{ij}^{\rm CKM} \bar u_i \gamma^{\mu} d_j 
	- \delta_{u_i} V_{ij}^{\rm CKM} \bar U_i \gamma^{\mu} d_j 
	+ \rm {h.c.} \right].
\end{equation}
The couplings of the $Z$ boson to quarks, including the corrections
from mixing between $Z$ and $Z^{\prime}$ and mixing between the TeV-scale
quarks and their SM partners, are
\begin{eqnarray}
	\mathcal{L}_{Z} &=&
	- \frac{g Z_{\mu}}{c_W} \left\{
	\left( J_3^{\mu} - s_W^2 J_Q^{\mu} \right)
	- \frac{1}{2} \delta_{u_i}^2 \bar u_i \gamma^{\mu} u_i
	- \frac{1}{2} \left[
		\delta_{u_i} \overline{U}_i \gamma^{\mu} u_i
		+ {\rm h.c.} \right] 
	\right.  \\ &+& \left.
	\frac{\delta_Z}{\sqrt{3 - 4 s_W^2}}
	\left[ \left( \frac{1}{2} - \frac{1}{3} s_W^2 \right)
		\left( \bar u_i \gamma^{\mu} u_i 
		+ \bar d_i \gamma^{\mu} d_i \right)
	- \frac{2}{3} s_W^2 \bar u_i^c \gamma^{\mu} u_i^c 
	+ \frac{1}{3} s_W^2 \bar d_i^c \gamma^{\mu} d_i^c \right]
	\right\},  \nonumber
\end{eqnarray}
where the leading-order coupling is given in terms of the usual fermion
currents $J_3$ and $J_Q$ defined in Eq.~(\ref{eq:currents}).
The $Z$ boson couples to pairs of heavy quarks $U_i$ at order one
through the electromagnetic current $J_Q$.  
The couplings of photons to fermions are given by the electromagnetic 
current as usual.

\subsection{Higgs potential}
\label{sec:Higgspotential}

In this section we describe the generation of the Higgs 
potential.\footnote{We thank Martin Schmaltz for very helpful discussions.}
Additional details can be found in Refs.~\cite{Schmaltznote,KRR}.
We start with the Coleman-Weinberg potential that is generated
by loops of gauge bosons and fermions in the running down from the cutoff scale
$\Lambda$.  Above the global symmetry breaking scale $f$, only operators
that are symmetric under the global [SU(3)$\times$U(1)]$^2$ symmetry are
generated by the running.  The three allowed operators up to dimension four
are
\begin{equation}
	\Phi_1^{\dagger} \Phi_1, \qquad
	\Phi_2^{\dagger} \Phi_2, \qquad
	|\Phi_1^{\dagger} \Phi_2 |^2.
\end{equation}
The first two of these operators are just constants and do not involve 
the Goldstone bosons.  We therefore focus on the third operator.  Expanding it
in terms of the Goldstone bosons to fourth order gives
\begin{equation}
	|\Phi_1^{\dagger} \Phi_2|^2 =
	f^4 s_{\beta}^2 c_{\beta}^2 - f^2 h^{\dagger} h
	+ \frac{1}{3 s_{\beta}^2 c_{\beta}^2} (h^{\dagger} h)^2
	+ \frac{3}{32 s_{\beta}^2 c_{\beta}^2} h^{\dagger} h \eta^2
	+ \mathcal{O}(\phi^6).
\end{equation}
Running below the global symmetry breaking scale $f$ can give contributions
to the Coleman-Weinberg potential that are not proportional to 
$|\Phi_1^{\dagger} \Phi_2|^2$.  These contributions will contain
logs of the ratios of masses-squared of $f$-scale particles and the
corresponding SM particles.  They will therefore be calculable, i.e.,
independent of cutoff-scale physics.

The Coleman-Weinberg potential from the $X^-$, $Y^0$ and $W^+$ gauge 
bosons is,
\begin{eqnarray}
	V_2 &=& \frac{3}{64 \pi^2} g^4 {\rm log} (\Lambda^2/M_X^2) f^2 
	(h^{\dagger} h) 
	\nonumber \\
	V_4 &=& \frac{3}{64 \pi^2} g^4 {\rm log} (\Lambda^2/M_X^2)
	\left[ -\frac{1}{3 s_{\beta}^2 c_{\beta}^2} (h^{\dagger} h)^2
	- \frac{3}{32 s_{\beta}^2 c_{\beta}^2} (h^{\dagger} h) \eta^2 \right]
	\nonumber \\ &&
	- \frac{3}{128 \pi^2} g^4 {\rm log} (M_X^2/M_W^2) (h^{\dagger} h)^2.
\end{eqnarray}
Here $V_2$ and the first line of $V_4$ come from running between $\Lambda$ 
and $M_X$ and are proportional to $|\Phi_1^{\dagger} \Phi_2|^2$, 
while the second line of $V_4$ comes from running between $M_X$ and $M_W$.
The running below $M_X$ contributes only a term involving
$(h^{\dagger} h)^2$.  It does not contribute any terms involving $\eta$
since there is no coupling of $W$ boson pairs to $h \eta$.

The Coleman-Weinberg potential from the $Z^{\prime}$ and $Z$ gauge bosons
is,
\begin{eqnarray}
	V_2 &=& \frac{3}{32 \pi^2} g^4 \frac{1 + t_W^2}{3 - t_W^2} 
	{\rm log} (\Lambda^2/M^2_{Z^{\prime}}) f^2 (h^{\dagger} h)
	\nonumber \\
	V_4 &=& \frac{3}{32 \pi^2} g^4 \frac{1 + t_W^2}{3 - t_W^2} 
	{\rm log} (\Lambda^2/M^2_{Z^{\prime}})
	\left[ -\frac{1}{3 s_{\beta}^2 c_{\beta}^2} (h^{\dagger} h)^2
	- \frac{3}{32 s_{\beta}^2 c_{\beta}^2} (h^{\dagger} h) \eta^2 \right]
	\nonumber \\ &&
	- \frac{3}{256 \pi^2} g^4 (1 + t_W^2)^2 
	{\rm log}(M^2_{Z^{\prime}}/M^2_Z)
	(h^{\dagger} h)^2.
\end{eqnarray}
Again, $V_2$ and the first line of $V_4$ come from running between $\Lambda$ 
and $M_{Z^{\prime}}$ and are proportional to $|\Phi_1^{\dagger} \Phi_2|^2$, 
while the second line of $V_4$ comes from running between $M_{Z^{\prime}}$ 
and $M_Z$.
The running below $M_{Z^{\prime}}$ contributes only a term involving
$(h^{\dagger} h)^2$.  It does not contribute any terms involving $\eta$
since there is no coupling of $Z$ boson pairs to $h \eta$.

The Coleman-Weinberg potential from the fermions can in principle come from
loops of any fermion with an order-one Yukawa coupling.  However, due to 
the feature of collective breaking in the model, the order-one Yukawa 
couplings that give mass to the neutrino partners and the quark
partners of the first two generations do not contribute to the terms of the
Coleman-Weinberg potential involving the Goldstone bosons (neglecting the tiny
Yukawa couplings of the quarks of the first two generations).  
The only significant contribution is then due to the top quark and its
partner $T$.  In what follows we neglect the mixing between quark generations.
The Coleman-Weinberg potential from the top quark and its partner $T$ is,
\begin{eqnarray}
	V_2 &=& - \frac{3}{8 \pi^2} \lambda_t^2 M_T^2 
	{\rm log} (\Lambda^2/M_T^2) (h^{\dagger} h)
	\nonumber \\
	V_4 &=& - \frac{3}{8 \pi^2} \lambda_t^2 \frac{M_T^2}{f^2}
	{\rm log} (\Lambda^2/M_T^2) 
	\left[ -\frac{1}{3 s_{\beta}^2 c_{\beta}^2} (h^{\dagger} h)^2
	- \frac{3}{32 s_{\beta}^2 c_{\beta}^2} (h^{\dagger} h) \eta^2 \right]
	\nonumber \\ &&
	+ \frac{3}{16 \pi^2} \lambda_t^4 {\rm log} (M_T^2/m_t^2) 
	(h^{\dagger} h)^2,
\end{eqnarray}
where $\lambda_t \equiv \lambda_1^t \lambda_2^t f / M_T \simeq \sqrt{2} m_t/v$.
Again, $V_2$ and the first line of $V_4$ come from running between $\Lambda$ 
and $M_T$ and are proportional to $|\Phi_1^{\dagger} \Phi_2|^2$, 
while the second line of $V_4$ comes from running between $M_T$
and $m_t$.
The running below $M_T$ contributes only a term involving
$(h^{\dagger} h)^2$.  It does not contribute any terms involving $\eta$
since there is no coupling of top quark pairs to $h \eta$ or $\eta^2$.

Collecting terms, we can write the Coleman-Weinberg potential as follows:
\begin{equation}
	V = - m^2 h^{\dagger} h + \lambda (h^{\dagger} h)^2
	+ \lambda^{\prime} h^{\dagger} h \eta^2,
\end{equation}
where
\begin{eqnarray}
	m^2 &=& \frac{3}{8 \pi^2} \left[
	\lambda_t^2 M_T^2 {\rm log} (\Lambda^2/M_T^2)
	- \frac{g^2}{4} M_X^2 {\rm log} (\Lambda^2/M_X^2)
	- \frac{g^2}{8} (1 + t_W^2) M^2_{Z^{\prime}} 
	{\rm log} (\Lambda^2/M^2_{Z^{\prime}}) \right]
	\nonumber \\
	\lambda &=& \frac{1}{3 s_{\beta}^2 c_{\beta}^2} \frac{m^2}{f^2}
	+ \frac{3}{16 \pi^2} \left[
	\lambda_t^4 {\rm log} (M_T^2/m_t^2)
	- \frac{g^4}{8} {\rm log} (M_X^2/M_W^2)
	- \frac{g^4}{16} (1 + t_W^2)^2 
	{\rm log} (M^2_{Z^{\prime}}/M^2_Z) \right]
	\nonumber \\
	\lambda^{\prime} &=& \frac{3}{32 s_{\beta}^2 c_{\beta}^2} 
	\frac{m^2}{f^2}.
\end{eqnarray}
In the expression for $m^2$, in principle
the cutoff $\Lambda$ in the term generated by quark loops 
can be different from the cutoff $\Lambda$ in the two terms generated 
by gauge boson loops, because the physics that cuts off the quark loops
can be different from the physics that cuts off the gauge boson loops.
After EWSB, $\eta$ gets a small positive mass-squared of 
order $m_H^2 v^2/f^2$ from the $\lambda^{\prime}$ term.
The Higgs vev and mass are given by
\begin{equation}
	v^2 = m^2 / \lambda = (246 \ {\rm GeV})^2, \qquad
	m^2_H = 2 m^2 = 2 \lambda v^2.
\end{equation}
It turns out that this $m_H$ is \emph{too small}, because the quartic
coupling $\lambda$ is not big enough compared to $m^2$.

Following Ref.~\cite{Schmaltznote}, 
this problem can be fixed by adding a new operator, 
$\Phi_1^{\dagger} \Phi_2 + {\rm h.c.}$, to the scalar potential with
a coefficient $-\mu^2$ set by hand.  This operator breaks
the global SU(3)$^2$ down to the diagonal SU(3) while preserving
the gauged SU(3).  Expanding this operator to fourth order in the Goldstone
bosons gives
\begin{equation}
	\Phi_1^{\dagger} \Phi_2 + {\rm h.c.} =
	2 f^2 s_{\beta} c_{\beta} + \frac{1}{f^2 s_{\beta} c_{\beta}} \left[
	- f^2 (h^{\dagger} h) - \frac{f^2 \eta^2}{2} 
	+ \frac{(h^{\dagger} h)^2}{12 s_{\beta}^2 c_{\beta}^2} 
	+ \frac{3 (h^{\dagger} h) \eta^2}{32 s_{\beta}^2 c_{\beta}^2} 
	+ \frac{\eta^4 }{48 s_{\beta}^2 c_{\beta}^2} \right].
\end{equation}
Because the $(h^{\dagger} h)$ and $(h^{\dagger} h)^2$ terms in this
operator have different relative coefficients than in the original operator
$|\Phi_1^{\dagger} \Phi_2|^2$, it can be used to cancel
off part of the $m^2 h^{\dagger} h$ term without canceling too much of 
the $\lambda (h^{\dagger} h)^2$ term.  Adding the term 
$- \mu^2 (\Phi_1^{\dagger} \Phi_2 + {\rm h.c.})$ to the potential gives
\begin{equation}
	V = - m^2_{\rm new} h^{\dagger} h + \frac{1}{2} m_{\eta}^2 \eta^2
	+ \lambda_{\rm new} (h^{\dagger} h)^2
	+ \lambda^{\prime}_{\rm new} h^{\dagger} h \eta^2
	+ \lambda^{\prime\prime}_{\rm new} \eta^4,
\end{equation}
where
\begin{eqnarray}
	m^2_{\rm new} &=& m^2 - \frac{\mu^2}{s_{\beta}c_{\beta}},
	\qquad \qquad \qquad
	m^2_{\eta} = \frac{\mu^2}{s_{\beta}c_{\beta}},
	\nonumber \\
	\lambda_{\rm new} &=& 
		\frac{1}{3 s_{\beta}^2 c_{\beta}^2} \frac{m^2_{\rm new}}{f^2}
		+ \frac{1}{4 s_{\beta}^3 c_{\beta}^3} \frac{\mu^2}{f^2}
		\nonumber \\ &&
		+ \frac{3}{16 \pi^2} \left[
		\lambda_t^4 {\rm log} (M_T^2/m_t^2)
		- \frac{g^4}{8} {\rm log} (m_X^2/m_W^2)
		- \frac{g^4}{16} (1 + t_W^2)^2 
		{\rm log} (m^2_{Z^{\prime}}/m^2_Z) \right],
	\nonumber \\
	\lambda^{\prime}_{\rm new} &=& 
		\frac{3}{32 s_{\beta}^2 c_{\beta}^2} \frac{m^2_{\rm new}}{f^2},
	\qquad \qquad \qquad
	\lambda^{\prime\prime}_{\rm new} = 
		- \frac{1}{48 s_{\beta}^3 c_{\beta}^3} \frac{\mu^2}{f^2}.
\end{eqnarray}
Note that this term has also given rise to a mass-squared term for $\eta$ 
and an $\eta^4$ coupling.  The $\eta$ mass $m_{\eta}$ is now of order 
$\mu$, parametrically larger than the $\eta$ mass term generated by EWSB.


\end{document}